\documentclass[aps,prb,10pt,twocolumn,showpacs,preprintnumbers,amsmath,superscriptaddress,letter]{revtex4-1}
\usepackage{lipsum}
\usepackage{lineno}
\usepackage{graphicx}
\usepackage{dcolumn}
\usepackage{bm}
\usepackage{amssymb}
\usepackage{amsmath}
\usepackage{booktabs}
\newcolumntype{C}{>{$}c<{$}}
\AtBeginDocument{
	\heavyrulewidth=.08em
	\lightrulewidth=.05em
	\cmidrulewidth=.03em
	\belowrulesep=.65ex
	\belowbottomsep=0pt
	\aboverulesep=.4ex
	\abovetopsep=0pt
	\cmidrulesep=\doublerulesep
	\cmidrulekern=.5em
	\defaultaddspace=.5em
}
\usepackage{blkarray, multirow, graphicx, diagbox, color, xcolor, colortbl}
\usepackage{bbm, bbold}
\usepackage{ifthen}
\usepackage{xkeyval}
\usepackage{moreverb}
\usepackage{rotating}
\usepackage{slashbox}
\usepackage{xspace}
\usepackage{nicefrac}
\usepackage[]{units}
\usepackage[]{natbib}

\usepackage{ifthen}
\usepackage{braket}

\usepackage[acronym,nohypertypes={acronym,notation}]{glossaries}
\usepackage{hyphenat}

\newacronym{1D}{1D}{one\hyp dimensional}
\newacronym{MPS}{MPS}{matrix\hyp product state}
\newacronym{MPO}{MPO}{matrix\hyp product operator}
\newacronym{SVD}{SVD}{singular\hyp value decomposition}
\newacronym{QCS}{QCS}{quantum\hyp computer simulator}
\newacronym{QC}{QC}{quantum computer}
\newacronym{FSM}{FSM}{finite\hyp state machine}
\newacronym{ACA}{ACA}{adaptive cross\hyp approximation}
\newacronym{CDW}{CDW}{charge\hyp density wave}
\newacronym{CDP}{CDP}{charge\hyp density pattern}
\newacronym{SDP}{SDP}{spin\hyp density pattern}
\newacronym{LL}{LL}{Luttinger liquid}
\newacronym{SDW}{SDW}{spin\hyp density wave}
\newacronym{ARPES}{ARPES}{angle-resolved photoemission spectroscopy}
\newacronym{OBC}{OBC}{open-boundary conditions}
\newacronym{PBC}{PBC}{periodic-boundary conditions}
\newacronym{TEBD}{TEBD}{time-evolution block-decimation}
\newacronym{iff}{iff}{if and only if}
\newacronym{DFT}{DFT}{density\hyp functional theory}
\newacronym{DMFT}{DMFT}{dynamical mean\hyp field theory}
\newacronym{DMRG}{DMRG}{density\hyp matrix renormalization group}
\newacronym{tDMRG}{tDMRG}{time\hyp dependent \gls{DMRG}}
\newacronym{PS-DMRG}{PS-DMRG}{pseudo\hyp site density\hyp matrix renormalization group}
\newacronym{PP-2DMRG}{PP-2DMRG}{projected purified two-site density\hyp matrix renormalization group}
\newacronym{DMRG3S+LBO}{DMRG3S+LBO}{strictly single-site density\hyp matrix renormalization group with local basis optimization}
\newacronym{QMC}{QMC}{quantum Monte Carlo}
\newacronym{AIM}{AIM}{Anderson impurity model}
\newacronym{SIAM}{SIAM}{single impurity Anderson model}
\newacronym{LDA}{LDA}{local\hyp density approximation}
\newacronym{LBNL}{LBNL}{Lawrence Berkeley National Laboratory}
\newacronym{VQE}{VQE}{variational\hyp quantum eigensolver}
\newacronym{ED}{ED}{exact diagonalization}
\newacronym{QPT}{QPT}{quantum phase transition}
\newacronym{QCP}{QCP}{quantum critical point}
\newacronym{ETH}{ETH}{eigenstate thermalization hypothesis}
\newacronym{AKLT}{AKLT}{Affleck\hyp Lieb\hyp Kennedy\hyp Tasaki}
\newglossaryentry{QR}{name={QR},description={QR decomposition}}
\newacronym{TNS}{TNS}{tensor\hyp network state}
\newacronym{OISTR}{OISTR}{optically induced spin transfer}

\usepackage[caption=false]{subfig}
\usepackage{tikz}
\usepackage{calc}
\usetikzlibrary{external}
\usepackage{pgffor}
\usepackage{pgfplots}
\pgfplotsset{compat=1.8}
\usepackage{pgfplotstable}
\usepgfplotslibrary{groupplots}

\definecolor{colorA}{rgb} {0.58,0,0.8275}
\definecolor{colorB}{rgb} {0.11,0.663,0.51}
\definecolor{colorC}{rgb} {0.3373,0.7059,0.9137}
\definecolor{colorD}{rgb} {0.902,0.6235,0}
\definecolor{colorE}{rgb} {0.9451,0.902,0.3255}

\usetikzlibrary
{
	calc,
	patterns,
	positioning,
	petri,
	arrows,
	decorations.markings,
	backgrounds,
	fit,
	graphs,
	shapes.geometric,
	decorations.pathmorphing,
	shapes.misc,
	shapes,
	tikzmark
}

\newcommand{\hopping}[0]{t_{\text{hop}}}
\newcommand{\nodagger}[0]{\phantom{\dagger}}

\pgfmathdeclarefunction{peierlspotential}{6}%
{% 	plot with: \addplot[black] {peierls_potential(R, e0, k0, t0, c, s)};
	\pgfmathparse
	{
		#2 / (#3 * #5) 
		* sin(deg(#1 * #3 - #3 * #5 * (x)))
		* exp(-( ( #1 - #5 * ((x) - #4) ) )^2.0 / ( #6^2.0 ) )
	}%
}
\pgfmathdeclarefunction{bandstructure}{4}% 1: Delta; 2: first sign; 3: second sign; 4: x
{%
	\pgfmathparse{#2*sqrt(2.0+pow(#1/2.0, 2)+#3*2.0*sqrt((pow(cos(deg(3.1415926535/4.0*#4)), 2)+pow(#1/2.0, 2))))}%
}

\newcommand{\bspWidth}{0.6}
\newcommand{\bspArrowSize}{0.4}
\newcommand{\Zeeman}{0.75}
\newcommand{\yshift}[1]
{
	ifthenelse(mod(#1,2)<1,0,1)*\Zeeman
}
\newcommand{\bspDelta}[1]
{
	\node[inner sep = 0, outer sep = 0] (m#1b) at (#1*\bspWidth,0) {};
	\node[inner sep = 0, outer sep = 0] (m#1t) at (#1*\bspWidth,\Zeeman) {};
	\draw[thick,|<->|] (m#1b.south) -- (m#1t.north) node[midway, left=-1pt] {$\Delta$};
	
	\node[xshift=-0.25em,anchor=east] at (m#1b) {$E_1$};
	\node[xshift=-0.25em,anchor=east] at (m#1t) {$E_2$};
}

\newcommand{\bspNone}[1]
{
	\draw (#1*\bspWidth*1.5,{\yshift{#1}}) -- coordinate[midway] (m#1) (#1*\bspWidth*1.5+\bspWidth,{\yshift{#1}});
}
\newcommand{\bspUp}[1]
{
	\draw (#1*\bspWidth*1.5,{\yshift{#1}}) --  (#1*\bspWidth*1.5+\bspWidth,{\yshift{#1}});
	\draw[->, thick] (#1*\bspWidth*1.5+0.5*\bspWidth,{\yshift{#1}-\bspArrowSize*0.5}) --  (#1*\bspWidth*1.5+0.5*\bspWidth,{\yshift{#1}+\bspArrowSize*0.5}) coordinate (m#1);
}
\newcommand{\bspDown}[1]
{
	\draw (#1*\bspWidth*1.5,{\yshift{#1}}) -- (#1*\bspWidth*1.5+\bspWidth,{\yshift{#1}});
	\draw[->, thick] (#1*\bspWidth*1.5+0.5*\bspWidth,{\yshift{#1}+\bspArrowSize*0.5}) coordinate (m#1) -- (#1*\bspWidth*1.5+0.5*\bspWidth,{\yshift{#1}-\bspArrowSize*0.5});
}
\newcommand{\bspUpDown}[1]
{
	\draw (#1*\bspWidth*1.5,{\yshift{#1}}) -- (#1*\bspWidth*1.5+\bspWidth,{\yshift{#1}});
	\draw[->, thick] (#1*\bspWidth*1.5+0.3*\bspWidth,{\yshift{#1}-\bspArrowSize*0.5}) -- (#1*\bspWidth*1.5+0.3*\bspWidth,{\yshift{#1}+\bspArrowSize*0.5}) coordinate (m#1a); 
	\draw[->, thick] (#1*\bspWidth*1.5+0.7*\bspWidth,{\yshift{#1}+\bspArrowSize*0.5}) coordinate (m#1b) -- (#1*\bspWidth*1.5+0.7*\bspWidth,{\yshift{#1}-\bspArrowSize*0.5});
	\node (m#1) at ($(m#1a)!0.5!(m#1b)$) {};
}

\newcommand{\bspUpH}[1]
{
	\draw (#1*\bspWidth*1.5,{\yshift{#1}}) --  (#1*\bspWidth*1.5+\bspWidth,{\yshift{#1}});
	\draw[->, thick, densely dotted] (#1*\bspWidth*1.5+0.5*\bspWidth,{\yshift{#1}-\bspArrowSize*0.5}) -- (#1*\bspWidth*1.5+0.5*\bspWidth,{\yshift{#1}+\bspArrowSize*0.5}) coordinate (m#1);
}
\newcommand{\bspDownH}[1] 
{
	\draw (#1*\bspWidth*1.5,{\yshift{#1}}) -- (#1*\bspWidth*1.5+\bspWidth,{\yshift{#1}});
	\draw[->, thick, densely dotted] (#1*\bspWidth*1.5+0.5*\bspWidth,{\yshift{#1}+\bspArrowSize*0.5}) coordinate (m#1) -- (#1*\bspWidth*1.5+0.5*\bspWidth,{\yshift{#1}-\bspArrowSize*0.5});
}

\usepackage{xparse}

\ExplSyntaxOn
\DeclareExpandableDocumentCommand \eval { m } { \fp_eval:n { #1 } }
\ExplSyntaxOff

\usepackage{ulem}
\usepackage[colorlinks, linkcolor = blue, citecolor = blue, filecolor = black, urlcolor = blue]{hyperref}

\usepackage{cleveref}
\Crefname{appendix}{Appendix}{Appendices}
\Crefname{equation}{Equation}{Equations}
\Crefname{figure}{Figure}{Figures}
\Crefname{section}{Section}{Sections}
\Crefname{tabular}{Tabular}{Tabulars}
\crefname{appendix}{App.}{Apps.}
\crefname{equation}{Eq.}{Eqs.}
\crefname{figure}{Fig.}{Figs.}
\crefname{section}{Sec.}{Secs.}
\crefname{tabular}{Tab.}{Tabs.}

\newboolean{buildtikzpics}
\setboolean{buildtikzpics}{true}

\newboolean{simpletrue}
\newboolean{simplefalse}
\setboolean{simpletrue}{true}
\setboolean{simplefalse}{false}

\begin{document}
\title{Formation of spatial patterns by spin-selective excitations of interacting fermions}

\author{T. K\"ohler}
\affiliation{Institut f\"ur Theoretische Physik, Georg-August-Universit\"at G\"ottingen, 37077 G\"ottingen, Germany}
\affiliation{Department of Physics and Astronomy, Uppsala University, Box 516, S-751 20 Uppsala, Sweden}
\author{S. Paeckel}
\affiliation{Institut f\"ur Theoretische Physik, Georg-August-Universit\"at G\"ottingen, 37077 G\"ottingen, Germany}
\affiliation{Department of Physics, Arnold Sommerfeld Center for Theoretical Physics (ASC),
Munich Center for Quantum Science and Technology (MCQST), Fakult\"{a}t f\"{u}r Physik,
Ludwig-Maximilians-Universit\"{a}t M\"{u}nchen, 80333 M\"{u}nchen, Germany}
\author{C. Meyer}
\affiliation{Institut f\"ur Theoretische Physik, Georg-August-Universit\"at G\"ottingen, 37077 G\"ottingen, Germany}
\author{S.R. Manmana}
\affiliation{Institut f\"ur Theoretische Physik, Georg-August-Universit\"at G\"ottingen, 37077 G\"ottingen, Germany}

\date{\today}

\begin{abstract}
We describe the formation of charge- and spin-density patterns induced by spin-selective photoexcitations of interacting fermionic systems in the presence of a microstructure. 
As an example, we consider a one-dimensional Hubbard-like system with a periodic magnetic microstructure, which has a uniform charge distribution in its ground state, and in which a long-lived charge-density pattern is induced by the spin-selective photoexcitation. 
Using tensor-network methods, we study the full quantum dynamics in the presence of electron-electron interactions and identify doublons as the main decay channel for the induced charge pattern. 
Our setup is compared to the \gls{OISTR} mechanism, in which ultrafast optically induced spin transfer in Heusler and magnetic compounds is associated to the difference of the local density of states of the different elements in the alloys. We find that applying a spin-selective excitation there induces spatially periodic patterns in local observables.
Implications for pump-probe experiments on correlated materials and experiments with ultracold gases on optical lattices are discussed.  
\end{abstract}

\maketitle

\section{Introduction}
The emergence of order in nonequilibrium quantum systems has inspired a lot of experimental and theoretical research.
Examples are the recent observation of so-called time-crystal phases in Floquet-driven systems~\cite{Zhang2017,Choi2017,PhysRevLett.116.250401,0034-4885-81-1-016401,PhysRevLett.117.090402,PhysRevLett.118.030401,PhysRevLett.118.269901,PhysRevB.94.085112,PhysRevX.7.011026,Moessner2017}, as well as the experimental finding of metastable, ordered states following a photoexcitation using ultrashort laser pulses in pump-probe setups \cite{RevModPhys.81.163,pumpprobe7,pumpprobe6,pumpprobe5, PhysRevLett.118.107402,PhysRevLett.118.116402,1367-2630-18-9-093028,Tao62,Rini2007,Hu2014,Ropers1,Fausti189,Mitrano2016,doi:10.1002/9780470022184.hmm315,ncomms10459,PRLChromium,Schmitt1649,doublondynamics,Rohwer2011,Hellmann2012,Mathias2016,Stojchevska177}.
In these experiments, the possible observation of transient superconducting states at elevated temperatures~\cite{Fausti189,Mitrano2016} or the transformation of \gls{CDW} states has been reported \cite{ncomms10459,PRLChromium,Schmitt1649,Rohwer2011,Hellmann2012,PhysRevLett.118.247401}.
This includes optically driven transitions between ordered states \cite{Ropers2}, enhancement of existent order \cite{PRLChromium}, or its melting \cite{Schmitt1649,PhysRevLett.102.066404,Rohwer2011,Hellmann2012,PhysRevB.84.180507} due to an excitation. 
Identifying theoretical mechanisms predicting the behavior in such nonequilibrium situations is a major challenge and topic of ongoing research~\cite{review_pumpprobetheory,review_trARPEStheory,epjst_072013_222_5_1065-1075,PhysRevLett.112.176404,PhysRevLett.120.246402,photoinducedchiralspinliquid,PhysRevB.94.035121,PhysRevLett.119.247601,PhysRevLett.116.086401,photoexcitation_ionichubbardmodel,Sentef2015,PhysRevB.98.035138,PhysRevB.89.045106,PhysRevB.88.165115,doi:10.7566/JPSJ.86.024711,PhysRevLett.109.197401,sc-noneq}.

Recently, a mechanism providing ultrafast control of magnetic subsystems via \gls{OISTR} has been proposed theoretically~\cite{oistr_orig_2,oistr_orig} and was verified experimentally in Heusler materials and ferromagnetic alloys~\cite{light_induced_spin_transfer,light_induced_spin_transfer_efficiency,light_induced_spin_transfer_FMalloys}.
Based on the existence of two magnetic sublattices and mutually different local densities of states (DOSs) for minority and majority spin directions, optical excitations cause spin currents between the subsystems on the time-scale of electronic dynamics, i.e., femtoseconds.
Notably, the different DOSs in the minority and majority spin directions for the different sublattices can be modeled in terms of lattice electrons by means of a magnetic microstructure~\cite{PhysRevB.97.235120}.

In this paper, we connect both aspects and ask for the possibility of inducing periodic spatial patterns in \gls{OISTR}-like setups and find that a spin-selective photoexcitation is a suitable way to realize this.
We study the underlying microscopic dynamics in the idealized framework of \gls{1D} tight-binding electrons in the presence of a microstructure, subject to a photoexcitation, and ask for the effect of strong electron-electron interactions.
We observe not only spin transfer, but also the formation of a spatially modulated charge distribution, which is stable on comparably long time scales.
Crucially, our microscopic picture does not only apply to magnetic microstructures, but also to systems with a modulation of local chemical potentials, as realized in charge\hyp transfer salts\cite{IonicHubbard} and, e.g., modeled by the so-called ionic Hubbard model\cite{IonicHubbard,IonicHubbardManmana,ionichubbardmodelEsslinger}. 
Also, the considerations are not restricted by dimensionality of the system, so that we expect our scenario to be valid for generic systems with a microstructure.

The paper is organized as follows.
In \cref{sec:formation}, we present the physical mechanism leading to the formation of periodic patterns and its general validity.
Furthermore, we introduce the spin-selective photoexcitation and present the general setup.
In \cref{sec:model}, we focus on a specific Hubbard-type model with a magnetic microstructure as an example system.
In \cref{sec:mechanism} we discuss the stability and the decay channels for the \gls{CDP} in the presence of interactions as well as alternative scenarios by considering spin-dependent shaking of an optical lattice and single\hyp photon excitations.
%
%Also the difference to the low photon density limit is discussed.
%
In \cref{sec:conclusion} we conclude.
The appendices show detailed calculations for an idealized kick excitation (\cref{app:ionic}), a mean-field decoupling of the Hubbard-type model at low fillings (\cref{app:meanfield}), and the definition of the $k$ modes in the presence of \gls{OBC} (\cref{app:Ham_obc}).

\section{Formation of periodic structures}
\label{sec:formation}
\begin{figure}
	\centering
	\ifthenelse{\boolean{buildtikzpics}}
	{
		\tikzsetnextfilename{magnetic_background_excitation}
%		\tikzset{external/export next=false}
		\begin{tikzpicture}
			\begin{scope} [node distance = 0.1 and 0.1]
				\node[inner sep = 0em] (text) {(a) $\qquad$ Magnetic background: $\quad \Delta \sum_j (-1)^j \hat S^z_j$};
				\node (start) [below = of text] 
				{
					\tikzset{external/export next=false}
					\begin{tikzpicture}[baseline, yshift=-0.75em]
						\renewcommand{\bspWidth}{0.5}
						\bspDelta{-1}
						\bspUp{0} 
						\bspDown{1} 
						\bspUp{2} 
						\bspDown{3}
					\end{tikzpicture}
					\tikzset{external/export next=false}
					\begin{tikzpicture}
						\begin{scope}
							[node distance = 0.05]
							\node[inner xsep = 0.75em] (arrow) {$\rightarrow$};
							\node[above=of arrow, inner sep = 0pt] {\includegraphics[width=1.25em]{blitz}\hspace{-0.75em}${}_{\sigma=\uparrow}$};
						\end{scope}
					\end{tikzpicture}
					\tikzset{external/export next=false}
					\begin{tikzpicture}[baseline, yshift=-0.75em]
						\renewcommand{\bspWidth}{0.5}
						\bspNone{0}
						\bspUpDown{1}
						\bspNone{2}
						\bspUpDown{3}
					\end{tikzpicture}
				};
			\end{scope}
		\end{tikzpicture} \\[2em]
	}
	{
		\includegraphics{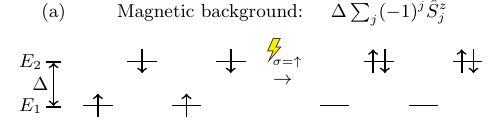}
	}

	\ifthenelse{\boolean{buildtikzpics}}
	{
		\tikzsetnextfilename{ionic_background_excitation}
%		\tikzset{external/export next=false} 
		\begin{tikzpicture}
			\begin{scope} [node distance = 0.1 and 0.1]
				\node[inner sep = 0em] (text) {(b) $\qquad$ Ionic background: $\quad \Delta \sum_j (-1)^j \hat n_j$};
				\node (start) [below = of text] 
				{
					\tikzset{external/export next=false}
					\begin{tikzpicture}[baseline, yshift=-0.75em]
						\renewcommand{\bspWidth}{0.5}
						\bspDelta{-1}
						\bspUpDown{0} 
						\bspNone{1} 
						\bspUpDown{2} 
						\bspNone{3}
					\end{tikzpicture}
					\tikzset{external/export next=false}
					\begin{tikzpicture}
						\begin{scope}
							[node distance = 0.05]
							\node[inner xsep = 0.75em] (arrow) {$\rightarrow$};
							\node[above=of arrow, inner sep = 0pt] {\includegraphics[width=1.25em]{blitz}\hspace{-0.75em}${}_{\sigma=\uparrow}$};
						\end{scope}
					\end{tikzpicture}
					\tikzset{external/export next=false}
					\begin{tikzpicture}[baseline, yshift=-0.75em]
						\renewcommand{\bspWidth}{0.5}
						\bspDown{0}
						\bspUp{1}
						\bspDown{2}
						\bspUp{3}
					\end{tikzpicture}
				};
			\end{scope}
		\end{tikzpicture}
	}
	{
		\includegraphics{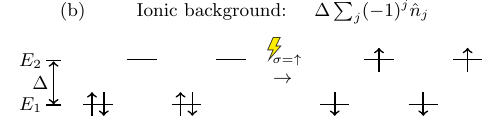}
	}
	\caption
	{
		Illustration of the effect of spin-selective excitations, in which only the spin-up electrons are affected. 
		In these simple illustrations, we assume the excitation induces a local current moving the spin-up electrons by one lattice site. 
		In (a), the initial state is a spin-density wave induced by a staggered magnetic field, which has a homogeneous distribution of the electron density. 
		The resulting state has zero magnetization per site and a periodically modulated electron density. 
		In (b), we start from a state with zero on-site magnetization, but a charge modulation caused by a staggered ionic potential. 
		The resulting state has a uniform charge distribution with staggered local magnetizations.
	} 
	\label{fig:sketch_excitations}
\end{figure}
We assume a system of itinerant electrons, in which different subsystems can be identified.
Simple situations, in which our setup is physically relevant, are lattice systems with a superlattice structure, which is either formed by spontaneous breaking of translational symmetry of the electronic system (e.g., a \gls{CDW} or a \gls{SDW}), or is present due to the general (lattice) properties of the system.
An illustration is given in Fig.~\ref{fig:sketch_excitations}, in which we display two simple scenarios, for which the proposed spin-selective excitation leads to a periodic pattern in the charge or spin degrees of freedom, respectively, which is absent in the initial state.
In these examples, we excite only the electrons with spin up.  

We first consider systems of noninteracting electrons with only nearest\hyp neighbor hopping on a lattice with a superstructure, which is modeled by a periodically modulated on-site potential.
In this case, the Hamiltonian can be written as
\begin{align}
	\hat{H} &= \hat{H}_{\uparrow} + \hat{H}_{\downarrow} \nonumber \\
	\hat{H}_{\sigma} &= - \hopping \sum_{\langle i,j\rangle}\left(\hat{c}^{\dagger}_{\sigma,i}\hat{c}^{\nodagger}_{\sigma,j} + \text{h.c.}\right) + \sum_j \Delta_{\sigma,j}\hat{n}_{\sigma,j} \; ,
\label{eq:U0}
\end{align}
with the usual annihilation (creation) operators $\hat{c}^{\left(\dagger\right)}_{\sigma,j}$ for electrons of spin $\sigma$ on lattice site $j$, the particle-density operator $\hat{n}_{\sigma,j} = \hat{c}^\dagger_{\sigma,j} \hat{c}^{\phantom{\dagger}}_{\sigma,j}$, hopping amplitude $\hopping$ and on-site potentials $\Delta_{\sigma,j}$ experienced by electrons with spin $\sigma$ on lattice site $j$.
Note that such a noninteracting system can be seen as two independent lattices of spinless fermions. 
Since we can write $\hat S^z_j = \nicefrac{1}{2} \left(\hat n_{j,\uparrow} - \hat n_{j,\downarrow}\right)$ for the local spin operator, we realize a Zeeman term by choosing $\Delta_{\downarrow,j} = - \Delta_{\uparrow,j}$. 
The case of a periodically modulated Zeeman term we refer to in the following as a magnetic microstructure.
Similarly, since the total density on lattice site $j$ is given by $\hat n_j = \hat n_{j,\uparrow} + \hat n_{j,\downarrow}$, the choice $\Delta_{\downarrow,j} =  \Delta_{\uparrow,j}$ realizes a local chemical potential or a local electrical field. 
In case of a periodic modulation, we call this an ionic potential, as realized, e.g., in the Ionic Hubbard model \cite{IonicHubbard,IonicHubbardManmana,ionichubbardmodelEsslinger}, which was introduced to model charge-transfer salts, or the Fermi-Hubbard-Harper model \cite{FabianSuperlattices}, which is studied in the context of topological phases in optical lattice systems. 
While generically $|\Delta_{\downarrow,j}| \neq |\Delta_{\uparrow,j}|$ is possible, we focus on the physically relevant cases of periodically modulated Zeeman fields or ionic potentials with  $|\Delta_{\downarrow,j}| = |\Delta_{\uparrow,j}|$ and consider systems in which the number of electrons for both spin directions is equal, $N_\uparrow = N_\downarrow$, so that we have no net magnetization. 
In the following, for the sake of simplicity we will focus on local observables, i.e., the local magnetization $\langle \hat S^z_j \rangle$ and the expectation value of the local density $\langle \hat n_j \rangle$ and leave the behavior of further interesting properties, e.g., possible topological features, to future work. 

As sketched in~\cref{fig:sketch_excitations}, for systems with magnetic microstructures, the local Zeeman terms usually lead to a periodic pattern in the spin density with the same periodicity as the on-site terms.
Such a modulation of the spins is not necessarily connected to a modulation of the charge density, so that situations can be realized, in which a constant charge density throughout the system is realized (up to boundary effects, which we neglect in the following), despite the periodic Zeeman term.
In complement to this, for periodic ionic potentials, a modulation of the charge density is expected, while the local magnetization can be zero throughout the system. 

Having these setups in mind, one can now ask for the effect of a spin-selective excitation. 
The simplest possible scenario sketched in~\cref{fig:sketch_excitations} is to assume an ad hoc excitation of the system that shifts all particles of the same spin direction by one lattice site.
In the presence of magnetic microstructures, the periodic spin density will be weakened, and at the same time a periodic charge modulation can be induced, whereas in the presence of ionic potentials, the periodic density distribution will be weakened, and a periodic modulation of the spin densities can be induced. 
Note that these considerations can be easily extended to systems with larger unit cells and in two or three dimensions. 
In the following we will show that spin-selective photoexcitations, realizable by circularly polarized monochromatic light (see, e.g., Ref.~¸\onlinecite{Bruno_moke}), induce this effect. 

\subsection{Photoexcitation in the noninteracting case}
\label{sec:photo}
We start modeling the photoexcitation by Peierls substitution \cite{Peierls1933,Mentink2015,PhysRevB.88.075135}, which we apply to noninteracting electrons to explore the general features sketched above. 
In this approach, the incident light is considered as a classical field and included in the Hamiltonian via minimal coupling \cite{Greiner2000}, which corresponds to a situation with large photon numbers or high intensities of the applied light. 
%
% In the usual modeling, this leads to a position- and time-dependent complex phase in the hopping amplitudes, and to additional local magnetic fields, which are often neglected.
%
In the following, we consider a generalization of the usual ansatz and assume that the effect of the light field can depend on the spin direction of the electrons.
This is motivated by the realization of spin-selective photoexcitations, e.g., in spin polarized \gls{ARPES} experiments \cite{Kirschner,ReviewSpinPolarizedPhotoemission,TrendsSpinResolvedPhotoelectronSpectroscopy,PRLSpinResolvedStudies} and through the tunability of parameters in experiments on optical lattices \cite{Bloch:2008p943}.
The tunneling amplitude in equilibrium $\hopping$ is multiplied by a complex phase factor, which depends on the position $j$, on time $\tau$, and on the spin direction $\sigma$, leading to
\begin{align}
	t^{\text{hop}}_{\sigma,j}(\tau) = e^{-i \alpha_{\sigma} \frac{e_{\text{el}} a_0}{2 \hbar}\left(A(j,\tau) + A(j+1,\tau)\right)} \, \hopping \, .
	\label{eq:peierls}
\end{align}
The coefficient $\alpha_\sigma$ takes either the value one, if the light field couples with the electrons of spin direction $\sigma$, or zero, if the coupling is suppressed. 
The vector potential $A(j,\tau)$ and magnetic field $B(j,\tau)$ are specified by 
\begin{align}\label{eq:vectorpotential}
	A(j, \tau) =	& \frac{E_0 \lambda}{2 \pi c} e^{-\frac{\left[a_0 j - c\left(\tau-\tau_0\right)\right]^2}{s^2}} \sin\left[\frac{2\pi}{\lambda} \left(a_0 j - c \tau \right) \right] \\
	B(j, \tau) =	& \frac{E_0 g_s \mu_B}{c} e^{-\frac{\left[a_0 j - c \left(\tau-\tau_0\right)\right]^2}{s^2}} \cos\left[ \frac{2\pi}{\lambda} \left(a_0 j - c \tau \right) \right] \;.
\end{align}
Here, $c$ denotes the speed of light, $\lambda$ the wavelength of the incoming light, which we assume for the sake of simplicity to be monochromatic, $e_{\rm el}$ is the charge of the electron, and $a_0$ is the lattice constant. 
We assume the light pulse to have a Gaussian envelope with amplitude $E_0$, peak at time $\tau = \tau_0$ and width $\frac{s}{c\sqrt 2}$ in the time domain.
We work in units, in which $\hbar = e_{\rm el} = a_0 = 1$, leading to the values displayed in \cref{tab:values}.
These values are used throughout the paper for the photoexcitation if not stated otherwise.

\begin{table}[!t]
	\caption
	{
		\label{tab:values}
		Wave package parameter resulting from $\hbar=e_{\text{el}}=\hopping=1$ and $a_0=1 \left[10^{-10} \text{~m}\right]$.
	}
	\begin{tabular}{lcclc}
		\toprule
		\multicolumn{2}{c}{$c=3374.85 \left[\frac{a_0 \cdot \hopping}{\hbar}\right]$}	& $\quad$	& \multicolumn{2}{c}{$g_s \mu_B=13.04 \left[\frac{e_\text{el} \hopping a_0^2}{\hbar}\right]$}		\\
		\midrule
		Wavelength						& $\lambda \approx 5000\left[a_0\right]$		&			& Amplitude								& $E_0=20 \left[\frac{\hopping}{e_{\text{el}}}\right]$		\\[0.5em]
		Width							& $s=6000 \left[a_0\right]$						&			& Maximum								& $\tau_0=10 \left[\frac{\hbar}{\hopping}\right]$			\\
		\bottomrule
	\end{tabular}
\end{table}

Having introduced such a spin-dependent hopping amplitude, we can now rewrite the Hamiltonian~\eqref{eq:U0} to 
\begin{align}
	\label{eq:model_excited}
	\hat{H}_{\sigma}(\tau)	=& - \sum_{\langle i,j\rangle} \left( t^{\text{hop}}_{\sigma,j}(\tau) \hat{c}^{\dagger}_{\sigma,i}\hat{c}^{\nodagger}_{\sigma,j} + \text{h.c.}\right) + \sum_j \Delta_{\sigma,j}\hat{n}_{\sigma,j} \nonumber \\
							&+\sum_j B(j,\tau) \, \hat{S}^{z}_{j}\; .
\end{align}

\subsection{``Kick''-like excitation}
We ask now for a situation, in which such an excitation is easy to treat, and for the lifetime of the realized state. 
For this, we consider a kicklike excitation, which can be regarded as an idealized `Peierls pulse'~\eqref{eq:peierls} with a width of the Gaussian envelope of the vector potential in Eqs.~\eqref{eq:vectorpotential} $s/c \to 0$, i.e., the pulse duration, which we denote by $\delta s$, is small compared to the time scale introduced by the microstructure $\delta s \cdot \Delta \ll 1$ (see \cref{app:ionic}).
Note that in the long wavelength limit $\lambda \gg c \cdot \delta s$ the Peierls pulse reduces to a simple Gaussian.
For small durations $\delta s$ or long enough wavelength $\lambda$ the pulse can be approximated by a box-shape, which is used in \cref{app:ionic} to treat the kick excitation.

For the sake of simplicity we treat one-dimensional systems and exploit the fact that the two spin directions for noninteracting electrons are completely decoupled. 
In the following, without loss of generality, we consider an excitation only on electrons with $\sigma = \uparrow$, a system with nearest\hyp neighbor hopping, and $N$ unit cells with each containing $M$ sites.
Based on~\cref{eq:model_excited}, assuming the induced magnetic fields are negligibly small, such an idealized excitation can be modeled by evolving the system with a modified hopping part of the Hamiltonian
\begin{align} 
	\hat{T}_{\varphi} 
	= 
	- \hopping \sum_{j,\sigma} & \left( e^{-i \varphi\delta_{\sigma,\uparrow}} \hat{c}_{\sigma,j}^\dagger \hat{c}_{\sigma,j+1}^{\nodagger} + \mathrm{h.c.} \right) \; .
	\label{eq:kick}
\end{align} 
The phase factor $\varphi$ is assumed to be constant due to the infinitesimally short pulse duration and the Kronecker delta $\delta_{\sigma,\uparrow}$ selects the spin direction.
We assume no spatial dependence of the perturbation or of $\hopping$. 

For illustrative purposes, we apply this perturbation now to an ionic system with a unit cell of $M=2$ sites with Hamiltonian
\begin{align}
	\hat{H}^{\rm ionic} = \hat T_0 + \Delta \sum_j (-1)^j \hat{n}_j \, ,
	\label{eq:ionic}
\end{align}
which is the noninteracting limit of the ionic Hubbard model\cite{IonicHubbard} and which is one of the scenarios sketched in \cref{fig:sketch_excitations}.
In the following, we consider half filling, so that the ground state for one spin direction is
\begin{align}
	\begin{split}
		\ket{\psi_{0,\sigma}} &= \prod_{k_n} \hat{a}^\dagger_{\sigma, -}(k_n) \ket{0} \, , \\
		k_n &= \frac{2 \pi n}{MN} \,, \; n = -N, \ldots, N-1
	\end{split}
	\label{eq:gs}
\end{align}
with operators $\hat{a}^{(\dagger)}_{\sigma, \mu}(k)$, $\mu=\pm$ labeling the upper and lower band and which diagonalize the Hamiltonian~\eqref{eq:ionic}, as discussed in~\cref{app:ionic}.
In the atomic limit $\nicefrac{\hopping}{\Delta} \to 0$, we can expand the results in the small parameter $\varepsilon \equiv \nicefrac{2 \hopping}{\Delta}$ to obtain compact expressions, giving us direct insights into the behavior of the system following the pulse~\eqref{eq:kick}.
As discussed in~\cref{app:ionic}, we obtain (labeling unit cells by $l=0,\ldots,N-1$ and the sites within the unit cells by $m=0, 1$)
\begin{align}
	\label{eq:ns0}
	\langle \hat n_{\sigma,l,0} \rangle &= 1 - \frac{\varepsilon^2}{8} \, , \\
	\langle \hat n_{\sigma,l,1} \rangle &= 1 - \langle \hat n_{\sigma,l,0} \rangle = \frac{\varepsilon^2}{8} \, ,
	\label{eq:ns1}
\end{align}
i.e., a \gls{CDP} with amplitude of the modulation $a_{\rm \text{\gls{CDP}}} := \langle \hat n_{\sigma,l,0} \rangle - \langle \hat n_{\sigma,l,1} \rangle = 1 - \frac{\varepsilon^2}{4}$.
Since we are populating the $k_n$ modes symmetrically around $k=0$, in the ground state the value of $\langle \hat n_{\sigma,l,0} \rangle$ is maximized, while $\langle \hat n_{\sigma,l,1} \rangle$ is minimal.
Note that from the discussions in~\cref{app:ionic} it follows that a finite $a_{\rm \text{\gls{CDP}}}$ is obtained for all values of $\Delta/\hopping > 0$ and not only in the atomic limit, which is related to the fact that there is no phase transition as a function of $\Delta$ for this system\cite{IonicHubbardManmana}.
We hence see from~\cref{eq:ns0,eq:ns1} that for the ionic system in the ground state the expectation values of the local densities $\braket{ \hat{n}_{\sigma,l,m} }$ will alternate within a unit cell, while the local magnetizations $\braket{ \hat S^z_{l,m} }$ are zero throughout the system.
The kick excitation will lead to a redistribution of particles to both bands, and since the densities were maximal (minimal) on the even (odd) sites within a unit cell, this means that the amplitude of the density pattern will decrease. 
More specifically, in the atomic limit $\varepsilon \ll 1$ we obtain for the local densities directly after the kick excitation
\begin{align}
	\braket{\hat n_{\sigma,l,0}}(\varphi) &= \braket{\hat n_{\sigma,l,0}} - \delta_{\sigma,\uparrow}\hopping^2 \frac{\delta s^2}{4}\sin^2({\nicefrac{\varphi}{2}}) \\
	\braket{\hat n_{\sigma,l,1}}(\varphi) &= \braket{\hat n_{\sigma,l,1}} + \delta_{\sigma,\uparrow}\hopping^2 \frac{\delta s^2}{4}\sin^2({\nicefrac{\varphi}{2}})
	\label{eq:ns_kick}
\end{align}
as discussed in more detail in \cref{app:ionic}. 
Importantly, due to the choice of a spin-selective excitation in \cref{eq:kick}, the redistribution happens only for the $\uparrow$-particles, while the $\downarrow$-particles remain unchanged.
Thus, the effect of the kick excitation is a weakening of the density modulation in each unit cell while at the same time a \gls{SDP} forms with amplitude
\begin{align}
	a_{\rm \text{\gls{SDP}}}
	\propto
(\hopping \delta s)^2 
\sin^2(\nicefrac{\varphi}{2})\; .
\end{align}
Note that the induced \gls{SDP} amplitude scales with $\hopping^2$ while the dependence on $\varepsilon$ is subleading ($\sim \mathcal O (\varepsilon^4)$).
In contrast, exciting both spin directions will not lead to a finite local magnetization, since the densities for both spin directions would remain equal.
Hence, we need a spin-selective excitation to induce a magnetization pattern. 

Similarly, one can treat a system with a simple magnetic microstructure by assuming a site-alternating Zeeman term,
\begin{align}
	\hat{H}^{\rm magnetic} = \hat T_0 + \Delta \sum_j (-1)^j \hat{S}^z_j 
	\label{eq:magnetic}
\end{align}
at quarter filling.
In this situation, in the ground state there is the same particle density on each lattice site, and a finite local magnetization caused by the local Zeeman term.
A similar reasoning as before shows that the initially present modulation of the spin density is weakened, while a finite density modulation is induced.
Again, this only happens in case of a spin-selective excitation, since otherwise the same change of particle densities in both spin directions would lead to an equal redistribution of both particle species.
These calculations on simple systems confirm the scenario depicted in~\cref{fig:sketch_excitations} for an idealized Peierls pulse.
Usually, the patterns induced in the charge and spin densities by such spin-selective excitations will correspond to the patterns of the on-site potentials $\Delta_{\sigma,j}$ in systems described by Hamiltonians of type~\eqref{eq:U0}. 
Such an excitation is, hence, a way to imprint or enhance this pattern to observables, even if their expectation values do not possess this pattern in the initial state. 
Similar considerations can also be made in higher dimensions, since the mechanism sketched above depends only on the existence of an underlying structure in the local potentials, and on exciting only one type of electrons. 

\label{repeat:IIb}
As we have shown in~\cref{app:ionic}, in general the kick-excitation leads to a redistribution of populations of the $k$ modes.
Since for the noninteracting system their number $\braket{ \hat{n}_{\sigma,\mu}(k) } = \braket{ \hat{a}^\dagger_{\sigma, \mu}(k) \hat{a}^{\nodagger}_{\sigma, \mu}(k) }$ is a conserved quantity, the electrons will not be able to relax back to the initial state, leading to a stable state with new spin and charge distributions.
Note that this state can still be time-dependent, as the excitation induces superpositions of eigenstates of the system, but it will not decay to a steady state due to the lack of scattering processes. 
Typically, this will lead to oscillations in time of the local observables, without destroying the generated charge and spin patterns. 

The above examples refer to idealized setups. 
The question arises for a more realistic scenario, which we treat in the following by applying Peierls pulses of duration $\sim 10$fs (using the units and values of~\cref{tab:values}), and for the role of interactions. 
For a realistic system, the two main modifications are due to electron-electron interactions, and phonons.
We postpone the investigation of the effect of phonons, which are difficult to treat numerically in fully-interacting electronic systems out-of equilibrium\cite{PhysRevB.91.104302,Stolpp2020,PhysRevB.102.165155,10.21468/SciPostPhys.10.3.058}.
Instead, we investigate the effect of electron-electron interactions on the possible realizations and lifetimes of the above mentioned patterns in a specific model with a magnetic microstructure.
We expect this to be a valid analysis, as long as the coupling between phonons and itinerant electrons is weak~\cite{Tezuka2007,Hardikar2007,Nocera2014} and leave the strong coupling regime for future work.
We also assume in the following the magnetic microstructure to be unaffected by the photoexcitation on the time scales treated by us.
This is justified by the \gls{OISTR} prediction, in which a timescale is identified, on which the equilibrium DOS can be used to describe the dynamics of the system \cite{oistr_orig,oistr_orig_2}.
As long as this is true, the magnetic microstructure is present and our considerations are applicable.
At later times, once the local DOS might be modified by further effects like spin-flips due to spin-orbit coupling, one would need to take these also into account.
However, as in the experimental observation of \gls{OISTR}~\cite{light_induced_spin_transfer,light_induced_spin_transfer_efficiency,light_induced_spin_transfer_FMalloys}, we expect an extended time window, in which such more complicated behavior is not active and in which the predicted effects should be observable.

\section{Hubbard model with magnetic microstructure and spin-selective photoexcitation}
\label{sec:model}
We now discuss in detail the interplay of a magnetic microstructure, spin-selective photoexcitations, and electron-electron interactions by considering a variant of the one-dimensional Fermi Hubbard model \cite{HubbBook,esslinger_review}, which was introduced in Ref.~\onlinecite{PhysRevB.97.235120} as a 1D toy model for manganates.   
We treat the Hamiltonian 
\begin{eqnarray}\label{eq:Ham}
	\hat{H}=\sum_{j}&\biggl\lbrace&	
	-\hopping \sum_\sigma \Bigl( \hat{c}^\dagger_{\sigma,j+1}\hat{c}^{\phantom{\dagger}}_{\sigma,j}+ \hat{c}^\dagger_{\sigma,j}\hat{c}^{\phantom{\dagger}}_{\sigma,j+1}\Bigr) \nonumber\\
	&+&U \hat{n}_{\uparrow,j}\hat{n}_{\downarrow, j} + \Delta_j \hat S^z_j 
	\biggr\rbrace \;  ,
\end{eqnarray}
where $U$ is the strength of the Hubbard interaction, and the sign of the spin-independent on-site potential $\Delta_j$ is alternating every two sites,  $(\Delta,\,\Delta,\,-\Delta,\,-\Delta)$, which defines a magnetic microstructure with a unit cell of four sites. 
\renewcommand{\yshift}[1]
{
	ifthenelse(mod(#1,4)<2,0,1)*\Zeeman
}
\begin{figure}
	\centering

	\subfloat[\label{subfig:chain}]
	{
		\hspace{0.6em}
		\tikzset{external/export next=false}
		\begin{tikzpicture}
			\bspDelta{-1}
			\bspDownH{0} 
			\bspDownH{1} 
			\bspUpH{2} 
			\bspUpH{3}
			\bspDownH{4} 
			\bspDownH{5} 
			\bspUpH{6} 
			\bspUpH{7}
		\end{tikzpicture}
	}

	\subfloat[\label{subfig:density}]
	{
		%head -n2 n_hub | tail -n1 | awk '{for (i=1;i<=NF;i++) print i-2 "\t" $i}' | tail -n+2 > N_gs
		\ifthenelse{\boolean{buildtikzpics}}
		{
			\hspace{0.15em}
			\tikzsetnextfilename{initial_densities}
%			\tikzset{external/export next=false}
			\begin{tikzpicture}
				\begin{axis}
				[
					ylabel				=	{{\color{colorA}$\braket{\hat n^{\phantom{z}}_j}-\nicefrac{1}{2}$}, {\color{colorB}$\braket{\hat S^z_j}$}},
					ylabel shift 		=	-2pt,
					xlabel				= 	{site $j$},
					width				=	0.46\textwidth, 
					height				=	0.175\textheight,
					xmin				=	15.75,
					xmax				=	23.125,
					ylabel near ticks, 
					tick label style={font=\footnotesize},
					colormap name	=	gnuplot,
					point meta min	=	0,
					point meta max	=	1,
				]
				\addplot
				[
%					scatter,
%					point meta	= {TeX code={\def\pgfplotspointmeta{\thisrowno{1}}}},	
					mark=square,
					unbounded coords=jump, 
					mark size = 2pt,
					thick,
					only marks,
					color = colorA,
				] 
					table
					[
						x expr = \thisrowno{0}, 
						y expr = \eval{\thisrowno{1}-0.5}
%						y expr = \thisrowno{1},
					]
						{../data/alpha_up0/alpha_down1/beta_up0/beta_down0/b_z8,8,-8,-8/U4/e020/sigma6000/t010/k00.00126/zeeman_coupling0/ChainLength40/TargetSymmetrySector20,0/RealTimeStepWidth0.01/rte/loc_observables/N_gs};
						
				\addplot
				[
					mark=o,
					unbounded coords=jump, 
					mark size = 2pt,
					thick,
					only marks,
					color = colorB,
				] 
					table
					[
						x expr = \thisrowno{0}, 
						y expr = \thisrowno{1}
					]
						{../data/alpha_up0/alpha_down1/beta_up0/beta_down0/b_z8,8,-8,-8/U4/e020/sigma6000/t010/k00.00126/zeeman_coupling0/ChainLength40/TargetSymmetrySector20,0/RealTimeStepWidth0.01/rte/loc_observables/Sz_gs};		
				\end{axis}
			\end{tikzpicture}
		}
		{
			\hspace{0.5em}
			\includegraphics{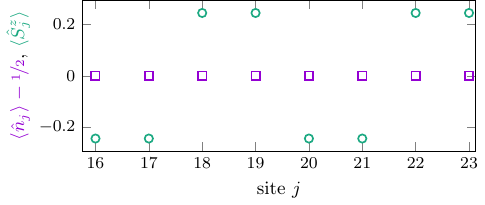}
		}
	}
	\caption
	{
		\label{fig:densities}
		\protect\subref{subfig:chain} Sketch of the distribution of the electrons on the magnetic microstructure in model~\eqref{eq:Ham} at quarter filling. 
		Note that the dashed arrows represent a local density of $\nicefrac 1 2$ for the corresponding spin.
		\protect\subref{subfig:density} \gls{DMRG} results for the local densities $\langle n_j \rangle$ and the local spin densities $\langle S^z_j \rangle$ in the ground state at $U/\hopping=4, \Delta/\hopping=8$ at quarter filling. 
	}
\end{figure}
This model is related to so-called block orbital-selective Mott insulators, which are discussed in the context of ladder systems like BaFe$_2$Se$_3$. 
These systems can be modeled by generalized Kondo-Heisenberg models and possess phases, in which the same magnetic microstructure is realized, but with a coupling to the conduction electrons via a Heisenberg exchange term rather than a Zeeman term \cite{Jacek_2020}.

Here, if not mentioned otherwise, we treat the system at quarter filling and zero net magnetization, i.e., the total number of electrons $N_\uparrow = N_\downarrow$, where the ground state can be seen as a crystal of Zener polarons, as further discussed in Ref.~\onlinecite{PhysRevB.97.235120}.
In the ground state, at finite values of $\Delta$, this results for all values of $U$ in a homogeneous charge distribution in the bulk and in a staggered local magnetization, which follows the pattern of the  microstructure.\cite{ole_masterarbeit}
This is illustrated in \cref{fig:densities}; \cref{subfig:chain} summarizes the expected bulk behavior at quarter filling, which is confirmed by \gls{MPS} results for a chain with $L=40$ sites, $U = 4 \hopping$ and $\Delta = 8 \hopping$ in \cref{subfig:density}.  
When applying the Peierls pulse Eqs.~\eqref{eq:peierls} and~\eqref{eq:vectorpotential}, we obtain a time-dependent Hamiltonian with spin-dependent hopping amplitude, which can be rewritten as 
\begin{align}
	\label{eq:model}
	\hat H(\tau) =	& - \sum_{\sigma,j} \left( t^{\text{hop}}_{\sigma,j}(\tau) \hat c^\dagger_{\sigma,j} \hat c^{\phantom{\dagger}}_{\sigma, j+1} + h.c.\right) \nonumber \\
				& +  \sum_{j} \left( U \hat n^{\phantom{\dagger}}_{\uparrow,j} \hat n^{\phantom{\dagger}}_{\downarrow,j} + \left[\Delta_j +				B(j,\tau)\right] \, \hat{S}^{z}_{j} \right)\, .
\end{align}
This models the interplay of the magnetic microstructure defined by $\Delta_j$, the electron-electron interaction $U$ and the spin-selective photoexcitation in the course of time.
Note that with the values of \cref{tab:values} the magnetic field induced by the pulse typically is $B(j,\tau) \ll \Delta$ and can therefore be safely disregarded.

In the following, we treat the time evolution of ground states of the Hamiltonian~\eqref{eq:Ham} subject to the time-dependent Hamiltonian~\eqref{eq:model} using \gls{tDMRG}\cite{
	PAECKEL2019167998,
	PhysRevB.85.205119,
	PhysRevB.91.165112,
	PhysRevB.94.165116,
	PhysRevLett.91.147902,
	PhysRevLett.93.040502,
	PhysRevLett.93.076401,
	PhysRevLett.93.207204,
	PhysRevLett.107.070601,
	Daley2004,
	1367-2630-8-12-305}.
If not stated otherwise, we use the time-evolution \gls{MPO} $W^{\text{II}}$ introduced in Ref.~\onlinecite{PhysRevB.91.165112} with a time step of $\Delta \tau = 0.01$.
The time-dependent \gls{MPO} taking into account the spin and particle number conservation is built from a finite-state machine as described in Ref.~\onlinecite{SciPostPhys.3.5.035}.
Due to the time dependence, a rebuilt of the \gls{MPO} is necessary at every time step in order to re-evaluate the function $A(j,\tau)$.
A maximum bond dimension $\chi = 500$ is sufficient to limit the discarded weight $\epsilon < 2 \cdot 10^{-7}$ if not stated otherwise.

\subsection{Charge-density patterns through spin-selective photoexcitation} % and doublon decay mechanisms}
\label{sec:CDW}
\begin{figure}[!t]	% DvariableU0_alphaup01
	\centering
	\newcommand{\maxTinLongtime}{20}

	\ifthenelse{\boolean{buildtikzpics}}
	{
	\tikzsetnextfilename{DvariableU0_alphaup01}
	\begin{tikzpicture}
		\pgfplotsset
		{
			/pgfplots/colormap={gnuplot}{rgb255=(0,0,0) rgb255=(46,0,53) rgb255=(65,0,103) rgb255=(80,0,149) rgb255=(93,0,189) rgb255=(104,1,220) rgb255=(114,2,242) rgb255=(123,3,253) rgb255=(131,4,253) rgb255=(139,6,242) rgb255=(147,9,220) rgb255=(154,12,189) rgb255=(161,16,149) rgb255=(167,20,103) rgb255=(174,25,53) rgb255=(180,31,0) rgb255=(186,38,0) rgb255=(191,46,0) rgb255=(197,55,0) rgb255=(202,64,0) rgb255=(208,75,0) rgb255=(213,87,0) rgb255=(218,100,0) rgb255=(223,114,0) rgb255=(228,130,0) rgb255=(232,147,0) rgb255=(237,165,0) rgb255=(241,185,0) rgb255=(246,207,0) rgb255=(250,230,0) rgb255=(255,255,0) }
		}
		\pgfplotsset
		{
			/pgfplots/colormap={temp}{
				rgb255=(36,0,217) 		% 0 "#2400d9"
				rgb255=(25,29,247) 		% 1 "#191df7"
				rgb255=(41,87,255) 		% 2 "#2957ff"
				rgb255=(61,135,255) 	% 3 "#3d87ff"
				rgb255=(87,176,255) 	% 4 "#57b0ff"
				rgb255=(117,211,255) 	% 5 "#75d3ff"
				rgb255=(153,235,255) 	% 6 "#99ebff"
				rgb255=(189,249,255) 	% 7 "#bdf9ff"
				rgb255=(235,255,255) 	% 8 "#d3ffff"
				rgb255=(255,255,235) 	% 9 "#ffffd9"
				rgb255=(255,242,189) 	% 10 "#fff2bd"
				rgb255=(255,214,153) 	% 11 "#ffd699"
				rgb255=(255,172,117) 	% 12 "#ffac75"
				rgb255=(255,120,87) 	% 13 "#ff7857"
				rgb255=(255,61,61) 		% 14 "#ff3d3d"
				rgb255=(247,40,54) 		% 15 "#f72836"
				rgb255=(217,22,48) 		% 16 "#d91630"
				rgb255=(166,0,33)		% 17 "#a60021"
			}
		}
		\begin{groupplot}
			[
				group style = 
				{
					group size 			=	3 by 5,
					vertical sep		=	3.5mm,
					horizontal sep		=	1.4mm,
					x descriptions at	=	edge bottom,
					y descriptions at	=	edge left
				},
				height			= 	0.15\textheight,
				width 			= 	0.22\textwidth,
				xmin 			= 	8, 
				xmax 			= 	32,
				ymin			=	0,
				ymax			=	\maxTinLongtime,	
				ylabel			= 	{time $\tau/\hopping$},
				ylabel shift 	=	-4pt,
			    point meta min	=	0,
			    point meta max	=	1
			]
				\nextgroupplot
				[
					enlargelimits		=	false,
					axis on top, 
					width				= 	0.15\textwidth,
					x dir				=	reverse,
					xmin				= 	-290,
					xmax				=	10,
					ymin				=	0,
					ymax				=	\maxTinLongtime,
					samples				=	2000,
					xticklabel pos=right,
					xlabel				=	{$A(j=0,\tau)$},					
					xlabel style		=	{yshift=-0.5em,color = blue},
					xtick				=	{-200, -50},
					xticklabels			=	{$0$, $10$},
					xticklabel style	=	{color = blue}
				]
				\addplot
				[
					color				=	black,
					unbounded coords	=	jump, 
					mark size			=	2pt,
				] 
				table
				[
					x expr = \thisrowno{1}, 
					y expr = \thisrowno{0},
				]
				{data/alpha_up0/alpha_down1/beta_up0/beta_down0/b_z0/U0/e020/sigma6000/t010/k00.00126/zeeman_coupling0/ChainLength40/TargetSymmetrySector20,0/RealTimeStepWidth0.01/rte/sim_data};
				\addplot
				[
					domain	=	0:\maxTinLongtime, 
					blue, 
				]
				({peierlspotential(0, 20, 0.00126, 10, 3374.85, 6000)*15-200}, {x});
				\nextgroupplot
				[
					enlargelimits	=	false,
					axis on top,
					title			=	{$\langle \hat n_j \rangle$},
					title style		=	{at = {(0.5,1)}, yshift = -0.7em},
					extra description/.code={\node[anchor=south east,white] at (0.99,0.01) {$U/\hopping=0$};},
				]
				\addplot graphics
				[ 
					xmin	=	0, 
					xmax	=	40, 
					ymin	=	0, 
					ymax	=	\maxTinLongtime
				]
				{data/alpha_up0/alpha_down1/beta_up0/beta_down0/b_z0/U0/e020/sigma6000/t010/k00.00126/zeeman_coupling0/ChainLength40/TargetSymmetrySector20,0/RealTimeStepWidth0.01/rte/loc_observables/plain_N.pdf};
				\nextgroupplot
				[
					enlargelimits	=	false,
					axis on top,
					colorbar right,
					colormap name	=	gnuplot,
					every colorbar/.append style =
					{
						height			=	2*\pgfkeysvalueof{/pgfplots/parent axis height} + 1*\pgfkeysvalueof{/pgfplots/group/vertical sep},
						ylabel			=	{$\langle \hat n_j \rangle$},
						width			=	2mm,
						ytick			= 	{0, 0.25, 0.5, 0.75, 1},
						yticklabels		=	{$0$, $\nicefrac{1}{4}$, $\nicefrac{1}{2}$, $\nicefrac{3}{4}$, $1$},
						ylabel shift 	=	-4pt,
					},
					extra description/.code =
					{
						\node[anchor = south east] at (0.99,0.01) {\(\Delta/\hopping=0\)};
					},
					title			=	{$\langle \hat S^z_j \rangle$},
					title style		=	{at = {(0.5,1)}, yshift = -0.7em}
				]
				\addplot graphics
				[ 
					xmin	=	0, 
					xmax	=	40, 
					ymin	=	0, 
					ymax	=	\maxTinLongtime
				]
				{data/alpha_up0/alpha_down1/beta_up0/beta_down0/b_z0/U0/e020/sigma6000/t010/k00.00126/zeeman_coupling0/ChainLength40/TargetSymmetrySector20,0/RealTimeStepWidth0.01/rte/loc_observables/plain_S.pdf};

				\nextgroupplot
				[
					enlargelimits	=	false,
					axis on top, 
					width			= 	0.15\textwidth,
					x dir			=	reverse,
					xmin			= 	-290,
					xmax			=	10,
					ymin			=	0,
					ymax			=	\maxTinLongtime,
					samples			=	2000,
				]
				\addplot
				[
					color				=	black,
					unbounded coords	=	jump, 
					mark size			=	2pt,
				] 
				table
				[
					x expr = \thisrowno{1}, 
					y expr = \thisrowno{0},
				]
				{data/alpha_up0/alpha_down1/beta_up0/beta_down0/b_z8,8,-8,-8/U0/e020/sigma6000/t010/k00.00126/zeeman_coupling0/ChainLength40/TargetSymmetrySector20,0/RealTimeStepWidth0.01/rte/sim_data};
				\addplot
				[
					domain	=	0:\maxTinLongtime, 
					blue, 
				]
				({peierlspotential(0, 20, 0.00126, 10, 3374.85, 6000)*15-200}, {x});
				\nextgroupplot
				[
					enlargelimits	=	false,
					axis on top,
					extra description/.code={\node[anchor=south east,white] at (0.99,0.01) {$U/\hopping=0$};},
				]
				\addplot graphics
				[ 
					xmin	=	0, 
					xmax	=	40, 
					ymin	=	0, 
					ymax	=	\maxTinLongtime
				]
				{data/alpha_up0/alpha_down1/beta_up0/beta_down0/b_z8,8,-8,-8/U0/e020/sigma6000/t010/k00.00126/zeeman_coupling0/ChainLength40/TargetSymmetrySector20,0/RealTimeStepWidth0.01/rte/loc_observables/plain_N.pdf};
				\nextgroupplot
				[
					enlargelimits	=	false,
					axis on top,
					extra description/.code =
					{
						\node[anchor = south east] at (0.99,0.01) {\(\Delta/\hopping=8\)};
					}
				]
				\addplot graphics
				[ 
					xmin	=	0, 
					xmax	=	40, 
					ymin	=	0, 
					ymax	=	\maxTinLongtime
				]
				{data/alpha_up0/alpha_down1/beta_up0/beta_down0/b_z8,8,-8,-8/U0/e020/sigma6000/t010/k00.00126/zeeman_coupling0/ChainLength40/TargetSymmetrySector20,0/RealTimeStepWidth0.01/rte/loc_observables/plain_S.pdf};

				\nextgroupplot
				[
					enlargelimits	=	false,
					axis on top, 
					width			= 	0.15\textwidth,
					x dir			=	reverse,
					xmin			= 	-290,
					xmax			=	10,
					ymin			=	0,
					ymax			=	\maxTinLongtime,
					samples			=	2000,
				]
				\addplot
				[
					color=black,
					unbounded coords=jump, 
					mark size = 2pt,
				] 
				table
				[
					x expr = \thisrowno{1}, 
					y expr = \thisrowno{0},
				]
				{data/alpha_up0/alpha_down1/beta_up0/beta_down0/b_z8,8,-8,-8/U4/e020/sigma6000/t010/k00.00126/zeeman_coupling0/ChainLength40/TargetSymmetrySector20,0/RealTimeStepWidth0.01/rte/sim_data};
				\addplot
				[
					domain		=	0:\maxTinLongtime, 
					blue, 
				]
				({peierlspotential(0, 20, 0.00126, 10, 3374.85, 6000)*15-200}, {x});

				\nextgroupplot
				[
					enlargelimits	=	false,
					axis on top,
					extra description/.code={\node[anchor=south east,white] at (0.99,0.01) {$U/\hopping=4$};},
				]
				\addplot graphics
				[ 
					xmin = 0, 
					xmax = 40, 
					ymin = 0, 
					ymax = \maxTinLongtime
				]
				{data/alpha_up0/alpha_down1/beta_up0/beta_down0/b_z8,8,-8,-8/U4/e020/sigma6000/t010/k00.00126/zeeman_coupling0/ChainLength40/TargetSymmetrySector20,0/RealTimeStepWidth0.01/rte/loc_observables/plain_N.pdf};

				\nextgroupplot
				[
					enlargelimits	=	false,
					axis on top,
					colorbar right,
	,			    point meta min	=	-0.5,
				    point meta max	=	0.5,
					every colorbar/.append style =
					{
						height			=	3.5mm + 2*\pgfkeysvalueof{/pgfplots/parent axis height} + 1*\pgfkeysvalueof{/pgfplots/group/vertical sep},
						ylabel			=	{$\langle \hat S^z_j \rangle$},
						width			=	2mm,				
						ytick			=	{-0.5, -0.25, 0, 0.25, 0.5},
						yticklabels		=	{$-\nicefrac{1}{2}$, $-\nicefrac{1}{4}$, $0$, $\nicefrac{1}{4}$, $\nicefrac{1}{2}$},
						ylabel shift	=	-12pt,
					},
					extra description/.code =
					{
						\node[anchor = south east] at (0.99,0.01) {\(\Delta/\hopping=8\)};
					}
				]
				\addplot graphics
				[ 
					xmin = 0, 
					xmax = 40, 
					ymin = 0, 
					ymax = \maxTinLongtime
				]
				{data/alpha_up0/alpha_down1/beta_up0/beta_down0/b_z8,8,-8,-8/U4/e020/sigma6000/t010/k00.00126/zeeman_coupling0/ChainLength40/TargetSymmetrySector20,0/RealTimeStepWidth0.01/rte/loc_observables/plain_S.pdf};

				\nextgroupplot[group/empty plot, y = 0.001pt]
				\nextgroupplot[group/empty plot, y = 0.001pt]
				\nextgroupplot[group/empty plot, y = 0.001pt]

				\nextgroupplot
				[
					enlargelimits	=	false,
					axis on top, 
					width			= 	0.15\textwidth,
					x dir			=	reverse,
					xmin			= 	-290,
					xmax			=	10,
					ymin			=	0,
					ymax			=	\maxTinLongtime,
					xtick distance	=	150,
					samples			=	2000,
					xlabel style	=	{yshift=0.155em},
					xlabel			= 	{energy},
					title			=	{Non-spin-selective Peierls pulse},
					title style 	= 	{at={(0,1.025)}, anchor=west}
				]
				\addplot
				[
					color				=	black,
					unbounded coords	=	jump, 
					mark size			=	2pt,
				] 
				table
				[
					x expr = \thisrowno{1}, 
					y expr = \thisrowno{0},
				]
				{data/alpha_up1/alpha_down1/beta_up0/beta_down0/b_z8,8,-8,-8/U4/e020/sigma6000/t010/k00.00126/zeeman_coupling0/ChainLength40/TargetSymmetrySector20,0/RealTimeStepWidth0.01/rte/sim_data};
				\addplot
				[
					domain	=	0:\maxTinLongtime, 
					blue, 
				]
				({peierlspotential(0, 20, 0.00126, 10, 3374.85, 6000)*15-200}, {x});
				\nextgroupplot
				[
					enlargelimits	=	false,
					axis on top,
					xlabel style	=	{yshift=0.3em},
					xlabel			= 	{site $j$},
					extra description/.code={\node[anchor=south east,white] at (0.99,0.01) {$U/\hopping=4$};},
				]
				\addplot graphics
				[ 
					xmin	=	0, 
					xmax	=	40, 
					ymin	=	0, 
					ymax	=	\maxTinLongtime
				]
				{data/alpha_up1/alpha_down1/beta_up0/beta_down0/b_z8,8,-8,-8/U4/e020/sigma6000/t010/k00.00126/zeeman_coupling0/ChainLength40/TargetSymmetrySector20,0/RealTimeStepWidth0.01/rte/loc_observables/plain_N.pdf};
				\nextgroupplot
				[
					enlargelimits	=	false,
					axis on top,
					xlabel			= 	{site $j$},
					xlabel style	=	{yshift=0.3em},
					extra description/.code =
					{
						\node[anchor = south east] at (0.99,0.01) {\(\Delta/\hopping=8\)};
					}
				]
				\addplot graphics
				[ 
					xmin	=	0, 
					xmax	=	40, 
					ymin	=	0, 
					ymax	=	\maxTinLongtime
				]
				{data/alpha_up1/alpha_down1/beta_up0/beta_down0/b_z8,8,-8,-8/U4/e020/sigma6000/t010/k00.00126/zeeman_coupling0/ChainLength40/TargetSymmetrySector20,0/RealTimeStepWidth0.01/rte/loc_observables/plain_S.pdf};
		\end{groupplot}
	\end{tikzpicture}
	}
	{
		\includegraphics{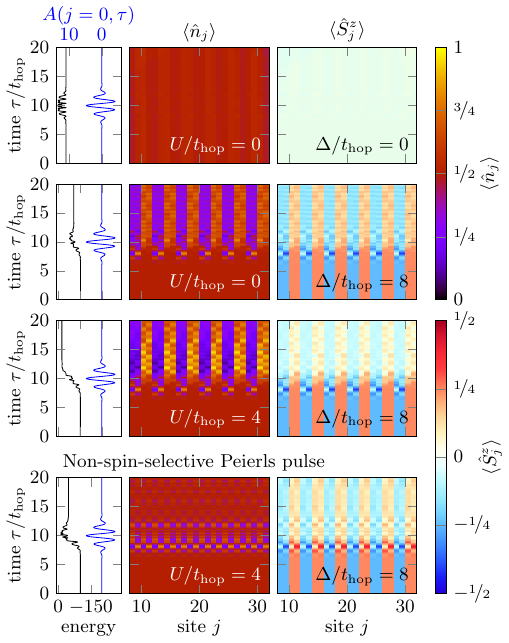}
	}
	\caption
	{
		Time evolution of system~\eqref{eq:model} with $L=40$ sites from \gls{tDMRG} at quarter filling excited by a pump pulse as discussed in the text. 
		First column: total energy of the system (black) and the modulation of the vector potential (blue).
		Second column: particle density $\langle \hat{n}_j \rangle(\tau)$ in the bulk (sites 8-32).
		Third column: local magnetizations $\langle \hat{S}^z_j \rangle(\tau)$, also in the bulk. 
		The color bars on the right indicate the values for $\langle \hat{n}_j \rangle(\tau)$ and $\langle \hat{S}^z_j \rangle(\tau)$, respectively.
		The top row shows results for an excitation acting only on spin-down particles in the absence of a magnetic structure, $\Delta = 0$, and without interaction, $U=0$. 
		The second row shows results for the same excitation, but with $\Delta/\hopping = 8$ and $U=0$. 
		In contrast, the third row shows results for the same excitation and also $\Delta/\hopping = 8$ but $U=4$.
		The bottom row shows results for an excitation acting on both spin directions for $\Delta/\hopping = 8$ and $U=4$. 
	}
	\label{fig:DvariableU0_alphaup01}
\end{figure}
We first discuss the time evolution of the total energy of the system and of local observables in real space, in particular the particle and spin densities $\langle \hat n_{j} \rangle(\tau)$ and $\langle \hat S^z_j \rangle (\tau)$, respectively.
In \cref{fig:DvariableU0_alphaup01}, the first two rows display results at $U = 0$, and $\Delta/\hopping = 0$ and $8$, which show the generic behavior when applying a spin-selective photoexcitation discussed in~\cref{sec:formation}.
In the top panel of \cref{fig:DvariableU0_alphaup01} ($U=\Delta = 0$), the ground state exhibits Friedel-like oscillations in the particle density \cite{giamarchi,Bedurftig:1998p457} but its overall time evolution is only weakly affected by the pulse and in particular there is no enhanced charge ordering.
Furthermore, we note that there is no visible energy absorption after the pulse has passed. 
We attribute these observations to our choice of parameters, for which the photoexcitation has approximately no site-dependence throughout the system, so that no significant change in the local observables can be expected.
For finite values of $\Delta$, a gap opens, so that the Friedel-like oscillations are strongly suppressed, leading to the constant charge density in the bulk of the system discussed above.
As expected, the ground state shows a periodic pattern in the local spin densities $\langle S^z_j\rangle$, which follows the magnetic microstructure. 
A finite amount of energy is absorbed by the system from the pulse, leading to a highly excited state. 
The values of the local observables are significantly modified and remain time-dependent also after the pulse has left the system. 
As discussed in~\cref{sec:formation}, a periodic pattern in the charge density is induced, which follows the periodicity of the Zeeman term, and is very stable on the time scales treated here.
This is also true at finite values of $U$, as shown in the third panel of \cref{fig:DvariableU0_alphaup01}, indicating that including scattering between electrons does not necessarily destroy the pattern, at least not on the time-scale shown.    
In all cases, the formation of this \gls{CDP} occurs together with the weakening of the spin pattern. 

In the bottom panel of \cref{fig:DvariableU0_alphaup01}, we present the time evolution with the photoexcitation coupling to both spin directions ($\Delta/\hopping=8$ and $U/\hopping=4$).
As expected from \gls{OISTR}, we observe a redistribution of magnetic moments, which lead to a weakening of the spin pattern.
Note that even though a significant energy absorption takes place, no stable charge pattern is obtained. 
Hence, the spin-selective excitation is essential to create the \gls{CDP}.

\begin{figure}[!ht]	% D8U0_8_quarter_filling_8888
	\centering	
	\newcommand{\maxTinLongtime}{20}
%	\immediate\write18{gnuplot -e "tmax=\maxTinLongtime" -e "FILE=\string\"data/alpha_up0/alpha_down1/beta_up0/beta_down0/b_z8,-8/U0/e020/sigma6000/t010/k00.00126/zeeman_coupling0/ChainLength40/TargetSymmetrySector20,0/RealTimeStepWidth0.01/RealTimeCheckpointInterval10/RealTime100/rte/loc_observables/n_hub\string\"" -e "OUTPUT=\string\"data/alpha_up0/alpha_down1/beta_up0/beta_down0/b_z8,-8/U0/e020/sigma6000/t010/k00.00126/zeeman_coupling0/ChainLength40/TargetSymmetrySector20,0/RealTimeStepWidth0.01/RealTimeCheckpointInterval10/RealTime100/rte/loc_observables/plain_N.pdf\string\"" data/plain_N.plt}
%	\immediate\write18{gnuplot -e "tmax=\maxTinLongtime" -e "FILE=\string\"data/alpha_up0/alpha_down1/beta_up0/beta_down0/b_z8,-8/U8/e020/sigma6000/t010/k00.00126/zeeman_coupling0/ChainLength40/TargetSymmetrySector20,0/RealTimeStepWidth0.01/RealTimeCheckpointInterval10/RealTime100/rte/loc_observables/n_hub\string\"" -e "OUTPUT=\string\"data/alpha_up0/alpha_down1/beta_up0/beta_down0/b_z8,-8/U8/e020/sigma6000/t010/k00.00126/zeeman_coupling0/ChainLength40/TargetSymmetrySector20,0/RealTimeStepWidth0.01/RealTimeCheckpointInterval10/RealTime100/rte/loc_observables/plain_N.pdf\string\"" data/plain_N.plt}
	\ifthenelse{\boolean{buildtikzpics}} %buildtikzpics
	{
		\tikzsetnextfilename{D8U0_8_quarter_filling_8}
		\begin{tikzpicture}
			\pgfplotsset
			{
				/pgfplots/colormap={gnuplot}{rgb255=(0,0,0) rgb255=(46,0,53) rgb255=(65,0,103) rgb255=(80,0,149) rgb255=(93,0,189) rgb255=(104,1,220) rgb255=(114,2,242) rgb255=(123,3,253) rgb255=(131,4,253) rgb255=(139,6,242) rgb255=(147,9,220) rgb255=(154,12,189) rgb255=(161,16,149) rgb255=(167,20,103) rgb255=(174,25,53) rgb255=(180,31,0) rgb255=(186,38,0) rgb255=(191,46,0) rgb255=(197,55,0) rgb255=(202,64,0) rgb255=(208,75,0) rgb255=(213,87,0) rgb255=(218,100,0) rgb255=(223,114,0) rgb255=(228,130,0) rgb255=(232,147,0) rgb255=(237,165,0) rgb255=(241,185,0) rgb255=(246,207,0) rgb255=(250,230,0) rgb255=(255,255,0) }
			}
			\begin{groupplot}
				[
					group style = 
					{
						group size 			=	2 by 2,
						vertical sep		=	3mm,
						horizontal sep		=	1mm,
						x descriptions at	=	edge bottom,
						y descriptions at	=	edge left
					},
					height 			= 	0.15\textheight,
					width 			= 	0.31\textwidth,
					xmin 			= 	8, 
					xmax 			= 	32,
					ymin			=	0,
					ymax			=	\maxTinLongtime,	
					ylabel			= 	{time $\tau/\hopping$},
				    point meta min	=	0,
				    point meta max	=	1
				]
					\nextgroupplot
					[
						enlargelimits	=	false,
						axis on top, 
						width			= 	0.175\textwidth,
						x dir			=	reverse,
						xmin			= 	-290,
						xmax			=	10,
						ymin			=	0,
						ymax			=	\maxTinLongtime,
						samples			=	2000,
						xticklabel pos=right,
						xlabel={$A(j=0,\tau)$},					
						xlabel style		=	{yshift=-0.5em,color = blue},
						xtick	= {-200, -50},
						xticklabels={$0$, $10$},
						xticklabel style={color = blue}
					]
					\addplot
					[
						color				=	black,
						unbounded coords	=	jump, 
						mark size			=	2pt,
						restrict y to domain				=	0:\maxTinLongtime,
					] 
					table
					[
						x expr = \thisrowno{1}, 
						y expr = \thisrowno{0},
					]
					{data/alpha_up0/alpha_down1/beta_up0/beta_down0/b_z8,-8/U0/e020/sigma6000/t010/k00.00126/zeeman_coupling0/ChainLength40/TargetSymmetrySector20,0/RealTimeStepWidth0.01/RealTimeCheckpointInterval10/RealTime100/rte/sim_data};
					\addplot
					[
						domain	=	0:\maxTinLongtime, 
						blue, 
					]
					({peierlspotential(0, 20, 0.00126, 10, 3374.85, 6000)*15-200}, {x});
					\nextgroupplot
					[
						enlargelimits	=	false,
						axis on top,
						colorbar right,
						every colorbar/.append style =
						{
							height	=	2*\pgfkeysvalueof{/pgfplots/parent axis height} + 1*\pgfkeysvalueof{/pgfplots/group/vertical sep},
							ylabel	=	{$\langle \hat n_j \rangle$},
							width	=	2mm
						},
						extra description/.code =
						{
							\node[anchor = south east, white] at (0.99,0.01) {\(U=0\)};
						},
						title			=	{$\langle \hat n_j \rangle$},
						title style		=	{at={(0.5,1)}, yshift=-0.7em}
					]
					\addplot graphics
					[ 
						xmin	=	0, 
						xmax	=	40, 
						ymin	=	0, 
						ymax	=	\maxTinLongtime
					]
					{data/alpha_up0/alpha_down1/beta_up0/beta_down0/b_z8,-8/U0/e020/sigma6000/t010/k00.00126/zeeman_coupling0/ChainLength40/TargetSymmetrySector20,0/RealTimeStepWidth0.01/RealTimeCheckpointInterval10/RealTime100/rte/loc_observables/plain_N.pdf};
					\nextgroupplot
					[
						enlargelimits	=	false,
						axis on top, 
						width			= 	0.175\textwidth,
						x dir			=	reverse,
						xmin			= 	-290,
						xmax			=	10,
						ymin			=	0,
						ymax			=	\maxTinLongtime,
						xtick distance	=	150,
						samples			=	2000,
						xlabel			= 	{energy},
					]
					\addplot
					[
						color				=	black,
						unbounded coords	=	jump, 
						mark size			=	2pt,
						restrict y to domain				=	0:\maxTinLongtime,
					] 
					table
					[
						x expr = \thisrowno{1}, 
						y expr = \thisrowno{0},
					]
					{data/alpha_up0/alpha_down1/beta_up0/beta_down0/b_z8,-8/U8/e020/sigma6000/t010/k00.00126/zeeman_coupling0/ChainLength40/TargetSymmetrySector20,0/RealTimeStepWidth0.01/RealTimeCheckpointInterval10/RealTime100/rte/sim_data};
					\addplot
					[
						domain	=	0:\maxTinLongtime, 
						blue, 
					]
					({peierlspotential(0, 20, 0.00126, 10, 3374.85, 6000)*15-200}, {x});
					\nextgroupplot
					[
						enlargelimits	=	false,
						axis on top,
						xlabel			= 	{site $j$},
						extra description/.code =
						{
							\node[anchor = south east, white] at (0.99,0.01) {\(U/\hopping=8\)};
						}
					]
					\addplot graphics
					[ 
						xmin	=	0, 
						xmax	=	40, 
						ymin	=	0, 
						ymax	=	\maxTinLongtime
					]
					{data/alpha_up0/alpha_down1/beta_up0/beta_down0/b_z8,-8/U8/e020/sigma6000/t010/k00.00126/zeeman_coupling0/ChainLength40/TargetSymmetrySector20,0/RealTimeStepWidth0.01/RealTimeCheckpointInterval10/RealTime100/rte/loc_observables/plain_N.pdf};
			\end{groupplot}
		\end{tikzpicture}
	}
	{
		\includegraphics{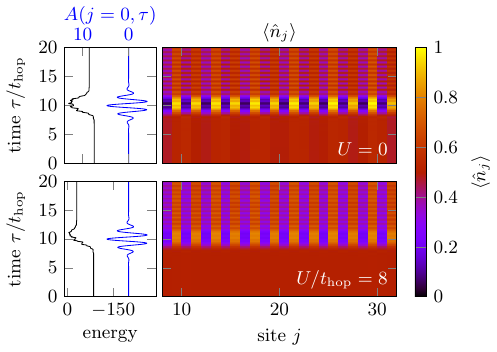}
	}
	\caption
	{
		Time evolution of system~\eqref{eq:model} with $L=40$ sites from \gls{tDMRG} with a unit cell of two sites, i.e., $\Delta=\{8,-8\}$ at (top) $U=0$ and (bottom) $U/\hopping=8$ excited by a spin-selective pump pulse as discussed in the text. 
		First column: total energy of the system (black) and the modulation of the vector potential (blue).
		Second column: particle density $\langle \hat{n}_j \rangle(\tau)$ in the bulk (sites 8-32).
	}
	\label{fig:D8U0_8_quarter_filling_8}
\end{figure}

Similar behavior is also obtained at other than quarter filling, and when changing the periodicity of the magnetic microstructure.
For example, a \gls{CDP} is also formed for N\'eel-type single site magnetic structure, see \cref{fig:D8U0_8_quarter_filling_8}.  

\begin{figure}[!t]	% D8U4_half_vs_quarter_filling
	\centering
	\newcommand{\maxTinLongtime}{20}
	\ifthenelse{\boolean{buildtikzpics}}
	{
	\tikzsetnextfilename{D8U4_half_vs_quarter_filling}
	\begin{tikzpicture}
		\pgfplotsset
		{
			/pgfplots/colormap={gnuplot}{rgb255=(0,0,0) rgb255=(46,0,53) rgb255=(65,0,103) rgb255=(80,0,149) rgb255=(93,0,189) rgb255=(104,1,220) rgb255=(114,2,242) rgb255=(123,3,253) rgb255=(131,4,253) rgb255=(139,6,242) rgb255=(147,9,220) rgb255=(154,12,189) rgb255=(161,16,149) rgb255=(167,20,103) rgb255=(174,25,53) rgb255=(180,31,0) rgb255=(186,38,0) rgb255=(191,46,0) rgb255=(197,55,0) rgb255=(202,64,0) rgb255=(208,75,0) rgb255=(213,87,0) rgb255=(218,100,0) rgb255=(223,114,0) rgb255=(228,130,0) rgb255=(232,147,0) rgb255=(237,165,0) rgb255=(241,185,0) rgb255=(246,207,0) rgb255=(250,230,0) rgb255=(255,255,0) }
		}
		\pgfplotsset
		{
			/pgfplots/colormap={temp}{
				rgb255=(36,0,217) 		% 0 "#2400d9"
				rgb255=(25,29,247) 		% 1 "#191df7"
				rgb255=(41,87,255) 		% 2 "#2957ff"
				rgb255=(61,135,255) 	% 3 "#3d87ff"
				rgb255=(87,176,255) 	% 4 "#57b0ff"
				rgb255=(117,211,255) 	% 5 "#75d3ff"
				rgb255=(153,235,255) 	% 6 "#99ebff"
				rgb255=(189,249,255) 	% 7 "#bdf9ff"
				rgb255=(235,255,255) 	% 8 "#d3ffff"
				rgb255=(255,255,235) 	% 9 "#ffffd9"
				rgb255=(255,242,189) 	% 10 "#fff2bd"
				rgb255=(255,214,153) 	% 11 "#ffd699"
				rgb255=(255,172,117) 	% 12 "#ffac75"
				rgb255=(255,120,87) 	% 13 "#ff7857"
				rgb255=(255,61,61) 		% 14 "#ff3d3d"
				rgb255=(247,40,54) 		% 15 "#f72836"
				rgb255=(217,22,48) 		% 16 "#d91630"
				rgb255=(166,0,33)		% 17 "#a60021"
			}
		}
		\begin{groupplot}
			[
				group style = 
				{
					group size 			=	3 by 2,
					vertical sep		=	3.5mm,
					horizontal sep		=	1mm,
					x descriptions at	=	edge bottom,
					y descriptions at	=	edge left
				},
				height 	= 	0.15\textheight,
				width 	= 	0.22\textwidth,
				xmin 	= 	8, 
				xmax 	= 	32,
				ymin	=	0,
				ymax	=	\maxTinLongtime,	
				ylabel	= 	{time $\tau/\hopping$},
				ylabel shift = -4pt,
			    point meta min=0,
			    point meta max=1
			]
				\nextgroupplot
				[
					enlargelimits	=	false,
					axis on top, 
					width			= 	0.15\textwidth,
					x dir			=	reverse,
					xmin			= 	-290,
					xmax			=	10,
					ymin			=	0,
					ymax			=	\maxTinLongtime,
					samples			=	2000,
					xticklabel pos=right,
					xlabel={$A(j=0,\tau)$},					
					xlabel style		=	{yshift=-0.5em,color = blue},
					xtick	= {-200, -50},
					xticklabels={$0$, $10$},
					xticklabel style={color = blue}
				]
				\addplot
				[
					color=black,
					unbounded coords=jump, 
					mark size = 2pt,
				] 
				table
				[
					x expr = \thisrowno{1}, 
					y expr = \thisrowno{0},
				]
				{data/alpha_up0/alpha_down1/beta_up0/beta_down0/b_z8,8,-8,-8/U4/e020/sigma6000/t010/k00.00126/zeeman_coupling0/ChainLength40/TargetSymmetrySector20,0/RealTimeStepWidth0.01/rte/sim_data};
				\addplot
				[
					domain		=	0:\maxTinLongtime, 
					blue, 
				]
				({peierlspotential(0, 20, 0.00126, 10, 3374.85, 6000)*15-200}, {x});

				\nextgroupplot
				[
					enlargelimits	=	false,
					axis on top,
					extra description/.code={\node[anchor=south east,white] at (0.99,0.01) {quarter filling};},
					title			=	{$\langle \hat n_j \rangle$},
					title style		=	{at={(0.5,1)}, yshift=-0.7em}
				]
				\addplot graphics
				[ 
					xmin = 0, 
					xmax = 40, 
					ymin = 0, 
					ymax = \maxTinLongtime
				]
				{data/alpha_up0/alpha_down1/beta_up0/beta_down0/b_z8,8,-8,-8/U4/e020/sigma6000/t010/k00.00126/zeeman_coupling0/ChainLength40/TargetSymmetrySector20,0/RealTimeStepWidth0.01/rte/loc_observables/plain_N_extCB.pdf};
				\nextgroupplot
				[
					enlargelimits	=	false,
					axis on top,
%					xlabel			= 	{site $i$},
					title			=	{$\langle \hat S^z_j \rangle$},
					title style		=	{at={(0.5,1)}, yshift=-0.7em},
					xtick	= {10, 20, 30},
					xticklabels={$10$, $20$, $30$},
					xticklabel style={color = white},
					colorbar right,
					colormap name=gnuplot,
				    point meta min=0,
				    point meta max=2,
					every colorbar/.append style =
					{
						height	=	1*\pgfkeysvalueof{/pgfplots/parent axis height} + 0*\pgfkeysvalueof{/pgfplots/group/vertical sep},
						ylabel	=	{$\langle \hat n_j \rangle$},
						width	=	2mm,				
						ytick	= {0, 0.5, 1, 1.5, 2},
						yticklabels={$0$, $\nicefrac{1}{2}$, $1$, $\nicefrac{3}{2}$, $2$},
						ylabel shift = -4pt,
					}
				]
				\addplot graphics
				[ 
					xmin = 0, 
					xmax = 40, 
					ymin = 0, 
					ymax = \maxTinLongtime
				]
				{data/alpha_up0/alpha_down1/beta_up0/beta_down0/b_z8,8,-8,-8/U4/e020/sigma6000/t010/k00.00126/zeeman_coupling0/ChainLength40/TargetSymmetrySector20,0/RealTimeStepWidth0.01/rte/loc_observables/plain_S.pdf};

				\nextgroupplot
				[
					enlargelimits	=	false,
					axis on top, 
					width			= 	0.15\textwidth,
					x dir			=	reverse,
					xmin			= 	-290,
					xmax			=	10,
					ymin			=	0,
					ymax			=	\maxTinLongtime,
					xtick distance	=	150,
					samples			=	2000,
					xlabel style	=	{yshift=0.155em},
					xlabel			= 	{energy},
				]
				\addplot
				[
					color=black,
					unbounded coords=jump, 
					mark size = 2pt,
				] 
				table
				[
					x expr = \thisrowno{1}, 
					y expr = \thisrowno{0},
				]
				{data/alpha_up0/alpha_down1/beta_up0/beta_down0/b_z8,8,-8,-8/U4/e020/sigma6000/t010/k00.00126/zeeman_coupling0/ChainLength40/TargetSymmetrySector40,0/RealTimeStepWidth0.01/rte/sim_data};
				\addplot
				[
					domain		=	0:\maxTinLongtime, 
					blue, 
				]
				({peierlspotential(0, 20, 0.00126, 10, 3374.85, 6000)*15-200}, {x});

				\nextgroupplot
				[
					enlargelimits	=	false,
					axis on top,
					xtick	= {10, 20, 30},
					xticklabels={$10$, $20$, $30$},
					xlabel			= 	{site $j$},
					xlabel style	=	{yshift=0.3em},
					extra description/.code={\node[anchor=south east,white] at (0.99,0.01) {half filling};}
				]
				\addplot graphics
				[ 
					xmin = 0, 
					xmax = 40, 
					ymin = 0, 
					ymax = \maxTinLongtime
				]
				{data/alpha_up0/alpha_down1/beta_up0/beta_down0/b_z8,8,-8,-8/U4/e020/sigma6000/t010/k00.00126/zeeman_coupling0/ChainLength40/TargetSymmetrySector40,0/RealTimeStepWidth0.01/rte/loc_observables/plain_N_extCB.pdf};
				\nextgroupplot
				[
					enlargelimits	=	false,
					axis on top,
					xtick	= {10, 20, 30},
					xticklabels={$10$, $20$, $30$},
					xlabel			= 	{site $j$},
					xlabel style	=	{yshift=0.3em},
					colorbar right,
				    point meta min=-0.5,
				    point meta max=0.5,
					every colorbar/.append style =
					{
						height	=	1*\pgfkeysvalueof{/pgfplots/parent axis height} + 0*\pgfkeysvalueof{/pgfplots/group/vertical sep},
						ylabel	=	{$\langle \hat S^z_j \rangle$},
						width	=	2mm,				
						ytick	= {-0.5, -0.25, 0, 0.25, 0.5},
						yticklabels={$-\nicefrac{1}{2}$, $-\nicefrac{1}{4}$, $0$, $\nicefrac{1}{4}$, $\nicefrac{1}{2}$},
						ylabel shift = -12pt,
					}
				]
				\addplot graphics
				[ 
					xmin = 0, 
					xmax = 40, 
					ymin = 0, 
					ymax = \maxTinLongtime
				]
				{data/alpha_up0/alpha_down1/beta_up0/beta_down0/b_z8,8,-8,-8/U4/e020/sigma6000/t010/k00.00126/zeeman_coupling0/ChainLength40/TargetSymmetrySector40,0/RealTimeStepWidth0.01/rte/loc_observables/plain_S.pdf};
		\end{groupplot}
	\end{tikzpicture}
	}
	{
		\includegraphics{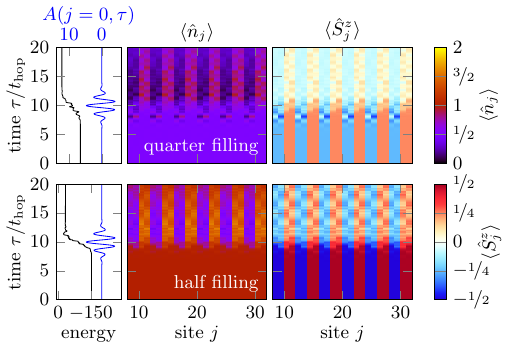}
	}
	\caption
	{
		Time evolution of system~\eqref{eq:model} with $L=40$ sites from \gls{tDMRG} with $U/\hopping=4$ and  $\Delta/\hopping=8$ at (top) quarter filling and (bottom) half filling excited by a spin-selective pump pulse as discussed in the text. 
		First column: total energy of the system (black) and the modulation of the vector potential (blue).
		Second column: particle density $\langle \hat{n}_j \rangle(\tau)$ in the bulk (sites 8-32).
		Third column: local magnetizations $\langle \hat{S}^z_j \rangle(\tau)$, also in the bulk. 
		The color bars on the right indicate the values for $\langle \hat{n}_j \rangle(\tau)$ and $\langle \hat{S}^z_j \rangle(\tau)$, respectively.
		Note that the maximum of the color bar for the particle density is doubled compared to the other figures to capture the results at half filling.
	}
	\label{fig:D8U4_half_vs_quarter_filling}
\end{figure}

As shown in \cref{fig:D8U4_half_vs_quarter_filling} for $U/\hopping = 4$ and $\Delta/\hopping = 8$ at quarter and at half filling, on the time scale displayed, at both values of the filling a \gls{CDP} is obtained.
However, as discussed below, the dominant formation of doublons at half filling leads to a faster decay of the \gls{CDP} in this case. 

\subsection{Effect of the wavelength of the incident light}
\label{sec:LambdaDependence}
\begin{figure}[!h]	% DeltaE_lambda
	\centering
%	\immediate\write18{php data/get_DeltaE_k0.php data/alpha_up0/alpha_down1/beta_up0/beta_down0/b_z8,8,-8,-8/U0/e020/sigma6000/t010/ > data/deltaE_k0_U0.dat}
%	\immediate\write18{php data/get_DeltaE_k0.php data/alpha_up0/alpha_down1/beta_up0/beta_down0/b_z8,8,-8,-8/U4/e020/sigma6000/t010/ > data/deltaE_k0_U4.dat}
	\ifthenelse{\boolean{buildtikzpics}} %buildtikzpics
	{
	\tikzsetnextfilename{DeltaE_lambda}
	\begin{tikzpicture}
		\begin{axis}
		[
			legend style		=	{at = {(0.075,1.0)}, anchor = north west, draw = none, fill = none, legend columns = 2},
			xlabel				=	{wavelength \(\lambda\) [nm]},
			ylabel				=	{\(\Delta E\)},
			width				=	0.501\textwidth, 
			height				=	0.225\textheight,
			legend cell align	=	left,
			xmin				=	138,
			xmax				=	605,
			ymin				=	0,
			ymax				=	97.5,
			smooth
		]
		\addplot
		[
			color=colorA,
			mark=o,
			unbounded coords=jump, 
			mark size = 2pt,
			mark repeat = 4
		] 
		table
		[
			col sep = tab,
			x expr = 2.0*3.1415926/(10.0*\thisrowno{0}), 
			y expr = abs(\thisrowno{1})
		]
		{data/deltaE_k0_U0.dat};
		\addplot
		[
			color=colorB,
			mark=square,
			unbounded coords=jump, 
			mark size = 2pt,
			mark repeat = 4
		] 
		table
		[
			col sep = tab,
			x expr = 2.0*3.1415926/(10.0*\thisrowno{0}), 
			y expr = abs(\thisrowno{1})
		]
		{data/deltaE_k0_U4.dat};
		\legend
		{
			\(U=0\), 
			\(U/\hopping=4\)
		};
		\end{axis}
	\end{tikzpicture}
	}
	{
		\includegraphics{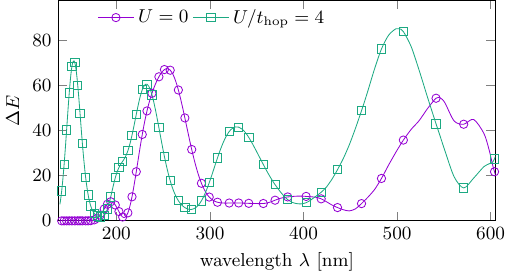}
	}
	\caption
	{
		Absorbed energy $\Delta E = \lvert E_0 - E_{t=20} \lvert$ as function of the wavelength $\lambda$. 
		Here we used the parameters $\Delta/\hopping=8$, $U/\hopping=0, 4$ with quarter filling. 
		The Peierls phase is only acting on the spin\hyp down direction. 
		Most of the calculations in this paper are performed at $\lambda \sim 500$~nm, so that a substantial amount of energy is absorbed.
		Note that for clarity not all computed data points are shown, furthermore a spline interpolation (over all computed data points) is used as guide for the eye.
	}
	\label{fig:DeltaE_lambda}
\end{figure}

In \cref{fig:DeltaE_lambda} the amount of absorbed energy $\Delta E$ as a function of the wavelength $\lambda$ is shown for two interaction strengths $U/\hopping=0$ and $4$.
One obtains a nontrivial dependence of the absorption from the parameters of the system. 
However, at the wavelength of our choice (see \cref{tab:values}) $\lambda=500$~nm, a significant energy absorption in both cases takes place so that we expect the effects studied at this value of $\lambda$ to be representative for the wavelengths, at which absorption takes place.
\section{Stability of the charge\hyp density patterns in various scenarios}
\label{sec:mechanism}
As we have seen in \cref{sec:formation} we generically expect the formation of spatial patterns for spin\hyp selective excitations.
Without interactions the band populations are conserved quantities.
Thus, excitations cannot relax back to the lowest band, and the \gls{CDP} remains stable up to arbitrary times, up to the aforementioned oscillations.
This is in agreement with the findings of \cref{fig:DvariableU0_alphaup01} for $U=0$ and $\Delta/\hopping=8$ for a Peierls pulse of finite duration on the time scales accessible to the \gls{tDMRG}. 
\begin{figure}[!ht]	% omegaU0_doublon_cleaned_2d
	\centering
\makeatletter
\makeatother
	\ifthenelse{\boolean{buildtikzpics}} %buildtikzpics
	{
	\tikzsetnextfilename{omegaU0_doublon_cleaned_2d}
	\begin{tikzpicture}
		\begin{axis}
		[
			legend style		=	{at = {(1.015,1.0)}, anchor = north east, draw = none, fill = none, legend columns = 1, font=\scriptsize},
			legend image post style={scale=0.75},
			xlabel				=	{$\omega$},
			ylabel				=	{$\lvert F(\omega) \rvert$},
			ylabel style		=	{yshift=-3.1047pt},
			width				=	0.51\textwidth, 
			height				=	0.225\textheight,
			legend cell align	=	left,
			xmin				=	0,
			xmax				=	15,
			ymin				=	0,
			ymax				=	10,
			smooth,
			clip mode=individual
		]
			\xdef\firstbandsum{0}
			\xdef\secondbandsum{0}
			\xdef\thirdbandsum{0}
			\xdef\fourthbandsum{0}
			\xdef\sumcounter{0}
			\foreach \i in {-2,-1.99,...,1.99} 
			{
			    \xdef\firstbandsum{\firstbandsum-bandstructure(8,-1,1,\i)}
			    \xdef\secondbandsum{\secondbandsum-bandstructure(8,-1,-1,\i)}
			    \xdef\thirdbandsum{\thirdbandsum+bandstructure(8,1,-1,\i)}
			    \xdef\fourthbandsum{\fourthbandsum+bandstructure(8,1,1,\i)}
				\xdef\sumcounter{\sumcounter+1}
			}
			
			\pgfmathparse{(\firstbandsum)/(\sumcounter)}\let\firstbandsum\pgfmathresult
			\pgfmathparse{(\secondbandsum)/(\sumcounter)}\let\secondbandsum\pgfmathresult
			\pgfmathparse{(\thirdbandsum)/(\sumcounter)}\let\thirdbandsum\pgfmathresult
			\pgfmathparse{(\fourthbandsum)/(\sumcounter)}\let\fourthbandsum\pgfmathresult
			\pgfmathparse{\firstbandsum-\secondbandsum}\let\two\pgfmathresult;
			\pgfmathparse{\secondbandsum+\thirdbandsum}\let\six\pgfmathresult;
			\pgfmathparse{\firstbandsum+\thirdbandsum}\let\eight\pgfmathresult;
			\pgfmathparse{\firstbandsum+\fourthbandsum}\let\ten\pgfmathresult;

			\pgfmathprintnumberto[precision=2]{\two}{\roundedtwo};
			\draw [densely dotted,red,thick] (axis cs:\two,0) -- (axis cs:\two,10) node[above] {{\tiny $\roundedtwo$}};
			\pgfmathprintnumberto[precision=2]{\six}{\roundedsix};
			\draw [dash dot,blue,thick] (axis cs:\six,0) -- (axis cs:\six,10) node[above] {{\tiny $\roundedsix$}};
			\pgfmathprintnumberto[precision=2]{\eight}{\roundedeight};
			\draw [dashed,green,thick] (axis cs:\eight,0) -- (axis cs:\eight,10) node[above] {{\tiny $\roundedeight$}};
			\pgfmathprintnumberto[precision=2]{\ten}{\roundedten};
			\draw [dash dot dot,purple,thick] (axis cs:\ten,0) -- (axis cs:\ten,10) node[above] {{\tiny $\roundedten$}};

			\addplot
			[
				color=colorA,
				unbounded coords=jump, 
				mark size = 2pt,
				mark repeat = 15,
				thick
			] 
			table
			[
				col sep = space,
				x expr = \thisrowno{2}/0.05, 
				y expr = sqrt(\thisrowno{3}*\thisrowno{3} + \thisrowno{4}*\thisrowno{4})
			]
			{data/alpha_up0/alpha_down1/beta_up0/beta_down0/b_z8,8,-8,-8/U0/e020/sigma6000/t010/k00.00126/zeeman_coupling0/ChainLength40/TargetSymmetrySector20,0/RealTimeStepWidth0.01/RealTimeCheckpointInterval10/RealTime1000/ChiMax5000/rte/loc_observables/result};
			\addplot
			[
				color=colorB,
				unbounded coords=jump, 
				mark size = 2pt,
				mark repeat = 15,
				thick
			] 
			table
			[
				col sep = space,
				x expr = \thisrowno{2}/0.05, 
				y expr = sqrt(\thisrowno{3} * \thisrowno{3} + \thisrowno{4} * \thisrowno{4})
			]
			{data/alpha_up0/alpha_down1/beta_up0/beta_down0/b_z8,8,-8,-8/U4/e020/sigma6000/t010/k00.00126/zeeman_coupling0/ChainLength40/TargetSymmetrySector20,0/RealTimeStepWidth0.01/RealTimeCheckpointInterval10/RealTime1000/ChiMax5000/rte/loc_observables/result};
			\addplot
			[
				color=colorC,
				unbounded coords=jump, 
				mark size = 2pt,
				mark repeat = 15,
				thick
			] 
			table
			[
				col sep = space,
				x expr = \thisrowno{2}/0.05, 
				y expr = sqrt(\thisrowno{3} * \thisrowno{3} + \thisrowno{4} * \thisrowno{4})
			]
			{data/alpha_up0/alpha_down1/beta_up0/beta_down0/b_z8,8,-8,-8/U100/e020/sigma6000/t010/k00.00126/zeeman_coupling0/ChainLength40/TargetSymmetrySector20,0/RealTimeStepWidth0.01/RealTimeCheckpointInterval10/RealTime1000/ChiMax5000/rte/loc_observables/result};
	
	        \addplot
			[
				unbounded coords=jump, 
				mark size = 0.1pt,
				only marks
			] 
		        table
				[
					col sep = space,
					x expr = \thisrowno{2}/0.05, 
					y expr = sqrt(\thisrowno{3} * \thisrowno{3} + \thisrowno{4}*\thisrowno{4})
				]
				{data/alpha_up0/alpha_down1/beta_up0/beta_down0/b_z8,8,-8,-8/U0/e020/sigma6000/t010/k00.00126/zeeman_coupling0/ChainLength40/TargetSymmetrySector20,0/RealTimeStepWidth0.01/RealTimeCheckpointInterval10/RealTime1000/ChiMax5000/rte/loc_observables/result};
	        \addplot
			[
				unbounded coords=jump, 
				mark size = 0.1pt,
				only marks
			] 
		        table
				[
					col sep = space,
					x expr = \thisrowno{2}/0.05, 
					y expr = sqrt(\thisrowno{3}*\thisrowno{3} + \thisrowno{4}*\thisrowno{4})
				]
				{data/alpha_up0/alpha_down1/beta_up0/beta_down0/b_z8,8,-8,-8/U4/e020/sigma6000/t010/k00.00126/zeeman_coupling0/ChainLength40/TargetSymmetrySector20,0/RealTimeStepWidth0.01/RealTimeCheckpointInterval10/RealTime1000/ChiMax5000/rte/loc_observables/result};        
			\addplot
			[
				unbounded coords=jump, 
				mark size = 0.1pt,
				only marks
			] 
		        table
				[
					col sep = space,
					x expr = \thisrowno{2}/0.05, 
					y expr = sqrt(\thisrowno{3}*\thisrowno{3} + \thisrowno{4}*\thisrowno{4})
				]
				{data/alpha_up0/alpha_down1/beta_up0/beta_down0/b_z8,8,-8,-8/U100/e020/sigma6000/t010/k00.00126/zeeman_coupling0/ChainLength40/TargetSymmetrySector20,0/RealTimeStepWidth0.01/RealTimeCheckpointInterval10/RealTime1000/ChiMax5000/rte/loc_observables/result};			
			\legend
			{
				$\nicefrac{U}{\hopping}=0$, 
				$\nicefrac{U}{\hopping}=4$,
				$\nicefrac{U}{\hopping}=100$ 
			};
			
		\end{axis}
		\begin{axis}
			[
				ylabel				=	{\(\epsilon_{\nu}(k)/t_{\text{hop}}\)},
				xticklabels			=	{,\(-X\),,\(\Gamma\),,\(X\)},
				width				=	0.18\textwidth, 
				height				=	0.16\textheight,
				xmin				=	-2,
				xmax				=	2,
				ymax				=	6,
				ymin				=	-6,
				samples				=	2000,
				ylabel shift 		= 	-10pt,
				at					=	{(0.11\textwidth,0.055\textheight)},
				axis background/.style	=	{fill=white},
				label style			=	{font=\scriptsize},
                tick label style	=	{font=\scriptsize} 				
			]
			\addplot[black] {bandstructure(8,-1,1,x)};
			\addplot[black] {bandstructure(8,-1,-1,x)};
			\addplot[black] {bandstructure(8,1,-1,x)};
			\addplot[black] {bandstructure(8,1,1,x)};

			\pgfmathparse{bandstructure(8,-1,1,0)}\let\zeroband\pgfmathresult;
			\pgfmathparse{bandstructure(8,-1,-1,0)}\let\firstband\pgfmathresult;
			\pgfmathparse{bandstructure(8,1,-1,0)}\let\secondband\pgfmathresult;
			\pgfmathparse{bandstructure(8,1,1,0)}\let\thirdband\pgfmathresult;

			\draw[->,red,densely dotted] (axis cs:0,\zeroband) to[bend right=0] (axis cs:0,\firstband);
			\draw[->,blue,dash dot] (axis cs:0,\zeroband) to[bend left=40] (axis cs:0,\secondband);
			\draw[->,purple,dash dot dot] (axis cs:0,\zeroband) to[bend right=40] (axis cs:0,\thirdband);
			\draw[->,green,dashed] (axis cs:0,\firstband) to[bend left=20] (axis cs:0,\secondband);
			\draw[->,blue,dash dot] (axis cs:0,\firstband) to[bend right=20] (axis cs:0,\thirdband);
			\draw[->,red,densely dotted] (axis cs:0,\secondband) to[bend right=0] (axis cs:0,\thirdband);
		\end{axis}
	\end{tikzpicture}
	}
	{
		\includegraphics{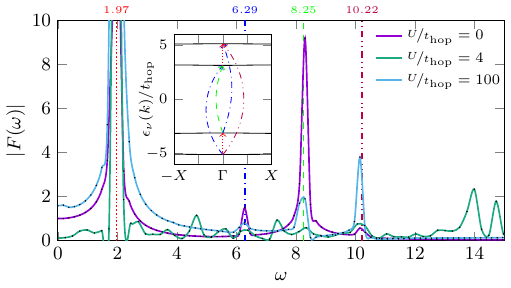}
	}
	\caption
	{
		Fourier-transformed dynamics of $\langle (\hat S^z_j)^2-(\hat S^z_{j+1})^2 \rangle = \langle \hat{n}_{j} - \hat{n}_{j+1} -2\left(\hat{n}_{\uparrow,j}\hat{n}_{\downarrow,j} - \hat{n}_{\uparrow,j+1}\hat{n}_{\downarrow,j+1}\right)\rangle$ with $j$ and $j+1$ on a dimer in the center of the system, for different values of $U$ with $\Delta/\hopping=8$.
		The vertical lines correspond to the average band gap between the first and the second band (red dotted), the first and the third band (blue, dash-dotted), the second and the third band (green, dashed), and the first and the fourth band (wine red, dash-dot-dotted).
		The inset shows the band structure in the noninteracting case (also for $\Delta/\hopping=8$).
		The transitions between the bands, corresponding to the vertical lines in the main figure, are shown with the same styles at $k=\Gamma$.
	}
	\label{fig:omegaU0_doublon_cleaned_2d}
\end{figure}
In \cref{fig:omegaU0_doublon_cleaned_2d} we further elucidate this scenario by considering the frequencies of the density oscillations on a dimer, which we obtain by first subtracting the double occupancies (see below) and then Fourier transforming the result. 
The noninteracting band structure at $\Delta/\hopping=8$ is shown in the inset, and the results are compared to those at finite $U$.
The oscillation frequencies of the particle density within a dimer and the associated averaged band gaps are marked and show excellent agreement. 
This analysis for the noninteracting case can now be used as a starting point to investigate the behavior in the interacting case.

\subsection{Effect of interactions}
The main effect of a finite Hubbard interaction is to induce scattering between the two fermion species and thereby to transfer energy between them, opening a decay channel for the \gls{CDP}. 
However, the Hubbard term also enforces the formation of local moments with finite $\hat{S}^{2}_{j}$, which lower the energy in the staggered field and in this way stabilize the periodic pattern for each fermion species.
Thus, localization of single fermions within the dimers is enforced by the repulsion.
Therefore, we expect the description in terms of the noninteracting system to give at least a qualitative understanding of the dynamics.
%
%\textcolor{red}{FIXME: MERGE BLUE AND GREEN}
%\textcolor{blue}{
%Indeed, at large $U$ a mean-field decoupling in the spins ($\hat S_j^z = \langle \hat S_j^z \rangle + \delta \hat S_j^z$) shows that the Zeeman interaction is shifted according to $\Delta \rightarrow \tilde \Delta = \frac{1}{2}\left(\Delta+4U\right)$ while a strong on-site potential $\propto U$ localizes the fermions and correlates the motion between the two species, see~\cref{app:meanfield}.
%%
%Thus, in the strong coupling limit, the single-particle dynamics for the excited dimers is also dominated by the frequencies of the noninteracting system, indicating a strong localization of single particles on the dimers and hence a stabilization of the \gls{CDP}.
%%
%From \cref{fig:omegaU0_doublon_cleaned_2d} we see that in the regime of intermediate interaction $U \approx \Delta$ there are more decay channels for single-particle excitations.
%%
%However, there is still the dominant contribution at $\omega \approx 2$, i.e., the low energy excitation of the noninteracting single-particle description.
%%
%}

In order to better understand the connection between the two limits we consider the mean field decoupling for $\hopping \ll U$ in more detail.
Decoupling in the spins ($\hat S_j^z = \langle \hat S_j^z \rangle + \delta \hat S_j^z$) shows that the Zeeman interaction is shifted according to $\Delta \rightarrow \tilde \Delta = \frac{1}{2}\left(\Delta+4U\right)$ while a strong on-site potential $\propto U$ localizes the fermions and correlates the motion between the two species, see~\cref{app:meanfield}.
Within this limit, a Peierls pulse redistributes the amplitude of the local moments $\left(\hat{S}^{z}_{i}\right)^{2}$ over the dimers.
The mean-field Hamiltonian (see \cref{app:meanfield}) essentially resembles a Heisenberg $XX$ chain with a strong, staggered magnetic field $\tilde \Delta$.
Thus, relaxation of the local moments after the quench is suppressed with $\tilde{\Delta}$.
The corresponding observable can be written in terms of the local particle densities via $\left(\hat{S}^{z}_{j}\right)^{2} = \frac{1}{4}(\hat{n}_{\uparrow,j}-\hat{n}_{\downarrow,j})^2\propto \hat{n}_{j}-2\hat{n}_{\uparrow,j}\hat{n}_{\downarrow,j}$.
Since the states obtained after the excitation are to a good approximation invariant under translation by one unit cell at all instances of time, the total number of particles in one unit cell can be considered to be conserved, so that we can identify the doublon density $\hat{n}_{\uparrow,j}\hat{n}_{\downarrow,j}$ and its dynamics as the dominating decay channel.
Subtracting the doublon density from the local density, we expect to obtain the single-particle dynamics.
Indeed, in \cref{fig:omegaU0_doublon_cleaned_2d} we see that on the time scales reached by our simulations, the doublon-purified density follows the single-particle dynamics for any value of the interaction strength.
The question arises how interaction effects during the pulse may correlate the two fermion species, thereby reducing the amplitude of the \gls{CDP}.
Note that a doublon consists of two particles of each spin direction and has energy $U$, irrespective of its position.
Thus, in the large $U$-limit, doublons can essentially move throughout the magnetic microstructure at no energy cost.
Therefore, we expect that in the long time limit this yields a homogeneous background particle density.
On top of this background, we expect the reappearance of the behavior of the noninteracting system. Hence, the motion of doublons is one important mechanism for the decay of the \gls{CDP} in the presence of interactions. 
Consequently, creating fewer doublons is beneficial for the strength of the \gls{CDP}.
In contrast, at half filling most of the absorbed energy is used to form doublons, so that the \gls{CDP} will vanish on their delocalization time scale. 
\label{app:LongtimeDensity}

\begin{figure*}[t]	% longtime_N
	\centering
	\newcommand{\maxTinLongtime}{40}
	\ifthenelse{\boolean{buildtikzpics}}
	{
	\tikzsetnextfilename{longtime_N}
	\begin{tikzpicture}
		\pgfplotsset
		{
			/pgfplots/colormap={gnuplot}{rgb255=(0,0,0) rgb255=(46,0,53) rgb255=(65,0,103) rgb255=(80,0,149) rgb255=(93,0,189) rgb255=(104,1,220) rgb255=(114,2,242) rgb255=(123,3,253) rgb255=(131,4,253) rgb255=(139,6,242) rgb255=(147,9,220) rgb255=(154,12,189) rgb255=(161,16,149) rgb255=(167,20,103) rgb255=(174,25,53) rgb255=(180,31,0) rgb255=(186,38,0) rgb255=(191,46,0) rgb255=(197,55,0) rgb255=(202,64,0) rgb255=(208,75,0) rgb255=(213,87,0) rgb255=(218,100,0) rgb255=(223,114,0) rgb255=(228,130,0) rgb255=(232,147,0) rgb255=(237,165,0) rgb255=(241,185,0) rgb255=(246,207,0) rgb255=(250,230,0) rgb255=(255,255,0) }
		}
		\begin{groupplot}
			[
				group style = 
				{
					group size 			=	5 by 3,
					vertical sep		=	4mm,
					horizontal sep		=	1mm,
					x descriptions at	=	edge bottom,
					y descriptions at	=	edge left
				},
				height 	= 	0.16\textheight,
				width 	= 	0.25\textwidth,
				xmin 	= 	8, 
				xmax 	= 	32,
				ymin	=	0,
				ymax	=	\maxTinLongtime,	
				xlabel	=	{site $j$},
				ylabel = {time $\tau/\hopping$},
				ylabel shift = -4pt,
				xlabel shift = -2pt,
			]
				\nextgroupplot
				[
					enlargelimits	=	false,
					axis on top, 
				]
				\addplot graphics
				[
					xmin	=	0,
					xmax	=	40,
					ymin	=	0,
					ymax	=	\maxTinLongtime
				]
				{data/alpha_up0/alpha_down1/beta_up0/beta_down0/b_z8,8,-8,-8/U0/e020/sigma6000/t010/k00.00126/zeeman_coupling0/ChainLength40/TargetSymmetrySector20,0/RealTimeStepWidth0.01/RealTimeCheckpointInterval10/RealTime1000/ChiMax5000/rte/loc_observables/plain_N};

				\nextgroupplot
				[
					enlargelimits	=	false,
					axis on top, 
				]
				\addplot graphics
				[
					xmin	=	0,
					xmax	=	40,
					ymin	=	0,
					ymax	=	\maxTinLongtime
				]
				{data/alpha_up0/alpha_down1/beta_up0/beta_down0/b_z8,8,-8,-8/U2/e020/sigma6000/t010/k00.00126/zeeman_coupling0/ChainLength40/TargetSymmetrySector20,0/RealTimeStepWidth0.01/RealTimeCheckpointInterval10/RealTime1000/ChiMax5000/rte/loc_observables/plain_N};
	
				\nextgroupplot
				[
					enlargelimits	=	false, 
					axis on top, 
				]
				\addplot graphics
				[	xmin	=	0,
					xmax	=	40,
					ymin	=	0,
					ymax	=	\maxTinLongtime
				]
				{data/alpha_up0/alpha_down1/beta_up0/beta_down0/b_z8,8,-8,-8/U4/e020/sigma6000/t010/k00.00126/zeeman_coupling0/ChainLength40/TargetSymmetrySector20,0/RealTimeStepWidth0.01/RealTimeCheckpointInterval10/RealTime1000/ChiMax5000/rte/loc_observables/plain_N};
	
				\nextgroupplot
				[
					enlargelimits	=	false,
					axis on top, 
				]
				\addplot graphics
				[
					xmin	=	0, 
					xmax	=	40,
					ymin	=	0,
					ymax	=	\maxTinLongtime
				]
				{data/alpha_up0/alpha_down1/beta_up0/beta_down0/b_z8,8,-8,-8/U20/e020/sigma6000/t010/k00.00126/zeeman_coupling0/ChainLength40/TargetSymmetrySector20,0/RealTimeStepWidth0.01/RealTimeCheckpointInterval10/RealTime1000/ChiMax5000/rte/loc_observables/plain_N};
	
				\nextgroupplot
				[
					enlargelimits	=	false,
					axis on top,
					colorbar right, 
				    point meta min=0,
				    point meta max=1,
					every colorbar/.append style =
					{
						height	=	1*\pgfkeysvalueof{/pgfplots/parent axis height} + 0*\pgfkeysvalueof{/pgfplots/group/vertical sep},
						width			=	2mm,
						ytick			= 	{0, 0.25, 0.5, 0.75, 1},
						yticklabels		=	{$0$, $\nicefrac{1}{4}$, $\nicefrac{1}{2}$, $\nicefrac{3}{4}$, $1$},
						ylabel shift 	=	-5pt,
						ylabel = {$\langle \hat n_j \rangle$},
					}
				]
				\addplot graphics
				[
					xmin	=	0, 
					xmax	=	40, 
					ymin	=	0, 
					ymax	=	\maxTinLongtime
				]
				{data/alpha_up0/alpha_down1/beta_up0/beta_down0/b_z8,8,-8,-8/U100/e020/sigma6000/t010/k00.00126/zeeman_coupling0/ChainLength40/TargetSymmetrySector20,0/RealTimeStepWidth0.01/RealTimeCheckpointInterval10/RealTime1000/ChiMax5000/rte/loc_observables/plain_N};

				\nextgroupplot
				[
					enlargelimits	=	false,
					axis on top, 
				]
				\addplot graphics
				[
					xmin	=	0,
					xmax	=	40,
					ymin	=	0,
					ymax	=	\maxTinLongtime
				]
				{data/alpha_up0/alpha_down1/beta_up0/beta_down0/b_z8,8,-8,-8/U0/e020/sigma6000/t010/k00.00126/zeeman_coupling0/ChainLength40/TargetSymmetrySector20,0/RealTimeStepWidth0.01/RealTimeCheckpointInterval10/RealTime1000/ChiMax5000/rte/loc_observables/doublon_density};

				\nextgroupplot
				[
					enlargelimits	=	false,
					axis on top, 
				]
				\addplot graphics
				[
					xmin	=	0,
					xmax	=	40,
					ymin	=	0,
					ymax	=	\maxTinLongtime
				]
				{data/alpha_up0/alpha_down1/beta_up0/beta_down0/b_z8,8,-8,-8/U2/e020/sigma6000/t010/k00.00126/zeeman_coupling0/ChainLength40/TargetSymmetrySector20,0/RealTimeStepWidth0.01/RealTimeCheckpointInterval10/RealTime1000/ChiMax5000/rte/loc_observables/doublon_density};
	
				\nextgroupplot
				[
					enlargelimits	=	false, 
					axis on top, 
				]
				\addplot graphics
				[	xmin	=	0,
					xmax	=	40,
					ymin	=	0,
					ymax	=	\maxTinLongtime
				]
				{data/alpha_up0/alpha_down1/beta_up0/beta_down0/b_z8,8,-8,-8/U4/e020/sigma6000/t010/k00.00126/zeeman_coupling0/ChainLength40/TargetSymmetrySector20,0/RealTimeStepWidth0.01/RealTimeCheckpointInterval10/RealTime1000/ChiMax5000/rte/loc_observables/doublon_density};
	
				\nextgroupplot
				[
					enlargelimits	=	false,
					axis on top, 
				]
				\addplot graphics
				[
					xmin	=	0, 
					xmax	=	40,
					ymin	=	0,
					ymax	=	\maxTinLongtime
				]
				{data/alpha_up0/alpha_down1/beta_up0/beta_down0/b_z8,8,-8,-8/U20/e020/sigma6000/t010/k00.00126/zeeman_coupling0/ChainLength40/TargetSymmetrySector20,0/RealTimeStepWidth0.01/RealTimeCheckpointInterval10/RealTime1000/ChiMax5000/rte/loc_observables/doublon_density};
	
				\nextgroupplot
				[
					enlargelimits	=	false,
					axis on top,
					colorbar right, 
				    point meta min=0,
				    point meta max=0.25,
					every colorbar/.append style =
					{
						height	=	1*\pgfkeysvalueof{/pgfplots/parent axis height} + 0*\pgfkeysvalueof{/pgfplots/group/vertical sep},
						width			=	2mm,
						ytick			= 	{0, 0.0625, 0.125, 0.1875, 0.25},
						yticklabels		=	{$0$, $\nicefrac{1}{16}$, $\nicefrac{1}{8}$, $\nicefrac{3}{16}$, $\nicefrac{1}{4}$},
						ylabel shift 	=	-5pt,
						ylabel = {$\langle \hat n_{j,\uparrow} \hat n_{j,\downarrow} \rangle$},
					}
				]
				\addplot graphics
				[
					xmin	=	0, 
					xmax	=	40, 
					ymin	=	0, 
					ymax	=	\maxTinLongtime
				]
				{data/alpha_up0/alpha_down1/beta_up0/beta_down0/b_z8,8,-8,-8/U100/e020/sigma6000/t010/k00.00126/zeeman_coupling0/ChainLength40/TargetSymmetrySector20,0/RealTimeStepWidth0.01/RealTimeCheckpointInterval10/RealTime1000/ChiMax5000/rte/loc_observables/doublon_density};

				\nextgroupplot
				[
					enlargelimits	=	false,
					axis on top, 
					extra description/.code={\node[anchor=south east,white] at (0.99,0.01) {$U/\hopping=0$};}
				]
				\addplot graphics
				[
					xmin	=	8,
					xmax	=	32,
					ymin	=	0,
					ymax	=	\maxTinLongtime
				]
				{data/alpha_up0/alpha_down1/beta_up0/beta_down0/b_z8,8,-8,-8/U0/e020/sigma6000/t010/k00.00126/zeeman_coupling0/ChainLength40/TargetSymmetrySector20,0/RealTimeStepWidth0.01/RealTimeCheckpointInterval10/RealTime1000/ChiMax5000/rte/loc_observables/doublon_cleaned_density};

				\nextgroupplot
				[
					enlargelimits	=	false,
					axis on top, 
					extra description/.code={\node[anchor=south east,white] at (0.99,0.01) {$U/\hopping=2$};}
				]
				\addplot graphics
				[
					xmin	=	8,
					xmax	=	32,
					ymin	=	0,
					ymax	=	\maxTinLongtime
				]
				{data/alpha_up0/alpha_down1/beta_up0/beta_down0/b_z8,8,-8,-8/U2/e020/sigma6000/t010/k00.00126/zeeman_coupling0/ChainLength40/TargetSymmetrySector20,0/RealTimeStepWidth0.01/RealTimeCheckpointInterval10/RealTime1000/ChiMax5000/rte/loc_observables/doublon_cleaned_density};
	
				\nextgroupplot
				[
					enlargelimits	=	false, 
					axis on top, 
					extra description/.code={\node[anchor=south east,white] at (0.99,0.01) {$U/\hopping=4$};}
				]
				\addplot graphics
				[	xmin	=	8,
					xmax	=	32,
					ymin	=	0,
					ymax	=	\maxTinLongtime
				]
				{data/alpha_up0/alpha_down1/beta_up0/beta_down0/b_z8,8,-8,-8/U4/e020/sigma6000/t010/k00.00126/zeeman_coupling0/ChainLength40/TargetSymmetrySector20,0/RealTimeStepWidth0.01/RealTimeCheckpointInterval10/RealTime1000/ChiMax5000/rte/loc_observables/doublon_cleaned_density};
	
				\nextgroupplot
				[
					enlargelimits	=	false,
					axis on top, 
					extra description/.code={\node[anchor=south east,white] at (0.99,0.01) {$U/\hopping=20$};}
				]
				\addplot graphics
				[
					xmin	=	8, 
					xmax	=	32,
					ymin	=	0,
					ymax	=	\maxTinLongtime
				]
				{data/alpha_up0/alpha_down1/beta_up0/beta_down0/b_z8,8,-8,-8/U20/e020/sigma6000/t010/k00.00126/zeeman_coupling0/ChainLength40/TargetSymmetrySector20,0/RealTimeStepWidth0.01/RealTimeCheckpointInterval10/RealTime1000/ChiMax5000/rte/loc_observables/doublon_cleaned_density};
	
				\nextgroupplot
				[
					enlargelimits	=	false,
					axis on top,
					colorbar right, 
					extra description/.code={\node[anchor=south east,white] at (0.99,0.01) {$U/\hopping=100$};},
				    point meta min=0,
				    point meta max=1,
					every colorbar/.append style =
					{
						height	=	1*\pgfkeysvalueof{/pgfplots/parent axis height} + 0*\pgfkeysvalueof{/pgfplots/group/vertical sep},
						width			=	2mm,
						ytick			= 	{0, 0.25, 0.5, 0.75, 1},
						yticklabels		=	{$0$, $\nicefrac{1}{4}$, $\nicefrac{1}{2}$, $\nicefrac{3}{4}$, $1$},
						ylabel shift 	=	-4pt,
						ylabel = {$\langle \hat n_j \rangle - 2 \langle \hat n_{j,\uparrow} \hat n_{j,\downarrow} \rangle$},
					}
				]
				\addplot graphics
				[
					xmin	=	8, 
					xmax	=	32, 
					ymin	=	0, 
					ymax	=	\maxTinLongtime
				]
				{data/alpha_up0/alpha_down1/beta_up0/beta_down0/b_z8,8,-8,-8/U100/e020/sigma6000/t010/k00.00126/zeeman_coupling0/ChainLength40/TargetSymmetrySector20,0/RealTimeStepWidth0.01/RealTimeCheckpointInterval10/RealTime1000/ChiMax5000/rte/loc_observables/doublon_cleaned_density};
		\end{groupplot}
	\end{tikzpicture}
	}
	{
		\includegraphics{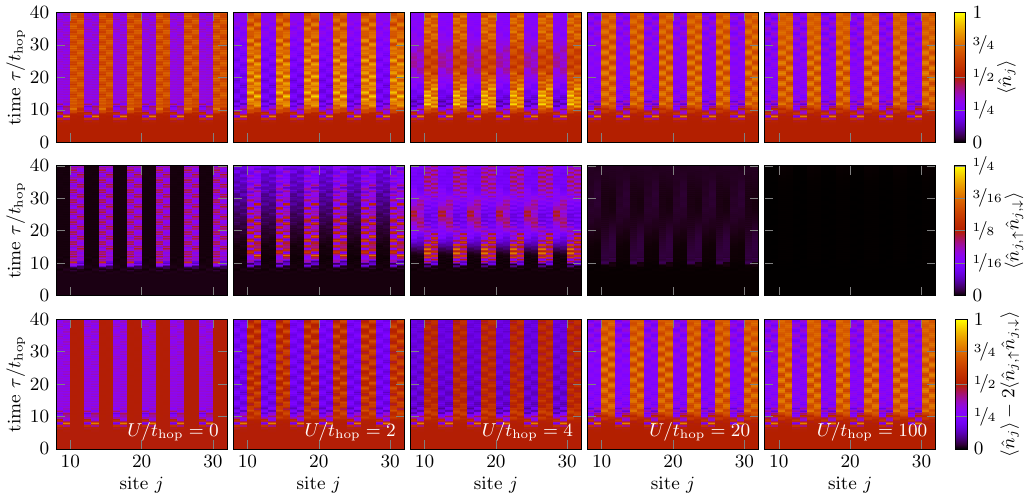}
	}
	\caption
	{
		Local particle density (top), local double occupancy (middle), and the local, doublon-cleaned particle density (bottom) for systems with $\Delta/\hopping=8$, $L=40$, and $U/\hopping=0,\, 2,\, 4,\, 20,\,\text{and } 100$.
		In the case of small but finite $U$ the \gls{CDP} seems to decay in the local particle density plots.
		Considering the double occupancy plots it becomes clear that this decay is due to the creation of the translationally invariant background.
		In the doublon-cleaned particle density the decay is essentially absent on the time scale shown.
		Note that for $U=0$ the double occupancy is trapped on the dimer to which the spin\hyp down¸ particles were moved and that at large $U$ nearly no double occupancies were created in the first place.
		In order to obtain longer times the bond dimension $\chi$ is increased by a factor of $10$, i.e., $\chi=5000$, to reach a discarded weight $\epsilon < 10^{-7}$.
	}
	\label{fig:longtime_N}
\end{figure*}
These considerations are further illustrated by~\cref{fig:longtime_N}, where we show the behavior of the particle density, the double occupancy and the particle density after removing the double occupancy for different values of the interaction $U/\hopping=0,\, 2,\, 4,\, 20,\,\text{and } 100$ for longer times up to $40 \nicefrac{1}{\hopping} \sim 50$fs for the parameters of \cref{tab:values}.% Ref.~\onlinecite{PhysRevB.97.235120}.
The first observation is that at $U=0$ all affected spin\hyp down¸ particles create double occupancies as the particle density on the particle rich dimers stays constant and the \gls{CDP} is only visible due to the particle poor dimers in the doublon-purified particle density.
At finite interaction this is no longer the case and the amplitude of the \gls{CDP} is increased.
On the other hand, the creation of double occupancies is suppressed with further increasing the value of the interaction $U/\hopping$; a superposition of particles each located on one of the sites of a dimer is preferred due to the repulsive interaction.

Hence, subtracting the doublon contribution from the charge density is insightful in the presence of interactions, as the double occupancy will eventually spread equally over the whole system as discussed above.
Therefore, the density after subtracting the double occupancy part is indicative for the long time behavior of the \gls{CDP}.

\subsection{Low photon densities}
\label{sec:singlephoton}
So far, we have considered Peierls pulses, which due to their semi-classical ansatz model a situation with a high photon number, as realized in ultrafast pump-probe setups. 
In contrast, e.g., photoexcitations by sunlight carry only a very small number of photons per unit of time, so that it would be interesting to check for the formation of the \gls{CDP} in this situation as well.
Also, it is an important test to see whether the observed effects are artifacts due to the generalized Peierls substitution.
We model such an excitation through the application of creation and annihilation operators in $k$-space, $\hat{a}^{\left(\dagger\right)}_{\sigma,\nu}\!\left(k\right)$ (see \cref{app:Ham_obc} for the definition of these operators for systems with \gls{OBC}), to the ground state of the system $\ket{\psi_0}$, where in the noninteracting case, with \gls{PBC}, the operators can be chosen such that the excitation over a band gap without transfer of momentum is modeled. 

\begin{figure}[!t]
		\newcommand{\maxTinLongtime}{20}

	\ifthenelse{\boolean{buildtikzpics}}
	{
		\def\Emin{-112.5}
		\def\Emax{-17.5}
		\def\DeltaE{45}
		\tikzsetnextfilename{DvariableU0_alphaup01_single_photon}
%		\tikzset{external/export next=false}
		\begin{tikzpicture}
		\pgfplotsset
		{
			/pgfplots/colormap={gnuplot}{rgb255=(0,0,0) rgb255=(46,0,53) rgb255=(65,0,103) rgb255=(80,0,149) rgb255=(93,0,189) rgb255=(104,1,220) rgb255=(114,2,242) rgb255=(123,3,253) rgb255=(131,4,253) rgb255=(139,6,242) rgb255=(147,9,220) rgb255=(154,12,189) rgb255=(161,16,149) rgb255=(167,20,103) rgb255=(174,25,53) rgb255=(180,31,0) rgb255=(186,38,0) rgb255=(191,46,0) rgb255=(197,55,0) rgb255=(202,64,0) rgb255=(208,75,0) rgb255=(213,87,0) rgb255=(218,100,0) rgb255=(223,114,0) rgb255=(228,130,0) rgb255=(232,147,0) rgb255=(237,165,0) rgb255=(241,185,0) rgb255=(246,207,0) rgb255=(250,230,0) rgb255=(255,255,0) }
		}
		\pgfplotsset
		{
			/pgfplots/colormap={temp}{
				rgb255=(36,0,217) 		% 0 "#2400d9"
				rgb255=(25,29,247) 		% 1 "#191df7"
				rgb255=(41,87,255) 		% 2 "#2957ff"
				rgb255=(61,135,255) 	% 3 "#3d87ff"
				rgb255=(87,176,255) 	% 4 "#57b0ff"
				rgb255=(117,211,255) 	% 5 "#75d3ff"
				rgb255=(153,235,255) 	% 6 "#99ebff"
				rgb255=(189,249,255) 	% 7 "#bdf9ff"
				rgb255=(235,255,255) 	% 8 "#d3ffff"
				rgb255=(255,255,235) 	% 9 "#ffffd9"
				rgb255=(255,242,189) 	% 10 "#fff2bd"
				rgb255=(255,214,153) 	% 11 "#ffd699"
				rgb255=(255,172,117) 	% 12 "#ffac75"
				rgb255=(255,120,87) 	% 13 "#ff7857"
				rgb255=(255,61,61) 		% 14 "#ff3d3d"
				rgb255=(247,40,54) 		% 15 "#f72836"
				rgb255=(217,22,48) 		% 16 "#d91630"
				rgb255=(166,0,33)		% 17 "#a60021"
			}
		}
		\begin{groupplot}
		[
		group style = 
		{
			group size 			=	3 by 5,
			vertical sep		=	3.5mm,
			horizontal sep		=	1.4mm,
			x descriptions at	=	edge bottom,
			y descriptions at	=	edge left
		},
		height			= 	0.15\textheight,
		width 			= 	0.22\textwidth,
		xmin 			= 	8, 
		xmax 			= 	32,
		ymin			=	-5,
		ymax			=	\maxTinLongtime,	
		ylabel			= 	{time $\tau/\hopping$},
		ylabel shift 	=	-4pt,
		point meta min	=	0,
		point meta max	=	1
		]
		\nextgroupplot
		[
		enlargelimits		=	false,
		axis on top, 
		width				= 	0.15\textwidth,
		x dir				=	reverse,
		xmin				= 	\Emin,
		xmax				=	\Emax,
		xtick distance		=	\DeltaE,
		ymin				=	-5,
		ymax				=	\maxTinLongtime,
		samples				=	20,
		xticklabel pos=right,
		]
			\addplot
			[
				color				=	black,
				unbounded coords	=	jump, 
				mark size			=	2pt,
			] 
				table
				[
					x expr = \thisrowno{2}/\thisrowno{1}, 
					y expr = \thisrowno{0},
				]
					{data/low_photon_numbers/obc_n_40_u_0_delta_0_q_excitation_momentum_space_kin_19_kout_0_cdw/pump/global_operators_real};
			\pgfplotstableread[header=false]{data/low_photon_numbers/obc_n_40_u_0_delta_0_q_excitation_momentum_space_kin_19_kout_0_cdw/groundstatesearch/gs_energy}{\DataGS}%
			\pgfplotstablegetelem{0}{0}\of\DataGS%
			\pgfmathsetmacro{\gs}{\pgfplotsretval}%
			\addplot+
			[
				color				=	black,
				unbounded coords	=	jump, 
				mark size			=	2pt,
				domain	= -5:0,
				no marks,
			] (\gs,x);
		
		\nextgroupplot
		[
			enlargelimits	=	false,
			axis on top,
			title			=	{$\langle \hat n_j \rangle$},
			title style		=	{at = {(0.5,1)}, yshift = -0.7em},
			extra description/.code={\node[anchor=south east,white] at (0.99,0.01) {$U/\hopping=0$};},
		]
			\addplot graphics
			[ 
				xmin	=	0, 
				xmax	=	40, 
				ymin	=	-5, 
				ymax	=	1
			]
				{data/low_photon_numbers/obc_n_40_u_0_delta_0_q_excitation_momentum_space_kin_19_kout_0_cdw/groundstatesearch/plain_N_gs.pdf};
			\addplot graphics
			[ 
				xmin	=	0, 
				xmax	=	40, 
				ymin	=	0, 
				ymax	=	\maxTinLongtime
			]
				{data/low_photon_numbers/obc_n_40_u_0_delta_0_q_excitation_momentum_space_kin_19_kout_0_cdw/pump/plain_N.pdf};

		\nextgroupplot
		[
			enlargelimits	=	false,
			axis on top,
			colorbar right,
			colormap name	=	gnuplot,
			every colorbar/.append style =
			{
				height			=	2*\pgfkeysvalueof{/pgfplots/parent axis height} + 1*\pgfkeysvalueof{/pgfplots/group/vertical sep},
				ylabel			=	{$\langle \hat n_j \rangle$},
				width			=	2mm,
				ytick			= 	{0, 0.25, 0.5, 0.75, 1},
				yticklabels		=	{$0$, $\nicefrac{1}{4}$, $\nicefrac{1}{2}$, $\nicefrac{3}{4}$, $1$},
				ylabel shift 	=	-4pt,
			},
			extra description/.code =
			{
				\node[anchor = south east] at (0.99,0.01) {\(\Delta/\hopping=0\)};
			},
			title			=	{$\langle \hat S^z_j \rangle$},
			title style		=	{at = {(0.5,1)}, yshift = -0.7em}
		]
			\addplot graphics
			[ 
				xmin	=	0, 
				xmax	=	40, 
				ymin	=	-5, 
				ymax	=	1
			]
				{data/low_photon_numbers/obc_n_40_u_0_delta_0_q_excitation_momentum_space_kin_19_kout_0_cdw/groundstatesearch/plain_S_gs.pdf};
			\addplot graphics
			[ 
				xmin	=	0, 
				xmax	=	40, 
				ymin	=	0, 
				ymax	=	\maxTinLongtime
			]
				{data/low_photon_numbers/obc_n_40_u_0_delta_0_q_excitation_momentum_space_kin_19_kout_0_cdw/pump/plain_S.pdf};
		
		\nextgroupplot
		[
			enlargelimits	=	false,
			axis on top, 
			width			= 	0.15\textwidth,
			x dir			=	reverse,
			xmin				= 	\Emin,
			xmax				=	\Emax,
			xtick distance		=	\DeltaE,
			ymin			=	-5,
			ymax			=	\maxTinLongtime,
			samples			=	2000,
		]
			\addplot
			[
				color				=	black,
				unbounded coords	=	jump, 
				mark size			=	2pt,
			] 
				table
				[
					x expr = \thisrowno{2}/\thisrowno{1}, 
					y expr = \thisrowno{0},
				]
				{data/low_photon_numbers/obc_n_40_u_0_delta_8_q_excitation_momentum_space_kin_19_kout_0_cdw/pump/global_operators_real};
			\pgfplotstableread[header=false]{data/low_photon_numbers/obc_n_40_u_0_delta_8_q_excitation_momentum_space_kin_19_kout_0_cdw/groundstatesearch/gs_energy}{\DataGS}%
			\pgfplotstablegetelem{0}{0}\of\DataGS%
			\pgfmathsetmacro{\gs}{\pgfplotsretval}%
			\addplot+
			[
				color				=	black,
				unbounded coords	=	jump, 
				mark size			=	2pt,
				domain	= -5:0,
				no marks,
			] (\gs,x);
		
		\nextgroupplot
		[
			enlargelimits	=	false,
			axis on top,
			extra description/.code={\node[anchor=south east,white] at (0.99,0.01) {$U/\hopping=0$};},
		]
			\addplot graphics
			[ 
				xmin	=	0, 
				xmax	=	40, 
				ymin	=	-5, 
				ymax	=	1
			]
			{data/low_photon_numbers/obc_n_40_u_0_delta_8_q_excitation_momentum_space_kin_19_kout_0_cdw/groundstatesearch/plain_N_gs.pdf};
			\addplot graphics
			[ 
				xmin	=	0, 
				xmax	=	40, 
				ymin	=	0, 
				ymax	=	\maxTinLongtime
			]
			{data/low_photon_numbers/obc_n_40_u_0_delta_8_q_excitation_momentum_space_kin_19_kout_0_cdw/pump/plain_N.pdf};

		\nextgroupplot
		[
			enlargelimits	=	false,
			axis on top,
			extra description/.code =
			{
				\node[anchor = south east] at (0.99,0.01) {\(\Delta/\hopping=8\)};
			}
		]
			\addplot graphics
			[ 
				xmin	=	0, 
				xmax	=	40, 
				ymin	=	-5, 
				ymax	=	1
			]
				{data/low_photon_numbers/obc_n_40_u_0_delta_8_q_excitation_momentum_space_kin_19_kout_0_cdw/groundstatesearch/plain_S_gs.pdf};
			\addplot graphics
			[ 
				xmin	=	0, 
				xmax	=	40, 
				ymin	=	0, 
				ymax	=	\maxTinLongtime
			]
				{data/low_photon_numbers/obc_n_40_u_0_delta_8_q_excitation_momentum_space_kin_19_kout_0_cdw/pump/plain_S.pdf};
		
		\nextgroupplot
		[
			enlargelimits	=	false,
			axis on top, 
			width			= 	0.15\textwidth,
			x dir			=	reverse,
			xmin				= 	\Emin,
			xmax				=	\Emax,
			xtick distance		=	\DeltaE,
			ymin			=	-5,
			ymax			=	\maxTinLongtime,
			samples			=	2000,
		]
			\addplot
			[
				color				=	black,
				unbounded coords	=	jump, 
				mark size			=	2pt,
			] 
				table
				[
					x expr = \thisrowno{2}/\thisrowno{1}, 
					y expr = \thisrowno{0},
				]
				{data/low_photon_numbers/obc_n_40_u_4_delta_8_q_excitation_momentum_space_kin_19_kout_0_cdw/pump/global_operators_real};
			\pgfplotstableread[header=false]{data/low_photon_numbers/obc_n_40_u_4_delta_8_q_excitation_momentum_space_kin_19_kout_0_cdw/groundstatesearch/gs_energy}{\DataGS}%
			\pgfplotstablegetelem{0}{0}\of\DataGS%
			\pgfmathsetmacro{\gs}{\pgfplotsretval}%
			\addplot+
			[
				color				=	black,
				unbounded coords	=	jump, 
				mark size			=	2pt,
				domain	= -5:0,
				no marks,
			] (\gs,x);

		\nextgroupplot
		[
			enlargelimits	=	false,
			axis on top,
			extra description/.code={\node[anchor=south east,white] at (0.99,0.01) {$U/\hopping=4$};},
		]
			\addplot graphics
			[ 
				xmin	=	0, 
				xmax	=	40, 
				ymin	=	-5, 
				ymax	=	1
			]
			{data/low_photon_numbers/obc_n_40_u_4_delta_8_q_excitation_momentum_space_kin_19_kout_0_cdw/groundstatesearch/plain_N_gs.pdf};
			\addplot graphics
			[ 
				xmin	=	0, 
				xmax	=	40, 
				ymin	=	0, 
				ymax	=	\maxTinLongtime
			]
			{data/low_photon_numbers/obc_n_40_u_4_delta_8_q_excitation_momentum_space_kin_19_kout_0_cdw/pump/plain_N.pdf};
		\nextgroupplot
		[
			enlargelimits	=	false,
			axis on top,
			colorbar right,
			,			    point meta min	=	-0.5,
			point meta max	=	0.5,
			every colorbar/.append style =
			{
				height			=	3.5mm + 2*\pgfkeysvalueof{/pgfplots/parent axis height} + 1*\pgfkeysvalueof{/pgfplots/group/vertical sep},
				ylabel			=	{$\langle \hat S^z_j \rangle$},
				width			=	2mm,				
				ytick			=	{-0.5, -0.25, 0, 0.25, 0.5},
				yticklabels		=	{$-\nicefrac{1}{2}$, $-\nicefrac{1}{4}$, $0$, $\nicefrac{1}{4}$, $\nicefrac{1}{2}$},
				ylabel shift	=	-12pt,
			},
			extra description/.code =
			{
				\node[anchor = south east] at (0.99,0.01) {\(\Delta/\hopping=8\)};
			}
		]
			\addplot graphics
			[ 
				xmin	=	0, 
				xmax	=	40, 
				ymin	=	-5, 
				ymax	=	1
			]
				{data/low_photon_numbers/obc_n_40_u_4_delta_8_q_excitation_momentum_space_kin_19_kout_0_cdw/groundstatesearch/plain_S_gs.pdf};
			\addplot graphics
			[ 
				xmin	=	0, 
				xmax	=	40, 
				ymin	=	0, 
				ymax	=	\maxTinLongtime
			]
				{data/low_photon_numbers/obc_n_40_u_4_delta_8_q_excitation_momentum_space_kin_19_kout_0_cdw/pump/plain_S.pdf};
		\nextgroupplot[group/empty plot, y = 0.001pt]
		\nextgroupplot[group/empty plot, y = 0.001pt]
		\nextgroupplot[group/empty plot, y = 0.001pt]
		\nextgroupplot
		[
			enlargelimits	=	false,
			axis on top, 
			width			= 	0.15\textwidth,
			x dir			=	reverse,
			xmin			= 	\Emin,
			xmax			=	\Emax,
			xtick distance	=	\DeltaE,
			ymin			=	-5,
			ymax			=	\maxTinLongtime,
			samples			=	2000,
			xlabel style	=	{yshift=0.155em},
			xlabel			= 	{energy},
			title			=	{Non-spin-selective excitation},
			title style 	= 	{at={(0,1.025)}, anchor=west}
		]
			\addplot
			[
				color				=	black,
				unbounded coords	=	jump, 
				mark size			=	2pt,
			] 
				table
				[
					x expr = \thisrowno{2}/\thisrowno{1}, 
					y expr = \thisrowno{0},
				]
				{data/low_photon_numbers/obc_n_40_u_4_delta_8_q_excitation_momentum_space_kin_19_kout_0_cdw_non_spin_selective/pump/global_operators_real};
			\pgfplotstableread[header=false]{data/low_photon_numbers/obc_n_40_u_4_delta_8_q_excitation_momentum_space_kin_19_kout_0_cdw_non_spin_selective/groundstatesearch/gs_energy}{\DataGS}%
			\pgfplotstablegetelem{0}{0}\of\DataGS%
			\pgfmathsetmacro{\gs}{\pgfplotsretval}%
			\addplot+
			[
				color				=	black,
				unbounded coords	=	jump, 
				mark size			=	2pt,
				domain	= -5:0,
				no marks,
			] (\gs,x);
		\nextgroupplot
		[
			enlargelimits	=	false,
			axis on top,
			xlabel style	=	{yshift=0.3em},
			xlabel			= 	{site $j$},
			extra description/.code={\node[anchor=south east,white] at (0.99,0.01) {$U/\hopping=4$};},
		]
			\addplot graphics
			[ 
				xmin	=	0, 
				xmax	=	40, 
				ymin	=	-5, 
				ymax	=	1
			]
			{data/low_photon_numbers/obc_n_40_u_4_delta_8_q_excitation_momentum_space_kin_19_kout_0_cdw_non_spin_selective/groundstatesearch/plain_N_gs.pdf};
			\addplot graphics
			[ 
				xmin	=	0, 
				xmax	=	40, 
				ymin	=	0, 
				ymax	=	\maxTinLongtime
			]
			{data/low_photon_numbers/obc_n_40_u_4_delta_8_q_excitation_momentum_space_kin_19_kout_0_cdw_non_spin_selective/pump/plain_N.pdf};
		\nextgroupplot
		[
			enlargelimits	=	false,
			axis on top,
			xlabel			= 	{site $j$},
			xlabel style	=	{yshift=0.3em},
			extra description/.code =
			{
				\node[anchor = south east] at (0.99,0.01) {\(\Delta/\hopping=8\)};
			}
		]
			\addplot graphics
			[ 
				xmin	=	0, 
				xmax	=	40, 
				ymin	=	-5, 
				ymax	=	1
			]
				{data/low_photon_numbers/obc_n_40_u_4_delta_8_q_excitation_momentum_space_kin_19_kout_0_cdw_non_spin_selective/groundstatesearch/plain_S_gs.pdf};
			\addplot graphics
			[ 
				xmin	=	0, 
				xmax	=	40, 
				ymin	=	0, 
				ymax	=	\maxTinLongtime
			]
				{data/low_photon_numbers/obc_n_40_u_4_delta_8_q_excitation_momentum_space_kin_19_kout_0_cdw_non_spin_selective/pump/plain_S.pdf};
		\end{groupplot}
		\end{tikzpicture}
	}
	{
		\includegraphics{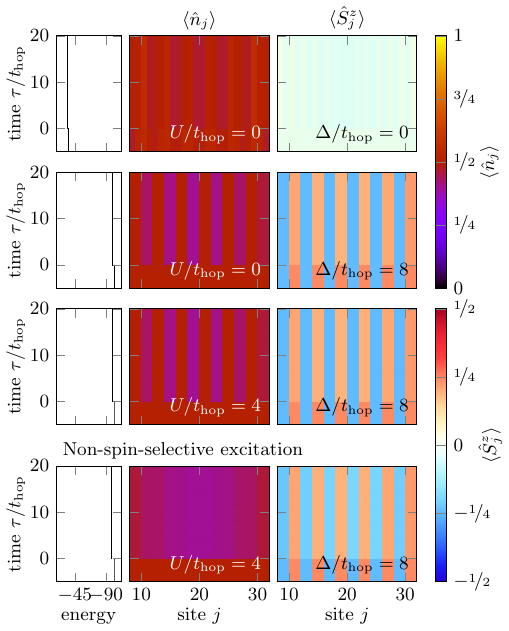}
	}
	
	\caption
	{
		\label{fig:singlephoton}
		Time evolution of system~\eqref{eq:Ham} with $L=40$ sites from \gls{tDMRG} at quarter filling following the operator-based spin-selective excitation discussed in the text and \cref{app:Ham_obc} at time $\tau = 0$. 
		Analogous discussion to \cref{fig:DvariableU0_alphaup01}.
		First column: total energy of the system.
		Second column: particle density $\langle \hat{n}_j \rangle(\tau)$ in the bulk (sites 8-32).
		Third column: local magnetizations $\langle \hat{S}^z_j \rangle(\tau)$, also in the bulk. 
		The color bars on the right indicate the values for $\langle \hat{n}_j \rangle(\tau)$ and $\langle \hat{S}^z_j \rangle(\tau)$, respectively.
		The top row shows results for an excitation acting only on spin-up particles in the absence of a magnetic structure, $\Delta = 0$, and without interaction, $U=0$. 
		The second row shows results for the same excitation, but with $\Delta/\hopping = 8$ (and, thus, for adjusted operators) and $U=0$. 
		In contrast, the third row shows results for the same excitation and also $\Delta/\hopping = 8$ but $U=4$.
		The bottom row shows results for an excitation acting on both spin directions for $\Delta/\hopping = 8$ and $U=4$.
		The time evolutions are obtained employing the time-dependent variational principle (TDVP) in its two-site implementation with a time step $\Delta \tau = 0.05$, a maximum bond dimension $\chi = 1000$, and a discarded weight of $\epsilon < 10^{-12}$. 
		For better comparison to Fig.~\ref{fig:DvariableU0_alphaup01}, we show for times $\tau < 0$ the expectation values for the unperturbed ground state.
	}
\end{figure}

We present our results in \cref{fig:singlephoton}, where a system of $L=40$ sites is excited via
\begin{align}
\label{eq:low_photon_ex}
	\ket{\psi(\tau=0)} = \hat{a}^\dagger_{\uparrow,2}\!\left(\nicefrac{\pi}{41}\right) \hat{a}^{\vphantom{\dagger}}_{\uparrow,1}\!\left(\nicefrac{\pi}{41}\right)\ket{\psi_0} \, .
\end{align}
In order to better compare our findings, we discuss the same setups as in \cref{fig:DvariableU0_alphaup01} (without loss of generality we apply the excitation here to $\uparrow$ electrons).
In the top row of \cref{fig:singlephoton} we first show results for the noninteracting system in the absence of a magnetic field, i.e., $U = \Delta = 0$. 
As expected, the Friedel-like oscillations in the particle density induced by the \gls{OBC}\cite{Bedurftig:1998p457} are not destroyed; however, for this choice of parameters, they are slightly enhanced, and due to an absorption of energy become weakly visible also in the local spin densities.
For finite values of $\Delta/\hopping$ as in the second row of \cref{fig:singlephoton}, the findings are very similar to the ones of \cref{fig:DvariableU0_alphaup01}:
The periodic pattern in the local spin densities $\braket{ \hat S^{z}_{j} }$, which follows the magnetic microstructure, is weakened after the excitation and, again, a long-lived \gls{CDP} in the charge density $\langle \hat n_{j}\rangle$ is induced following the periodicity of the Zeeman term. 
In particular, this is also obtained at finite values of $U/\hopping$ as is depicted in the third panel.
In contrast, similar to the findings of \cref{fig:DvariableU0_alphaup01}, exciting $\uparrow$ electrons and $\downarrow$ electrons at the same time, does not lead to a charge modulation, but solely to a weaker spin pattern, as seen in the bottom panel of \cref{fig:singlephoton}. 
We hence conclude that also for very weak excitations, we find qualitatively the same behavior as in the case of a Peierls substitution.
Due to the lower energy density of the excitations the emerging structures are less pronounced; however, this indicates that the observed features are not an artifact of the generalized Peierls substitution ansatz \cref{eq:peierls}, but seem generic to spin-selective photoexcitations.

Note that in the case of low photon densities details of the \gls{CDP} depend on peculiarities of the excitation, e.g., whether it is at the edge of the Brillouin-zone or close to the center as discussed in this section, and also on the number of electrons excited. 
This is left for future research.

\subsection{Alternative mechanism: periodic modulation of the lattice in cold-gases experiments}
\label{sec:experiments}
\begin{figure}[!h]	% RealHopping
	\centering	
	\newcommand{\maxTinLongtime}{20}
	\ifthenelse{\boolean{buildtikzpics}}
	{
	\tikzsetnextfilename{RealHopping}
	\begin{tikzpicture}
		\pgfplotsset
		{
			/pgfplots/colormap={gnuplot}{rgb255=(0,0,0) rgb255=(46,0,53) rgb255=(65,0,103) rgb255=(80,0,149) rgb255=(93,0,189) rgb255=(104,1,220) rgb255=(114,2,242) rgb255=(123,3,253) rgb255=(131,4,253) rgb255=(139,6,242) rgb255=(147,9,220) rgb255=(154,12,189) rgb255=(161,16,149) rgb255=(167,20,103) rgb255=(174,25,53) rgb255=(180,31,0) rgb255=(186,38,0) rgb255=(191,46,0) rgb255=(197,55,0) rgb255=(202,64,0) rgb255=(208,75,0) rgb255=(213,87,0) rgb255=(218,100,0) rgb255=(223,114,0) rgb255=(228,130,0) rgb255=(232,147,0) rgb255=(237,165,0) rgb255=(241,185,0) rgb255=(246,207,0) rgb255=(250,230,0) rgb255=(255,255,0) }
		}
		\pgfplotsset
		{
			/pgfplots/colormap={temp}{
				rgb255=(36,0,217) 		% 0 "#2400d9"
				rgb255=(25,29,247) 		% 1 "#191df7"
				rgb255=(41,87,255) 		% 2 "#2957ff"
				rgb255=(61,135,255) 	% 3 "#3d87ff"
				rgb255=(87,176,255) 	% 4 "#57b0ff"
				rgb255=(117,211,255) 	% 5 "#75d3ff"
				rgb255=(153,235,255) 	% 6 "#99ebff"
				rgb255=(189,249,255) 	% 7 "#bdf9ff"
				rgb255=(235,255,255) 	% 8 "#d3ffff"
				rgb255=(255,255,235) 	% 9 "#ffffd9"
				rgb255=(255,242,189) 	% 10 "#fff2bd"
				rgb255=(255,214,153) 	% 11 "#ffd699"
				rgb255=(255,172,117) 	% 12 "#ffac75"
				rgb255=(255,120,87) 	% 13 "#ff7857"
				rgb255=(255,61,61) 		% 14 "#ff3d3d"
				rgb255=(247,40,54) 		% 15 "#f72836"
				rgb255=(217,22,48) 		% 16 "#d91630"
				rgb255=(166,0,33)		% 17 "#a60021"
			}
		}
		\begin{groupplot}
			[
				group style = 
				{
					group size 			=	3 by 2,
					vertical sep		=	3.5mm,
					horizontal sep		=	1mm,
					x descriptions at	=	edge bottom,
					y descriptions at	=	edge left
				},
				height 			= 	0.15\textheight,
				width 			= 	0.22\textwidth,
				xmin 			= 	8, 
				xmax 			= 	32,
				ymin			=	0,
				ymax			=	\maxTinLongtime,	
			    point meta min	=	0,
			    point meta max	=	1,
				ylabel			= 	{time $\tau/\hopping$},
				ylabel shift	=	-3pt,
			]
				\nextgroupplot
				[
					enlargelimits	=	false,
					axis on top, 
					width			= 	0.15\textwidth,
					x dir			=	reverse,
					xmin			=	-300,
					xmax			=	-50,
					ymin			=	0,
					ymax			=	\maxTinLongtime,
					xtick distance	=	160,
				]
				\addplot
				[
					color				=	black,
					unbounded coords	=	jump, 
					mark size			=	2pt,
				] 
				table
				[
					x expr = \thisrowno{1}, 
					y expr = \thisrowno{0},
				]
				{data/real_hopping/alpha_up0/alpha_down1/beta_up0/beta_down0/b_z8,8,-8,-8/U0/e020/sigma6000/t010/k00.00126/zeeman_coupling0/ChainLength40/TargetSymmetrySector20,0/RealTimeStepWidth0.01/RealTime100/rte/sim_data};		

				\nextgroupplot
				[
					enlargelimits	=	false,
					axis on top,
					extra description/.code =
					{
						\node[anchor = south east, white] at (0.99,0.01) {$U=0$};
					},
					title			=	{$\langle \hat n_j \rangle$},
					title style		=	{at = {(0.5,1)}, yshift = -0.7em}
				]
				\addplot graphics
				[ 
					xmin	=	0, 
					xmax	=	40, 
					ymin	=	0, 
					ymax	=	\maxTinLongtime
				]
				{data/real_hopping/alpha_up0/alpha_down1/beta_up0/beta_down0/b_z8,8,-8,-8/U0/e020/sigma6000/t010/k00.00126/zeeman_coupling0/ChainLength40/TargetSymmetrySector20,0/RealTimeStepWidth0.01/RealTime100/rte/loc_observables/plain_N.pdf};
				\nextgroupplot
				[
					enlargelimits		=	false,
					axis on top,
					title				=	{$\langle \hat S^z_j \rangle$},
					title style			=	{at={(0.5,1)}, yshift=-0.7em},
					xtick				=	{10, 20, 30},
					xticklabels			=	{$10$, $20$, $30$},
					xticklabel style	=	{color = white},
					colorbar right,
					colormap name		=	gnuplot,
				    point meta min		=	0,
				    point meta max		=	1,
					every colorbar/.append style =
					{
						height			=	1*\pgfkeysvalueof{/pgfplots/parent axis height} + 0*\pgfkeysvalueof{/pgfplots/group/vertical sep},
						ylabel			=	{$\langle \hat n_j \rangle$},
						width			=	2mm,				
						ytick			=	{0, 0.25, 0.5, 0.75, 1.0},
						yticklabels		=	{$0$, $\nicefrac{1}{4}$, $\nicefrac{1}{2}$, $\nicefrac{3}{4}$, $1$},
						ylabel shift	=	-4pt,
					}
				]
				\addplot graphics
				[ 
					xmin = 0, 
					xmax = 40, 
					ymin = 0, 
					ymax = \maxTinLongtime
				]
				{data/real_hopping/alpha_up0/alpha_down1/beta_up0/beta_down0/b_z8,8,-8,-8/U0/e020/sigma6000/t010/k00.00126/zeeman_coupling0/ChainLength40/TargetSymmetrySector20,0/RealTimeStepWidth0.01/RealTime100/rte/loc_observables/plain_S.pdf};

				\nextgroupplot
				[
					enlargelimits	=	false,
					axis on top, 
					width			= 	0.15\textwidth,
					x dir			=	reverse,
					xmin			=	-300,
					xmax			=	-50,
					ymin			=	0,
					ymax			=	\maxTinLongtime,
					xtick distance	=	160,
					xlabel			= 	{energy$\vphantom{\text{site }j}$},
				]
				\addplot
				[
					color				=	black,
					unbounded coords	=	jump, 
					mark size			=	2pt,
				] 
				table
				[
					x expr = \thisrowno{1}, 
					y expr = \thisrowno{0},
				]
				{data/real_hopping/alpha_up0/alpha_down1/beta_up0/beta_down0/b_z8,8,-8,-8/U8/e020/sigma6000/t010/k00.00126/zeeman_coupling0/ChainLength40/TargetSymmetrySector20,0/RealTimeStepWidth0.01/RealTime100/rte/sim_data};

				\nextgroupplot
				[
					enlargelimits			=	false,
					axis on top,
					xlabel					= 	{site $j\vphantom{\text{energy}}$},
					extra description/.code	=	{\node[anchor = south east, white] at (0.99,0.01) {$U=8$};}
				]
				\addplot graphics
				[ 
					xmin = 0, 
					xmax = 40, 
					ymin = 0, 
					ymax = \maxTinLongtime
				]
				{data/real_hopping/alpha_up0/alpha_down1/beta_up0/beta_down0/b_z8,8,-8,-8/U8/e020/sigma6000/t010/k00.00126/zeeman_coupling0/ChainLength40/TargetSymmetrySector20,0/RealTimeStepWidth0.01/RealTime100/rte/loc_observables/plain_N.pdf};
				\nextgroupplot
				[
					enlargelimits	=	false,
					axis on top,
					xlabel			= 	{site $j$},
					colorbar right,
				    point meta min	=	-0.5,
				    point meta max	=	0.5,
					every colorbar/.append style =
					{
						height			=	1*\pgfkeysvalueof{/pgfplots/parent axis height} + 0*\pgfkeysvalueof{/pgfplots/group/vertical sep},
						ylabel			=	{$\langle \hat S^z_j \rangle$},
						width			=	2mm,				
						ytick			=	{-0.5, -0.25, 0, 0.25, 0.5},
						yticklabels		=	{$-\nicefrac{1}{2}$, $-\nicefrac{1}{4}$, $0$, $\nicefrac{1}{4}$, $\nicefrac{1}{2}$},
						ylabel shift	=	-12pt,
					}
				]
				\addplot graphics
				[ 
					xmin = 0, 
					xmax = 40, 
					ymin = 0, 
					ymax = \maxTinLongtime
				]
				{data/real_hopping/alpha_up0/alpha_down1/beta_up0/beta_down0/b_z8,8,-8,-8/U8/e020/sigma6000/t010/k00.00126/zeeman_coupling0/ChainLength40/TargetSymmetrySector20,0/RealTimeStepWidth0.01/RealTime100/rte/loc_observables/plain_S.pdf};
		\end{groupplot}
	\end{tikzpicture}
	}
	{
		\includegraphics{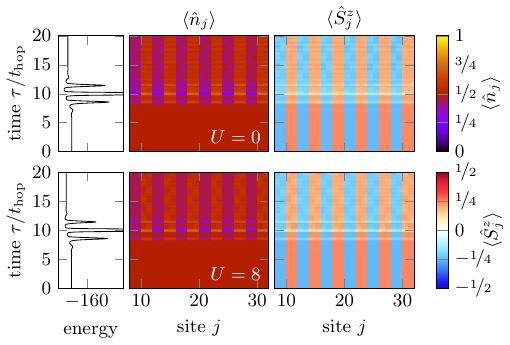}
	}
	\caption
	{
		Periodic modulation of the lattice of only one species of particles also leads to a \gls{CDP}.
		Left: Energy of systems  with $\Delta/\hopping=8$ and (top) $U/\hopping=0$ and (bottom) $U/\hopping=8$.
		Center: Particle density for these systems. 
		Right: Spin density for the systems above.
		In the interacting case ($U/\hopping = 8$), the discarded weight $\epsilon$ grows rapidly and reaches a value of $\epsilon \sim 10^{-5}$ at the end of the simulation.
	}
	\label{fig:RealHopping}
\end{figure}
The question arises, if these findings can be realized also for even more generic excitations.
In \cref{fig:RealHopping} we present our results, in which the spin-selective photoexcitation is emulated by a periodic modulation of the lattice of only one fermionic species, which can be realized in experiments with ultracold quantum gases on optical lattices \cite{modulationspectroscopy1,modulationspectroscopy2,modulationspectroscopy3,modulationspectroscopy4}  with a superlattice\cite{superlattice_orig1,superlattice_orig2,S.Trotzky01182008,Anderlini:2007p1571,ionichubbardmodelEsslinger,superlattice_magneticfield,superlattice_magneticfield_PRL,SpinDependentOptLat,ThoulessPumpSuperlattice,SpinPumpSuperlattice}.
As can be seen, the \gls{CDP} emerges also in this setup, indicating that the details of the excitation are not crucial, as long as it is acting on only one fermion species. 
This opens the path for studying the formation of periodic charge and spin patterns in non-equilibrium situations in state-of-the-art experiments with ultracold gases.

\section{Conclusion and outlook}
\label{sec:conclusion}
A spin-selective photoexcitation in the presence of a microstructure leads to the formation of periodic spin or charge patterns on femtosecond time scales.
Our scenario connects to the recently introduced \gls{OISTR} effect\cite{oistr_orig,oistr_orig_2}, in which ultrafast spin transfer is predicted theoretically using \textit{ab initio} methods, and was observed experimentally in Heusler and magnetic materials~\cite{light_induced_spin_transfer,light_induced_spin_transfer_efficiency,light_induced_spin_transfer_FMalloys}.
In the \gls{OISTR} setup, the site-dependent difference of the local DOS of the minority and majority spins in equilibrium is found to be the reason for ultrafast transfer of magnetic moments.
Such a difference in the local DOS corresponds to the microstructures treated by us on the level of tight-binding and Hubbard-type models.
We propose that in these setups when applying a spin-selective excitation in addition the formation of stable spatial patterns will be obtained. 

We used \gls{MPS} to study the dynamics of a Hubbard-type model with a magnetic microstructure.
For this system, we modeled high intensity light pulses of duration $\sim 10$fs by a generalized Peierls substitution ansatz.
The patterns are formed during the application of the pulse, and have a lifetime much longer than the pulse duration afterwards.
For noninteracting systems, we study analytically an idealized ``kick'' excitation for systems with magnetic and ionic backgrounds.
In these systems stable charge or spin patterns, respectively, are induced:
a light pulse leads to a redistribution of $k$ modes, which here are conserved quantities, so that without further effects no decay mechanism could lead to a destruction of the spatial patterns.
This is a generic picture valid also beyond one dimension. 
We find that electron-electron interactions can induce a decay channel for these patterns, which for our Hubbard-type system is realized via freely propagating doublons.
However, we find that on the time scales treated by us, a `backbone' remains.
To better estimate the lifetime of the patterns further effects, like phonons, need to be included, which is left for future research.

We find a similar formation of spatial patterns also for other excitations, e.g., a single-photon excitation, in which at fixed $k$ one particle is excited over a band-gap, and also for a periodic shaking of the lattice.
However, in all cases treated, it was crucial to apply a spin-dependent excitation in order to induce the spatial patterns; otherwise, the absorbed energy leads to an unstructured redistribution of the particles over the lattice. 

In experiments on materials, spin-selective excitations can be realized via circularly polarized monochromatic light~\cite{Bruno_moke}.
We thus suggest to test for the possible formation of periodic patterns on suitable materials showing \gls{OISTR}.
In addition, it would be interesting to study the transformation of density modulations following such a spin-selective photoexcitation, e.g., in charge-transfer salts \cite{IonicHubbard,chargetransfersalts_Becca} or in cuprates \cite{cupratesCDW1,cupratesCDW2,cupratesCDW3,cupratesCDW4,cupratesCDW5,cupratesCDW6,cuprates_intertwinnedorder1,Edkins976,cuprates_CDW_science,condmat2030026}, which possess locally alternating chemical potentials or \gls{CDW} states, respectively.
For the latter case, we envisage that our scenario will hold also on initial states, in which a spontaneous breaking of translational symmetry leads to a true \gls{CDW}.
An alternative realization is in cold-gas experiments, where the underlying magnetic (or ionic) pattern can be realized by a superlattice \cite{superlattice_orig1,superlattice_orig2,S.Trotzky01182008,Anderlini:2007p1571,ionichubbardmodelEsslinger,superlattice_magneticfield,superlattice_magneticfield_PRL,SpinDependentOptLat,ThoulessPumpSuperlattice,SpinPumpSuperlattice}.
For the spin-selective photoexcitation one can treat a more simplified situation, in which the lattice of only one species is shaken \cite{modulationspectroscopy1,modulationspectroscopy2,modulationspectroscopy3,modulationspectroscopy4}, and the subsequent dynamics can be investigated using quantum-gas microscopes \cite{Bakr:2009p2641,PhysRevLett.114.213002,Cheuk2015,quantumgasmicroscope_BlochPRL, quantumgasmicroscope_review,quantumgasmicroscope_spinresolved}.
It would be interesting to study these effects also for systems with more than only two species of particles, e.g., SU($N$) systems\cite{Gorshkov:2010p1052NO}.
Our scenario hence opens the path to study the formation of spatial charge and spin patterns in ongoing experiments. 

\begin{acknowledgments}
	We thank L. Cevolani, C. R\"uegg, J. Herbrych, F. Heidrich-Meisner, S. Marten, O. Schumann, J. Stolpp, and J. Walowski for fruitful discussions.
	This work is  
	%%%%%%%%%%%%%%%%%%%%%%%%%%
	% According to Regina Vinnen, this wording must not be changed. 
	funded by the Deutsche Forschungsgemeinschaft (DFG, German Research Foundation) -- 217133147/SFB 1073, project B03 and 207383564/Research Unit FOR 1807, project P7.
	%%%%%%%%%%%%%%%%%%%%%%%%%%
%
	TK acknowledges financial support by the ERC Starting Grant from the European Union's Horizon 2020 research and innovation program under Grant No. 758935.
	SP acknowledges support by the Deutsche Forschungsgemeinschaft (DFG, German Research Foundation) under Germany’s Excellence Strategy-426 EXC-2111-390814868.
	Computational resources made available by the Department of Applied Theoretical Physics, Clausthal Technical University, as well as the GWDG Scientific Compute Cluster are gratefully acknowledged. 
	At the GWDG, we wish to express our gratitude to S. Krey and A. Khuziyakhmetov for IT support. 
	All data were produced using the \textsc{SymMPS} toolkit \cite{symmps}. 
\end{acknowledgments}
\appendix
\section{Kick excitation for the Ionic chain}
\label{app:ionic}
Introducing operators
\begin{align}
\hat{b}^{\dagger}_{\sigma,m}(k) := \frac{1}{\sqrt{N}} \sum_{l=0}^{N-1} e^{ikr_l} \hat{c}^\dagger_{\sigma,m,l} \, 
\end{align}
with $r_l=a M l$ and $m\in\left\{0,\ldots,M-1 \right\}$ we can rewrite Hamiltonians of the type~\eqref{eq:U0} with periodic boundary conditions in the presence of $N=L/M$ unit cells with each possessing $M$ sites to
\begin{align}
	\hat{H} = \frac{1}{N} \sum_{k,\sigma}  \sum_{m,m'=0}^{M-1} h^{\nodagger}_{\sigma,m,m'}(k) \, \hat{b}^{\dagger}_{\sigma,m}(k) \hat{b}^{\nodagger}_{\sigma,m'}(k) \, ,
\end{align}
with spacing of $k$ points $\Delta_k = \frac{2\pi}{aNM}$, and $k \in [-\nicefrac{\pi}{Ma}, \nicefrac{\pi}{Ma})$.
For the ionic chain with the staggered potential as given in~\eqref{eq:ionic} we set $M=2$.
The Hamiltonian for a single unit cell and $k$ point is $h_{\sigma,m,m'}(k)$, and for the ionic model~\eqref{eq:ionic} one obtains
\begin{align}
	\underline{\underline{h}}_{\uparrow}(k) =&\underline{\underline{h}}_{\downarrow}(k) =   \nonumber \\
	&
	\left(
	\begin{matrix}
		-\Delta	& - 2\hopping \cos(ka) \\
		-2\hopping\cos(ka) & \Delta
	\end{matrix}
	\right) \, .
\end{align}
Diagonalizing this Hamiltonian, we find
\begin{align}
	\sum_{m^\prime = 0}^1 h_{\sigma,m,m^\prime}(k) T_{\sigma,m^\prime,\pm}(k) = \pm T_{\sigma,m,\pm}(k) \epsilon_{\sigma}(k) \, , \label{eq:Tmatrix}
\end{align}
with bands $\pm \epsilon_{\sigma}(k) = \pm \Delta \sqrt{1 + \varepsilon^2 \cos^2(ka)}$, which are the same for both spin directions and $\varepsilon = \frac{2 \hopping}{\Delta}$.
Introducing operators 
\begin{align}
	\hat{a}^\dagger_{\sigma,\pm}(k) = \sum_{m=0}^1 \hat{b}^{\dagger}_{\sigma,m}(k) T^{\nodagger}_{\sigma,m,\pm}(k) \, ,
	\label{eq:a_ops}
\end{align}
we finally diagonalize the Hamiltonian
\begin{equation}
	\hat{H}^{\rm ionic}
	=
	\sum_{\sigma,k} 
	\epsilon_{\sigma}(k) \left( \hat{a}^\dagger_{\sigma,+}(k) \hat{a}^{\nodagger}_{\sigma,+}(k) - \hat{a}^\dagger_{\sigma,-}(k) \hat{a}^{\nodagger}_{\sigma,-}(k) \right) \; .
\end{equation}
The eigenvectors $T_{\sigma, m, \pm}$ are given by
\begin{align}
	\underline T_{\sigma,\pm}(k)
	&=
	\frac{1}{\sqrt{2}} \left(\begin{array}{c} \mp \sqrt{1\mp\frac{\Delta}{\epsilon(k)}} \\ \sqrt{1\pm\frac{\Delta}{\epsilon(k)}} \end{array}\right) \; ,
\end{align}
so that the ground state expectation values for the density in $k$ space taking into account the different sublattices are given by
\begin{align}
	\braket{\hat n_{\sigma,0}(k)} &= \frac{1}{2} \left[ 1+\frac{\Delta}{\epsilon_\sigma(k)} \right] \\
	\braket{\hat n_{\sigma,1}(k)} &= \frac{1}{2} \left[ 1-\frac{\Delta}{\epsilon_\sigma(k)} \right] \; .
\end{align}
In the following, we treat the system in the atomic limit $\varepsilon \ll 1$ and expand the expectation values to first order in $\varepsilon^2$
\begin{align}
	\braket{\hat n_{\sigma,0}(k)} &=  1-\frac{\varepsilon^2}{4}\cos^2(ka) + \mathcal O(\varepsilon^4) \\
	\braket{\hat n_{\sigma,1}(k)} &=  \frac{\varepsilon^2}{4}\cos^2(ka) + \mathcal O(\varepsilon^4) \; .
\end{align}
Then, the occupation per sublattice is obtained in the thermodynamic limit at half filling by integrating the sublattice occupations over the first Brillouin zone 
\begin{align}
	\braket{\hat n_{\sigma,0}} 
	&= 
	\frac{a}{\pi}\int_{-\nicefrac{\pi}{2a}}^{\nicefrac{\pi}{2a}} \braket{\hat n_{\sigma,0}(k)} dk = 1-\frac{\varepsilon^2}{8} \\
	\braket{\hat n_{\sigma,1}} 
	&= 
	\frac{a}{\pi}\int_{-\nicefrac{\pi}{2a}}^{\nicefrac{\pi}{2a}} \braket{\hat n_{\sigma,1}(k)} dk = \frac{\varepsilon^2}{8} \;.
\end{align}
For the system being translationally invariant with respect to a shift of $M=2$ lattice sites these values yield the particle number densities in real space for the two sublattices.
Having in mind the Peierls substitution ansatz Eq.~\eqref{eq:peierls}, the kick excitation transforms the Hamiltonian of a unit cell in $k$ space into
\begin{align}
	\underline{\underline h}_{\sigma}(\varphi)
	&=
	\left(
	\begin{matrix}
		-\Delta	& - 2\hopping \cos(ka+\varphi) \\
		-2\hopping\cos(ka+\varphi) & \Delta
	\end{matrix}
	\right) \, .
\end{align}
Accordingly, the transformed eigenvectors are given by $\underline T_{\sigma,\pm}(k+\frac{\varphi}{a})$ with single-particle energies $\epsilon_{\sigma}(k+\frac{\varphi}{a})$.
From this, the time evolution with respect to the kick excitation can be evaluated by expanding $\underline{\underline h}_{\sigma}(\varphi)$ in its eigenbasis.
We model the excitation using a rectangular pulse shape $\varphi (t) = \varphi \left[ \theta(t-\delta s) - \theta(t) \right]$ on a finite time interval $t\in [0,\delta s)$.
Defining $k_\varphi = k+\frac{\varphi}{a}$ the action of the kick operator acting only on $\sigma = \uparrow$ electrons is given by 
\begin{align}
	\ket{\psi(\varphi)}
	&=
	e^{-\mathrm i \delta s \sum\limits_{k}\sum\limits_{\mu=\pm} \pm \epsilon(k_\varphi) \hat a^\dagger_{\uparrow,\mu}(k_\varphi)\hat a^{\nodagger}_{\uparrow,\mu}(k_\varphi)}\ket{\psi(0)} \; ,
\end{align}
where $\ket{\psi(0)} = \ket{\psi_{0,\uparrow}} \otimes \ket{\psi_{0,\downarrow}}$ is the unperturbed ground state of the entire system obtained via \cref{eq:gs}.
For a single $k$ mode of the perturbed ground state this evaluates to
\begin{align}
	e^{-\mathrm i \hat H_\varphi(k)}\ket{\psi(0)}
	=
	e^{-\mathrm i \delta s\sum\limits_{\mu=\pm} \pm \epsilon(k_\varphi) \hat a^\dagger_{\uparrow,\mu}(k_\varphi)\hat a^{\nodagger}_{\uparrow,\mu}(k_\varphi)}\ket{\psi(0)}  
\end{align}
yielding contributions
\begin{align}
	\prod_{\substack{k^\prime \neq k \\ \sigma=\uparrow,\downarrow}} \hat a^\dagger_{\sigma,-}(k^\prime)
	\left( 
		\sum_{\mu=\pm} e^{-\mathrm i \mu \epsilon_\uparrow(k_\varphi)\delta s} \alpha_\mu(k,\varphi) \hat a^\dagger_{\uparrow,\mu}(k_\varphi)
	\right) \ket{0}\;,
\end{align}
with the definition $\alpha_\mu(k,\varphi) = \underline T^\dagger_{\uparrow,\mu}(k_\varphi) \underline T^{\nodagger}_{\uparrow,-}(k)$.
The local density expectation values $\braket{\hat n_{\uparrow,m}(k)}(\varphi)$ after the kick excitation are then given by
\begin{widetext}
	\begin{align}
		\braket{\hat n_{\uparrow,m}(k)}(\varphi)
		&=
		\sum_{\mu=\pm} T_{\uparrow,\mu,m}(k_\varphi)\alpha_\mu(k,\varphi)
		\left[ T_{\uparrow,\mu,m}(k_\varphi)\alpha_\mu(k,\varphi) + \cos(2\epsilon_\uparrow(k_\varphi)\delta s)T_{\uparrow,\overline \mu,m}(k_\varphi)\alpha_{\overline \mu}(k,\varphi)\right]
		\; ;
	\end{align}
	with the definition $\overline \mu = -\mu$ and abbreviating $\xi_k = \frac{\Delta}{\epsilon_\uparrow(k)}$, this evaluates to
	\begin{align}
		\braket{\hat n_{\uparrow,0}(k)}(\varphi)
		&=
		\frac{1}{2}
		\left\{
			1 + \frac{1+\xi_{k_\varphi}\xi_k + \nicefrac{1}{2}\cos(2\epsilon_\uparrow(k_\varphi)\delta s) (\xi_{k_\varphi}-\xi_k)\xi_{k_\varphi}}{\left( 1+\xi^2_{k_\varphi} \right) \sqrt{1+\xi^2_k}}
		\right\} \; ,
	\end{align}
	and $\braket{\hat n_{\uparrow,1}(k)}(\varphi) = 1 - \braket{\hat n_{\uparrow,0}(k)}(\varphi)$, respectively.
	Expansion to first order in $\varepsilon^2$ yields
	\begin{align}
		\braket{\hat n_{\uparrow,0}(k)}(\varphi) &= 1-\frac{1}{2}\left[ \frac{\cos^2(ka)}{2} + 2\cos(ka+\varphi)\left(\cos(ka+\varphi) -\cos(ka) \right) \sin^2(\Delta\delta s) \right] \varepsilon^2 + \mathcal O(\varepsilon^4)\\
		\braket{\hat n_{\uparrow,1}(k)} &= \frac{1}{2}\left[ \frac{\cos^2(ka)}{2} + 2\cos(ka+\varphi)\left(\cos(ka+\varphi) -\cos(ka) \right) \sin^2(\Delta\delta s) \right] \varepsilon^2 + \mathcal O(\varepsilon^4) \; .
	\end{align}
\end{widetext}
Integrating over the first Brillouin zone we finally obtain for the $\sigma=\uparrow$ electrons the particle number density per sublattice site in the thermodynamic limit after the excitation:
\begin{align}
	\braket{\hat n_{\uparrow,0}}(\varphi) &= 1 - \frac{\varepsilon^2}{8}\left[ 1 + 2\sin^2(\Delta\delta s)\sin^2({\nicefrac{\varphi}{2}}) \right] \\
	\braket{\hat n_{\uparrow,1}}(\varphi) &= \frac{\varepsilon^2}{8}\left[ 1 + 2\sin^2(\Delta\delta s)\sin^2({\nicefrac{\varphi}{2}}) \right] \;.
\end{align}
Note that from the above expressions the maximal charge redistribution is reached if $\varphi = q \pi$ with $q \in \mathbb Z$.
Considering the limit of short pulse durations compared to the energy scale set by the microstructure $\Delta\delta s \ll 1$ we can expand these expressions
\begin{align}
	\braket{\hat n_{\uparrow,0}}(\varphi) &= 1 - \frac{\varepsilon^2}{8}\left[ 1 + 2(\Delta\delta s)^2\sin^2({\nicefrac{\varphi}{2}}) + \mathcal O (\Delta \delta s)^4\right] \\
	\braket{\hat n_{\uparrow,1}}(\varphi) &= \frac{\varepsilon^2}{8}\left[ 1 + 2(\Delta\delta s)^2\sin^2({\nicefrac{\varphi}{2}}) + \mathcal O (\Delta \delta s)^4 \right] \;.
\end{align}

The previous calculations allow us to compare the redistribution of particle and magnetization densities between the sublattices induced by the pulse.
We consider the magnetization density of the sublattices before $\braket{\hat S^z_m}$ and $\braket{\hat S^z_m}(\varphi)$ after the excitation.
Defining the enhanced staggered magnetization density induced by the spin-selective pulse acting on $\uparrow$ electrons only as
\begin{align}
	\delta S^z (\varphi)
	&=
	\left| \braket{\delta \hat S^z_0}(\varphi) - \braket{ \delta \hat S^z_1}(\varphi) \right| \notag \\
	&=
	\frac{1}{2}\left| \braket{\delta \hat n_{\uparrow,0}}(\varphi) - \braket{\delta  \hat n_{\uparrow,1}}(\varphi) \right|
\end{align}
using $\braket{\delta \hat S^z_m}(\varphi) = \braket{\hat S^z_m}(\varphi) - \braket{\hat S^z_m}$ and a similar definition for the electron densities we find
\begin{align}
	\delta S^z (\varphi)
	&=
	\frac{1}{2} - \frac{\varepsilon^2}{8}\left(1+ 2 ( \Delta \delta s)^2 \sin^2(\nicefrac{\varphi}{2})\right) \; .
\end{align}

\section{Mean-field decoupling at low fillings}
\label{app:meanfield}
We discuss the limit $\hopping/U \ll 1$ in which we can rewrite the Hubbard interaction in terms of local spin operators
\begin{align}
	\hat{n}^{\nodagger}_{\uparrow,j}\hat{n}^{\nodagger}_{\downarrow,j} &= -\frac{1}{2}\left( (\hat{n}^{\nodagger}_{\uparrow,j} - \hat{n}^{\nodagger}_{\downarrow,j})^{2} - (\hat{n}^{\nodagger}_{\uparrow,j} + \hat{n}^{\nodagger}_{\downarrow,j}) \right) \\
	&= \frac{1}{2}\hat{n}^{\nodagger}_{j}-2\left(\hat{S}^{z}_{j}\right)^{2}
\end{align}
\begin{align}
	\Rightarrow
	\hat{H} &= -\hopping\sum_{\sigma,j} \left(\hat{c}^{\dagger}_{\sigma,j}\hat{c}^{\nodagger}_{\sigma,j+1} + \text{h.c.} \right) \notag\\
	 &\phantom{=}\quad + \underbrace{\sum_{i} \Delta_{j}\hat{S}^{z}_{j}+U\left(\frac{1}{2}\hat{n}^{\nodagger}_{j} - 2\left(\hat{S}^{z}_{j}\right)^{2} \right)}_{\hat{H}_{\text{int}}} \;.
\end{align}
We note that for large $U$ the formation of local moments $\left(\hat{S}^{z}_{j}\right)^{2}$ with strong polarization in the expectation values of $\hat{S}^{z}_{j}$ minimizing the Zeeman coupling is beneficial.
Therefore we can perform a mean-field decoupling around the saturated local expectation values $\hat{S}^{z}_{j} = \langle \hat{S}^{z}_{j}\rangle + \delta\hat{S}^{z}_{j}$ and neglect contributions $\propto \delta^2$, leading to
\begin{align}
	\hat{H}_{\text{int}} &\stackrel{\sim}{=} \sum_{j} \delta\hat{S}^{z}_{j}\left(\Delta_{j} - 4U\langle\hat{S}^{z}_{j}\rangle \right) + \cdots \notag \\
	&\phantom{=} \quad + \langle\hat{S}^{z}_{j}\rangle\left( \Delta_{j}-2U\langle\hat{S}^{z}_{j}\rangle\right) + \frac U 2\hat{n}_{j} \, .
\end{align}
Inserting for the local expectation values the pattern of the Zeeman terms in a unit cell, $\langle \hat{S}^{z}_{j} \rangle = m\left( \nicefrac{-1}{2}, \nicefrac{-1}{2}, \nicefrac{1}{2}, \nicefrac{1}{2} \right)$, where $m$ is to be determined self consistently, we see that the second summand is only a constant.
Defining the renormalized Zeeman coupling $\tilde{\Delta} = \nicefrac{1}{2}\left(\Delta + 4U \right)$ and $\tilde{\Delta}_{\sigma,j} = \tilde{\Delta}_{j}\text{sgn}(\sigma)$ we obtain
\begin{align}
	\hat{H}_{\text{int}} &= \sum_{\sigma,j} \hat{n}_{\sigma,j}\left(\frac U 2 + \delta\tilde{\Delta}_{\sigma,j} \right) + \text{const}
\end{align}
and thereby (up to an on-site potential $\propto U$) we rewrote the interaction in terms of an effective Zeeman coupling for which we can use the noninteracting solution.

\section{Definition of $k$ modes for OBC}
\label{app:Ham_obc}

We illustrate the construction of operators $\hat{a}^{\left(\dagger\right)}_{\sigma,\nu}\!\left(k\right)$ diagonalizing the noninteracting, i.e., $U=0$, Hamiltonian \eqref{eq:Ham} for the case of \gls{OBC}.

For the excitation treated in Sec.~\ref{sec:singlephoton}, it appears convenient to employ the same procedure as outlined in \cref{app:ionic} obtaining again \cref{eq:a_ops} with adjusted tensors $T_{\sigma,j,\nu}(k)$. 
Indeed, similar calculations have been performed in Ref.~\onlinecite{PhysRevB.97.235120} yielding the operators $\hat{a}^{\left(\dagger\right)}_{\sigma,\nu}\!\left(k\right)$ for \gls{PBC}.
However, the occupation $\braket{ \hat n_k }$ of the so-obtained $k$ modes is not a constant of the motion when using OBC for the computation of the time evolution due to the different choice of boundary conditions.

Hence, to benefit from the significantly better scaling of \glspl{MPS} for systems with \gls{OBC}, we adapt the operators $\hat{a}^{\left(\dagger\right)}_{\sigma,\nu}\!\left(k\right)$ to \gls{OBC}.
In the absence of a magnetic field, i.e. $\Delta = 0$, this leads to the sine-transform $\hat{a}^{\left(\dagger\right)}_{\sigma}\!\left(k\right) = \sqrt{\frac{2}{L+1}}\sum_{j}\sin\!\left(kj\right)\hat{c}^{\left(\dagger\right)}_{\sigma,j}$ (see, e.g., Ref.~\onlinecite{PhysRevLett.92.256401}), with 
\begin{align}
\label{eq:momenta}
	k = \frac{\pi p}{L+1}, \quad p \in \left\{1, \dots , L\right\},
\end{align}
$L$ being the system size. 
This needs to be generalized to the present case with a four-site unit cell when $\Delta \neq 0$. For the sake of generality for further system types (e.g., in the presence of disorder), we do this here numerically.
We begin by noting that the noninteracting part of Hamiltonian \eqref{eq:Ham} can be expressed through 
\begin{align}
	\hat{H}_{0} = \sum_{\sigma} \sum_{i,j} H_{i,j}^{\sigma} \hat{c}^{\dagger}_{\sigma, i\vphantom{+1}}\hat{c}^{\vphantom{\dagger}}_{\sigma, j\vphantom{+1}}
\end{align}
with the Hamiltonian matrix $H^{\sigma}$, which we diagonalize via
\begin{align}
	H^{\sigma} = {P^{\sigma}} D^{\sigma} P^{\sigma\dagger} \Leftrightarrow H^{\sigma}_{i,j} = \sum_{m} P^{\sigma}_{i,m} D^{\sigma}_{m,m} P^{\sigma *}_{j,m}, \, 
\end{align} 
with $D$ being a diagonal matrix and $P$ hermitian.
Renaming $m \rightarrow k$ and $D_{m,m}^{\sigma} \rightarrow \epsilon_{\sigma}\!\left(k\right)$, and defining
\begin{align}
	\hat{a}^{\vphantom{\dagger}}_{\sigma,k} = \sum_{j}P_{j,k}^{\sigma *}\hat{c}^{\vphantom{\dagger}}_{\sigma, j\vphantom{+1}}, \quad \text{and} \quad \hat{a}^{\dagger}_{\sigma,k} = \sum_{j}P_{j,k}^{\sigma}\hat{c}^{\dagger}_{\sigma, j\vphantom{+1}}
\end{align}
we arrive at
\begin{align}
	\hat{H}_{0} = \sum_{\sigma, k} \epsilon_{\sigma}\!\left(k\right) \hat{a}^{\dagger}_{\sigma, k\vphantom{+1}}\hat{a}^{\vphantom{\dagger}}_{\sigma, k\vphantom{+1}}.
\end{align}
We note that this procedure gives the same eigenvalues $\epsilon_{\sigma}\!\left(k\right)$ as the sine transform for $\Delta = 0$.
The crucial difference is that this comes without any order of the eigenvalues, that is, \textit{a priori} it is not clear what eigenvalue $\epsilon_{\sigma}\!\left(k\right)$ belongs to what momentum $k$. 
Due to the existence of only nearest\hyp neighbor hopping in \eqref{eq:Ham}, we know that the dispersion relation (in the extended zone scheme) will be cosinelike. 
Hence, for $k \geq 0$ it is monotonously increasing allowing us to arrange the eigenvalues $\epsilon_{\sigma}\!\left(k\right)$ ascendingly and identify them as corresponding to the momenta from \eqref{eq:momenta}.

We further know that, because the unit cell of \eqref{eq:Ham} is composed of four sites, for $\Delta = 0$ the system will exhibit a four-band structure. 
Therefore, we may fold the operators back to the first Brillouin zone through
\begin{align}
\label{eq:fold_back}
	\hat{a}^{\left(\dagger\right)}_{\sigma,\nu}\!\left(k^{\prime}\right) =
	\begin{cases}
	\hat{a}^{\left(\dagger\right)}_{\sigma,k^{\prime}} ,& \nu = 1 \\
	\hat{a}^{\left(\dagger\right)}_{\sigma,\left(L/2+1\right)\pi/\left(L+1\right)-k^{\prime}} ,& \nu = 2 \\
	\hat{a}^{\left(\dagger\right)}_{\sigma,\left(L/2\right)\pi/\left(L+1\right)+k^{\prime}} ,& \nu = 3 \\
	\hat{a}^{\left(\dagger\right)}_{\sigma,\pi-k^{\prime}} ,& \nu = 4 \, ,\\
	\end{cases}
\end{align}
having introduced the band index $\nu$ and calculating the momenta $k^{\prime}$ with \eqref{eq:momenta} for $p^{\prime} \in \left\{1, \dots , L/4\right\}$.
This procedure, hence, allows us to define suitable operators for the photoexcitation treated in Sec.~\ref{sec:singlephoton}.

Note that the above defined scheme only diagonalizes Hamiltonian \eqref{eq:Ham} at $U=0$ exactly through the operators $\hat{a}^{\left(\dagger\right)}_{\sigma,\nu}\!\left(k^{\prime}\right)$. 
Nevertheless, we will use these operators for the interacting system \eqref{eq:Ham}. 
This is an approximation due to the presence of scattering between the bands induced by the finite interaction.
However, in particular for the case of quarter filling, it is to be expected that a finite interaction strength $U/\hopping > 0$ will only slightly affect the spectrum, since double occupancies -- and hence scattering processes of the electrons -- are strongly suppressed.
Hence, we expect this approximate description to be a useful modeling of the photoexcitation.

\bibliographystyle{prsty}
\bibliography{Literatur}

\begin{thebibliography}{100}

\bibitem{Zhang2017}
J. Zhang, P.~W. Hess, A. Kyprianidis, P. Becker, A. Lee, J. Smith, G. Pagano,
  I.-D. Potirniche, A.~C. Potter, A. Vishwanath, N.~Y. Yao, and C. Monroe,
  Nature {\bf 543},  217  (2017).

\bibitem{Choi2017}
S. Choi, J. Choi, R. Landig, G. Kucsko, H. Zhou, J. Isoya, F. Jelezko, S.
  Onoda, H. Sumiya, V. Khemani, C. von Keyserlingk, N.~Y. Yao, E. Demler, and
  M.~D. Lukin, Nature {\bf 543},  221  (2017).

\bibitem{PhysRevLett.116.250401}
V. Khemani, A. Lazarides, R. Moessner, and S.~L. Sondhi, Phys. Rev. Lett. {\bf
  116},  250401  (2016).

\bibitem{0034-4885-81-1-016401}
K. Sacha and J. Zakrzewski, Reports on Progress in Physics {\bf 81},  016401
  (2018).

\bibitem{PhysRevLett.117.090402}
D.~V. Else, B. Bauer, and C. Nayak, Phys. Rev. Lett. {\bf 117},  090402
  (2016).

\bibitem{PhysRevLett.118.030401}
N.~Y. Yao, A.~C. Potter, I.-D. Potirniche, and A. Vishwanath, Phys. Rev. Lett.
  {\bf 118},  030401  (2017).

\bibitem{PhysRevLett.118.269901}
N.~Y. Yao, A.~C. Potter, I.-D. Potirniche, and A. Vishwanath, Phys. Rev. Lett.
  {\bf 118},  269901  (2017).

\bibitem{PhysRevB.94.085112}
C.~W. von Keyserlingk, V. Khemani, and S.~L. Sondhi, Phys. Rev. B {\bf 94},
  085112  (2016).

\bibitem{PhysRevX.7.011026}
D.~V. Else, B. Bauer, and C. Nayak, Phys. Rev. X {\bf 7},  011026  (2017).

\bibitem{Moessner2017}
R. Moessner and S.~L. Sondhi, Nature Physics {\bf 13},  424  (2017).

\bibitem{RevModPhys.81.163}
F. Krausz and M. Ivanov, Rev. Mod. Phys. {\bf 81},  163  (2009).

\bibitem{pumpprobe7}
P. Baum, D.-S. Yang, and A.~H. Zewail, Science {\bf 318},  788  (2007).

\bibitem{pumpprobe6}
C.~W. Siders, A. Cavalleri, K. Sokolowski-Tinten, C. T{\'o}th, T. Guo, M.
  Kammler, M.~H.~v. Hoegen, K.~R. Wilson, D.~v.~d. Linde, and C.~P.~J. Barty,
  Science {\bf 286},  1340  (1999).

\bibitem{pumpprobe5}
E. Collet, M.-H. Lem{\'e}e-Cailleau, M. Buron-Le~Cointe, H. Cailleau, M. Wulff,
  T. Luty, S.-Y. Koshihara, M. Meyer, L. Toupet, P. Rabiller, and S. Techert,
  Science {\bf 300},  612  (2003).

\bibitem{PhysRevLett.118.107402}
R.~Y. Chen, S.~J. Zhang, M.~Y. Zhang, T. Dong, and N.~L. Wang, Phys. Rev. Lett.
  {\bf 118},  107402  (2017).

\bibitem{PhysRevLett.118.116402}
R. Mankowsky, B. Liu, S. Rajasekaran, H.~Y. Liu, D. Mou, X.~J. Zhou, R. Merlin,
  M. F\"orst, and A. Cavalleri, Phys. Rev. Lett. {\bf 118},  116402  (2017).

\bibitem{1367-2630-18-9-093028}
I. Avigo, S. Thirupathaiah, M. Ligges, T. Wolf, J. Fink, and U. Bovensiepen,
  New Journal of Physics {\bf 18},  093028  (2016).

\bibitem{Tao62}
Z. Tao, C. Chen, T. Szilv{\'a}si, M. Keller, M. Mavrikakis, H. Kapteyn, and M.
  Murnane, Science {\bf 353},  62  (2016).

\bibitem{Rini2007}
M. Rini, R. Tobey, N. Dean, J. Itatani, Y. Tomioka, Y. Tokura, R.~W.
  Schoenlein, and A. Cavalleri, Nature {\bf 449},  72  (2007).

\bibitem{Hu2014}
W. Hu, S. Kaiser, D. Nicoletti, C.~R. Hunt, I. Gierz, M.~C. Hoffmann, M.
  Le~Tacon, T. Loew, B. Keimer, and A. Cavalleri, Nature Materials {\bf 13},
  705  (2014).

\bibitem{Ropers1}
T. Eggebrecht, M. M\"oller, J.~G. Gatzmann, N. Rubiano~da Silva, A. Feist, U.
  Martens, H. Ulrichs, M. M\"unzenberg, C. Ropers, and S. Sch\"afer, Phys. Rev.
  Lett. {\bf 118},  097203  (2017).

\bibitem{Fausti189}
D. Fausti, R.~I. Tobey, N. Dean, S. Kaiser, A. Dienst, M.~C. Hoffmann, S. Pyon,
  T. Takayama, H. Takagi, and A. Cavalleri, Science {\bf 331},  189  (2011).

\bibitem{Mitrano2016}
M. Mitrano, A. Cantaluppi, D. Nicoletti, S. Kaiser, A. Perucchi, S. Lupi, P.
  Di~Pietro, D. Pontiroli, M. Ricc{\`o}, S.~R. Clark, D. Jaksch, and A.
  Cavalleri, Nature {\bf 530},  461  (2016).

\bibitem{doi:10.1002/9780470022184.hmm315}
P.~D. Johnson and G. G\"untherodt,  in {\em Handbook of Magnetism and Advanced
  Magnetic Materials} (Wiley, Chichester, UK, 2007), pp.\ 1635--1657.

\bibitem{ncomms10459}
L. Rettig, R. Cortés, J.-H. Chu, I.~R. Fisher, F. Schmitt, R.~G. Moore, Z.-X.
  Shen, P.~S. Kirchmann, M. Wolf, and U. Bovensiepen, Nature Communications
  {\bf 7},  10459  (2016).

\bibitem{PRLChromium}
A. Singer, S.~K.~K. Patel, R. Kukreja, V. Uhl\'{\i}\ifmmode~\check{r}\else
  \v{r}\fi{}, J. Wingert, S. Festersen, D. Zhu, J.~M. Glownia, H.~T. Lemke, S.
  Nelson, M. Kozina, K. Rossnagel, M. Bauer, B.~M. Murphy, O.~M. Magnussen,
  E.~E. Fullerton, and O.~G. Shpyrko, Phys. Rev. Lett. {\bf 117},  056401
  (2016).

\bibitem{Schmitt1649}
F. Schmitt, P.~S. Kirchmann, U. Bovensiepen, R.~G. Moore, L. Rettig, M. Krenz,
  J.-H. Chu, N. Ru, L. Perfetti, D.~H. Lu, M. Wolf, I.~R. Fisher, and Z.-X.
  Shen, Science {\bf 321},  1649  (2008).

\bibitem{doublondynamics}
M. Ligges, I. Avigo, D. Gole\ifmmode~\check{z}\else \v{z}\fi{}, H.~U.~R.
  Strand, Y. Beyazit, K. Hanff, F. Diekmann, L. Stojchevska, M. Kall\"ane, P.
  Zhou, K. Rossnagel, M. Eckstein, P. Werner, and U. Bovensiepen, Phys. Rev.
  Lett. {\bf 120},  166401  (2018).

\bibitem{Rohwer2011}
T. Rohwer, S. Hellmann, M. Wiesenmayer, C. Sohrt, A. Stange, B. Slomski, A.
  Carr, Y. Liu, L.~M. Avila, M. Kall{\"a}ne, S. Mathias, L. Kipp, K. Rossnagel,
  and M. Bauer, Nature {\bf 471},  490  (2011).

\bibitem{Hellmann2012}
S. Hellmann, T. Rohwer, M. Kall{\"a}ne, K. Hanff, C. Sohrt, A. Stange, A. Carr,
  M.~M. Murnane, H.~C. Kapteyn, L. Kipp, M. Bauer, and K. Rossnagel, Nature
  Communications {\bf 3},  1069  (2012).

\bibitem{Mathias2016}
S. Mathias, S. Eich, J. Urbancic, S. Michael, A.~V. Carr, S. Emmerich, A.
  Stange, T. Popmintchev, T. Rohwer, M. Wiesenmayer, A. Ruffing, S. Jakobs, S.
  Hellmann, P. Matyba, C. Chen, L. Kipp, M. Bauer, H.~C. Kapteyn, H.~C.
  Schneider, K. Rossnagel, M.~M. Murnane, and M. Aeschlimann, Nature
  Communications {\bf 7},  12902  (2016).

\bibitem{Stojchevska177}
L. Stojchevska, I. Vaskivskyi, T. Mertelj, P. Kusar, D. Svetin, S. Brazovskii,
  and D. Mihailovic, Science {\bf 344},  177  (2014).

\bibitem{PhysRevLett.118.247401}
C. Laulh\'e, T. Huber, G. Lantz, A. Ferrer, S.~O. Mariager, S. Gr\"ubel, J.
  Rittmann, J.~A. Johnson, V. Esposito, A. L\"ubcke, L. Huber, M. Kubli, M.
  Savoini, V.~L.~R. Jacques, L. Cario, B. Corraze, E. Janod, G. Ingold, P.
  Beaud, S.~L. Johnson, and S. Ravy, Phys. Rev. Lett. {\bf 118},  247401
  (2017).

\bibitem{Ropers2}
S. Vogelgesang, G. Storeck, J.~G. Horstmann, T. Diekmann, M. Sivis, S. Schramm,
  K. Rossnagel, S. Sch{\"a}fer, and C. Ropers, Nature Physics {\bf 14},  184
  (2017).

\bibitem{PhysRevLett.102.066404}
A. Tomeljak, H. Sch\"afer, D. St\"adter, M. Beyer, K. Biljakovic, and J.
  Demsar, Phys. Rev. Lett. {\bf 102},  066404  (2009).

\bibitem{PhysRevB.84.180507}
L. Stojchevska, P. Kusar, T. Mertelj, V.~V. Kabanov, Y. Toda, X. Yao, and D.
  Mihailovic, Phys. Rev. B {\bf 84},  180507  (2011).

\bibitem{review_pumpprobetheory}
J.~K. Freericks, O.~P. Matveev, W. Shen, A.~M. Shvaika, and T.~P. Devereaux,
  Physica Scripta {\bf 92},  034007  (2017).

\bibitem{review_trARPEStheory}
A.~F. Kemper, M.~A. Sentef, B. Moritz, T.~P. Devereaux, and J.~K. Freericks,
  Annalen der Physik {\bf 529},  1600235  (2017).

\bibitem{epjst_072013_222_5_1065-1075}
T. Tohyama, The European Physical Journal Special Topics {\bf 222},  1065
  (2013).

\bibitem{PhysRevLett.112.176404}
W. Shen, Y. Ge, A.~Y. Liu, H.~R. Krishnamurthy, T.~P. Devereaux, and J.~K.
  Freericks, Phys. Rev. Lett. {\bf 112},  176404  (2014).

\bibitem{PhysRevLett.120.246402}
Y. Wang, C.-C. Chen, B. Moritz, and T.~P. Devereaux, Phys. Rev. Lett. {\bf
  120},  246402  (2018).

\bibitem{photoinducedchiralspinliquid}
M. Claassen, H.-C. Jiang, B. Moritz, and T.~P. Devereaux, Nature Communications
  {\bf 8},  1192  (2017).

\bibitem{PhysRevB.94.035121}
D. Gole\ifmmode~\check{z}\else \v{z}\fi{}, P. Werner, and M. Eckstein, Phys.
  Rev. B {\bf 94},  035121  (2016).

\bibitem{PhysRevLett.119.247601}
Y. Murakami, D. Gole\ifmmode~\check{z}\else \v{z}\fi{}, M. Eckstein, and P.
  Werner, Phys. Rev. Lett. {\bf 119},  247601  (2017).

\bibitem{PhysRevLett.116.086401}
Y. Wang, B. Moritz, C.-C. Chen, C.~J. Jia, M. van Veenendaal, and T.~P.
  Devereaux, Phys. Rev. Lett. {\bf 116},  086401  (2016).

\bibitem{photoexcitation_ionichubbardmodel}
N. Dasari and M. Eckstein, Phys. Rev. B {\bf 98},  035113  (2018).

\bibitem{Sentef2015}
M.~A. Sentef, M. Claassen, A.~F. Kemper, B. Moritz, T. Oka, J.~K. Freericks,
  and T.~P. Devereaux, Nature Communications {\bf 6},  7047  (2015).

\bibitem{PhysRevB.98.035138}
M.~H. Kalthoff, G.~S. Uhrig, and J.~K. Freericks, Phys. Rev. B {\bf 98},
  035138  (2018).

\bibitem{PhysRevB.89.045106}
H. Schaefer, V.~V. Kabanov, and J. Demsar, Phys. Rev. B {\bf 89},  045106
  (2014).

\bibitem{PhysRevB.88.165115}
N. Tsuji and P. Werner, Phys. Rev. B {\bf 88},  165115  (2013).

\bibitem{doi:10.7566/JPSJ.86.024711}
K. Yonemitsu, Journal of the Physical Society of Japan {\bf 86},  024711
  (2017).

\bibitem{PhysRevLett.109.197401}
H. Lu, S. Sota, H. Matsueda, J. Bon\v{c}a, and T. Tohyama, Phys. Rev. Lett.
  {\bf 109},  197401  (2012).

\bibitem{sc-noneq}
S. Paeckel, B. Fauseweh, A. Osterkorn, T. K\"ohler, D. Manske, and S.~R.
  Manmana, Phys. Rev. B {\bf 101},  180507(R)  (2020).

\bibitem{oistr_orig_2}
P. Elliott, T. M{\"u}ller, J.~K. Dewhurst, S. Sharma, and E.~K.~U. Gross,
  Scientific Reports {\bf 6},  38911  (2016).

\bibitem{oistr_orig}
J.~K. Dewhurst, P. Elliott, S. Shallcross, E.~K.~U. Gross, and S. Sharma, Nano
  Letters {\bf 18},  1842  (2018), pMID: 29424230.

\bibitem{light_induced_spin_transfer}
P. Tengdin, C. Gentry, A. Blonsky, D. Zusin, M. Gerrity, L. Hellbr{\"u}ck, M.
  Hofherr, J. Shaw, Y. Kvashnin, E.~K. Delczeg-Czirjak, M. Arora, H. Nembach,
  T.~J. Silva, S. Mathias, M. Aeschlimann, H.~C. Kapteyn, D. Thonig, K.
  Koumpouras, O. Eriksson, and M.~M. Murnane, Science Advances {\bf 6},
  (2020).

\bibitem{light_induced_spin_transfer_efficiency}
D. Steil, J. Walowski, F. Gerhard, T. Kiessling, D. Ebke, A. Thomas, T. Kubota,
  M. Oogane, Y. Ando, J. Otto, A. Mann, M. Hofherr, P. Elliott, J.~K. Dewhurst,
  G. Reiss, L. Molenkamp, M. Aeschlimann, M. Cinchetti, M. M\"unzenberg, S.
  Sharma, and S. Mathias, Phys. Rev. Research {\bf 2},  023199  (2020).

\bibitem{light_induced_spin_transfer_FMalloys}
M. Hofherr, S. H{\"a}user, J.~K. Dewhurst, P. Tengdin, S. Sakshath, H.~T.
  Nembach, S.~T. Weber, J.~M. Shaw, T.~J. Silva, H.~C. Kapteyn, M. Cinchetti,
  B. Rethfeld, M.~M. Murnane, D. Steil, B. Stadtm{\"u}ller, S. Sharma, M.
  Aeschlimann, and S. Mathias, Science Advances {\bf 6},    (2020).

\bibitem{PhysRevB.97.235120}
T. K\"ohler, S. Rajpurohit, O. Schumann, S. Paeckel, F.~R.~A. Biebl, M.
  Sotoudeh, S.~C. Kramer, P.~E. Bl\"ochl, S. Kehrein, and S.~R. Manmana, Phys.
  Rev. B {\bf 97},  235120  (2018).

\bibitem{IonicHubbard}
J. Hubbard and J. Torrance, Phys. Rev. Lett. {\bf 47},  1750  (1981).

\bibitem{IonicHubbardManmana}
S.~R. Manmana, V. Meden, R.~M. Noack, and K. Sch\"onhammer, Phys. Rev. B {\bf
  70},  155115  (2004).

\bibitem{ionichubbardmodelEsslinger}
M. Messer, R. Desbuquois, T. Uehlinger, G. Jotzu, S. Huber, D. Greif, and T.
  Esslinger, Phys. Rev. Lett. {\bf 115},  115303  (2015).

\bibitem{FabianSuperlattices}
L. Stenzel, A.~L.~C. Hayward, C. Hubig, U. Schollw\"ock, and F.
  Heidrich-Meisner, Phys. Rev. A {\bf 99},  053614  (2019).

\bibitem{Bruno_moke}
P. Bruno, Y. Suzuki, and C. Chappert, Phys. Rev. B {\bf 53},  9214  (1996).

\bibitem{Peierls1933}
R.~E. Peierls, Z.Phys. {\bf 80},  763  (1933).

\bibitem{Mentink2015}
J.~H. Mentink, K. Balzer, and M. Eckstein, Nature Communications {\bf 6},  6708
   (2015).

\bibitem{PhysRevB.88.075135}
M. Eckstein and P. Werner, Phys. Rev. B {\bf 88},  075135  (2013).

\bibitem{Greiner2000}
W. Greiner, {\em Relativistic Quantum Mechanics. Wave equations}, 3 ed.
  (Springer, Berlin, Heidelberg, 2000).

\bibitem{Kirschner}
J. Kirschner, {\em Polarized Electrons at Surfaces}, Vol.~106 of {\em Springer
  Tracts in Modern Physics} (Springer, Berlin, Heidelberg, 1985).

\bibitem{ReviewSpinPolarizedPhotoemission}
P.~D. Johnson, Reports on Progress in Physics {\bf 60},  1217  (1997).

\bibitem{TrendsSpinResolvedPhotoelectronSpectroscopy}
T. Okuda, Journal of Physics: Condensed Matter {\bf 29},  483001  (2017).

\bibitem{PRLSpinResolvedStudies}
Z.-H. Zhu, C.~N. Veenstra, S. Zhdanovich, M.~P. Schneider, T. Okuda, K.
  Miyamoto, S.-Y. Zhu, H. Namatame, M. Taniguchi, M.~W. Haverkort, I.~S.
  Elfimov, and A. Damascelli, Phys. Rev. Lett. {\bf 112},  076802  (2014).

\bibitem{Bloch:2008p943}
I. Bloch, J. Dalibard, and W. Zwerger, Rev. Mod. Phys. {\bf 80},  885  (2008).

\bibitem{PhysRevB.91.104302}
F. Dorfner, L. Vidmar, C. Brockt, E. Jeckelmann, and F. Heidrich-Meisner, Phys.
  Rev. B {\bf 91},  104302  (2015).

\bibitem{Stolpp2020}
J. Stolpp, J. Herbrych, F. Dorfner, E. Dagotto, and F. Heidrich-Meisner,
  Physical Review B {\bf 101},    (2020).

\bibitem{PhysRevB.102.165155}
D. Jansen, J. Bon\ifmmode~\check{c}\else \v{c}\fi{}a, and F. Heidrich-Meisner,
  Phys. Rev. B {\bf 102},  165155  (2020).

\bibitem{10.21468/SciPostPhys.10.3.058}
T. Köhler, J. Stolpp, and S. Paeckel, SciPost Phys. {\bf 10},  58  (2021).

\bibitem{Tezuka2007}
M. Tezuka, R. Arita, and H. Aoki, Phys. Rev. B {\bf 76},  155114  (2007).

\bibitem{Hardikar2007}
R.~P. Hardikar and R.~T. Clay, Phys. Rev. B {\bf 75},  245103  (2007).

\bibitem{Nocera2014}
A. Nocera, M. Soltanieh-ha, C. Perroni, V. Cataudella, and A. Feiguin, Physical
  Review B {\bf 90},  195134  (2014).

\bibitem{HubbBook}
F.~H.~L. Essler, H. Frahm, F. G{\"o}hmann, A. Kl{\"u}mper, and V.~E. Korepin,
  {\em The One-Dimensional {H}ubbard Model} (Cambridge University Press,
  Cambridge, 2005).

\bibitem{esslinger_review}
T. Esslinger, Annual Review of Condensed Matter Physics {\bf 1},  129  (2010).

\bibitem{Jacek_2020}
J. Herbrych, G. Alvarez, A. Moreo, and E. Dagotto, Phys. Rev. B {\bf 102},
  115134  (2020).

\bibitem{ole_masterarbeit}
O. Schumann, Master's thesis, Georg-August-Universität Göttingen, 2016.

\bibitem{PAECKEL2019167998}
S. Paeckel, T. Köhler, A. Swoboda, S.~R. Manmana, U. Schollwöck, and C.
  Hubig, Annals of Physics {\bf 411},  167998  (2019).

\bibitem{PhysRevB.85.205119}
P.~E. Dargel, A. W\"ollert, A. Honecker, I.~P. McCulloch, U. Schollw\"ock, and
  T. Pruschke, Phys. Rev. B {\bf 85},  205119  (2012).

\bibitem{PhysRevB.91.165112}
M.~P. Zaletel, R.~S.~K. Mong, C. Karrasch, J.~E. Moore, and F. Pollmann, Phys.
  Rev. B {\bf 91},  165112  (2015).

\bibitem{PhysRevB.94.165116}
J. Haegeman, C. Lubich, I. Oseledets, B. Vandereycken, and F. Verstraete, Phys.
  Rev. B {\bf 94},  165116  (2016).

\bibitem{PhysRevLett.91.147902}
G. Vidal, Phys. Rev. Lett. {\bf 91},  147902  (2003).

\bibitem{PhysRevLett.93.040502}
G. Vidal, Phys. Rev. Lett. {\bf 93},  040502  (2004).

\bibitem{PhysRevLett.93.076401}
S.~R. White and A.~E. Feiguin, Phys. Rev. Lett. {\bf 93},  076401  (2004).

\bibitem{PhysRevLett.93.207204}
F. Verstraete, J.~J. Garc\'ia-Ripoll, and J.~I. Cirac, Phys. Rev. Lett. {\bf
  93},  207204  (2004).

\bibitem{PhysRevLett.107.070601}
J. Haegeman, J.~I. Cirac, T.~J. Osborne, I. Pi\ifmmode~\check{z}\else
  \v{z}\fi{}orn, H. Verschelde, and F. Verstraete, Phys. Rev. Lett. {\bf 107},
  070601  (2011).

\bibitem{Daley2004}
A.~J. Daley, C. Kollath, U. Schollw\"ock, and G. Vidal, J. Stat. Mech. {\bf
  2004},  P04005  (2004).

\bibitem{1367-2630-8-12-305}
J.~J. García-Ripoll, New Journal of Physics {\bf 8},  305  (2006).

\bibitem{SciPostPhys.3.5.035}
S. Paeckel, T. {K{\"o}hler}, and S.~R. Manmana, SciPost Phys. {\bf 3},  035
  (2017).

\bibitem{giamarchi}
T. Giamarchi, {\em Quantum Physics in One Dimension}, Vol.~121 of {\em
  International Series of Monographs on Physics} (Oxford University Press,
  Oxford, 2004).

\bibitem{Bedurftig:1998p457}
G. Bed{\"u}rftig, B. Brendel, H. Frahm, and R.~M. Noack, Phys. Rev. B {\bf 58},
   10225  (1998).

\bibitem{modulationspectroscopy1}
C. Kollath, A. Iucci, I.~P. McCulloch, and T. Giamarchi, Phys. Rev. A {\bf 74},
   041604  (2006).

\bibitem{modulationspectroscopy2}
D. Greif, L. Tarruell, T. Uehlinger, R. J\"ordens, and T. Esslinger, Phys. Rev.
  Lett. {\bf 106},  145302  (2011).

\bibitem{modulationspectroscopy3}
N. Strohmaier, D. Greif, R. J\"ordens, L. Tarruell, H. Moritz, T. Esslinger, R.
  Sensarma, D. Pekker, E. Altman, and E. Demler, Phys. Rev. Lett. {\bf 104},
  080401  (2010).

\bibitem{modulationspectroscopy4}
R. Sensarma, D. Pekker, E. Altman, E. Demler, N. Strohmaier, D. Greif, R.
  J\"ordens, L. Tarruell, H. Moritz, and T. Esslinger, Phys. Rev. B {\bf 82},
  224302  (2010).

\bibitem{superlattice_orig1}
S. Peil, J.~V. Porto, B.~L. Tolra, J.~M. Obrecht, B.~E. King, M. Subbotin,
  S.~L. Rolston, and W.~D. Phillips, Phys. Rev. A {\bf 67},  051603  (2003).

\bibitem{superlattice_orig2}
J. Sebby-Strabley, M. Anderlini, P.~S. Jessen, and J.~V. Porto, Phys. Rev. A
  {\bf 73},  033605  (2006).

\bibitem{S.Trotzky01182008}
S. Trotzky, P. Cheinet, S. Folling, M. Feld, U. Schnorrberger, A.~M. Rey, A.
  Polkovnikov, E.~A. Demler, M.~D. Lukin, and I. Bloch, Science {\bf 319},  295
   (2008).

\bibitem{Anderlini:2007p1571}
M. Anderlini, P.~J. Lee, B.~L. Brown, J. Sebby-Strabley, W.~D. Phillips, and
  J.~V. Porto, Nature {\bf 448},  452  (2007).

\bibitem{superlattice_magneticfield}
M. Aidelsburger, M. Atala, S. Nascimb{\`e}ne, S. Trotzky, Y.-A. Chen, and I.
  Bloch, Applied Physics B {\bf 113},  1  (2013).

\bibitem{superlattice_magneticfield_PRL}
M. Aidelsburger, M. Atala, S. Nascimb\`ene, S. Trotzky, Y.-A. Chen, and I.
  Bloch, Phys. Rev. Lett. {\bf 107},  255301  (2011).

\bibitem{SpinDependentOptLat}
B. Yang, H.-N. Dai, H. Sun, A. Reingruber, Z.-S. Yuan, and J.-W. Pan, Phys.
  Rev. A {\bf 96},  011602  (2017).

\bibitem{ThoulessPumpSuperlattice}
M. Lohse, C. Schweizer, O. Zilberberg, M. Aidelsburger, and I. Bloch, Nature
  Physics {\bf 12},  350  (2015), article.

\bibitem{SpinPumpSuperlattice}
C. Schweizer, M. Lohse, R. Citro, and I. Bloch, Phys. Rev. Lett. {\bf 117},
  170405  (2016).

\bibitem{chargetransfersalts_Becca}
R. Kaneko, L.~F. Tocchio, R. Valentí, and F. Becca, New Journal of Physics
  {\bf 19},  103033  (2017).

\bibitem{cupratesCDW1}
J. Chang, E. Blackburn, A.~T. Holmes, N.~B. Christensen, J. Larsen, J. Mesot,
  R. Liang, D.~A. Bonn, W.~N. Hardy, A. Watenphul, M.~v. Zimmermann, E.~M.
  Forgan, and S.~M. Hayden, Nature Physics {\bf 8},  871  (2012).

\bibitem{cupratesCDW2}
G. Ghiringhelli, M. Le~Tacon, M. Minola, S. Blanco-Canosa, C. Mazzoli, N.~B.
  Brookes, G.~M. De~Luca, A. Frano, D.~G. Hawthorn, F. He, T. Loew, M.~M. Sala,
  D.~C. Peets, M. Salluzzo, E. Schierle, R. Sutarto, G.~A. Sawatzky, E.
  Weschke, B. Keimer, and L. Braicovich, Science {\bf 337},  821  (2012).

\bibitem{cupratesCDW3}
A.~J. Achkar, R. Sutarto, X. Mao, F. He, A. Frano, S. Blanco-Canosa, M.
  Le~Tacon, G. Ghiringhelli, L. Braicovich, M. Minola, M. Moretti~Sala, C.
  Mazzoli, R. Liang, D.~A. Bonn, W.~N. Hardy, B. Keimer, G.~A. Sawatzky, and
  D.~G. Hawthorn, Phys. Rev. Lett. {\bf 109},  167001  (2012).

\bibitem{cupratesCDW4}
M. H\"ucker, N.~B. Christensen, A.~T. Holmes, E. Blackburn, E.~M. Forgan, R.
  Liang, D.~A. Bonn, W.~N. Hardy, O. Gutowski, M.~v. Zimmermann, S.~M. Hayden,
  and J. Chang, Phys. Rev. B {\bf 90},  054514  (2014).

\bibitem{cupratesCDW5}
T. Wu, H. Mayaffre, S. Kr{\"a}mer, M. Horvati{\v c}, C. Berthier, W.~N. Hardy,
  R. Liang, D.~A. Bonn, and M.-H. Julien, Nature {\bf 477},  191  (2011).

\bibitem{cupratesCDW6}
S. Gerber, H. Jang, H. Nojiri, S. Matsuzawa, H. Yasumura, D.~A. Bonn, R. Liang,
  W.~N. Hardy, Z. Islam, A. Mehta, S. Song, M. Sikorski, D. Stefanescu, Y.
  Feng, S.~A. Kivelson, T.~P. Devereaux, Z.-X. Shen, C.-C. Kao, W.-S. Lee, D.
  Zhu, and J.-S. Lee, Science {\bf 350},  949  (2015).

\bibitem{cuprates_intertwinnedorder1}
M.~H. {Hamidian}, S.~D. {Edkins}, K. {Fujita}, A. {Kostin}, A.~P. {Mackenzie},
  H. {Eisaki}, S. {Uchida}, M.~J. {Lawler}, E.-A. {Kim}, S. {Sachdev}, and
  J.~C. {S{\'e}amus Davis}, ArXiv e-prints 1508.00620  (2015).

\bibitem{Edkins976}
S.~D. Edkins, A. Kostin, K. Fujita, A.~P. Mackenzie, H. Eisaki, S. Uchida, S.
  Sachdev, M.~J. Lawler, E.-A. Kim, J.~C. S{\'e}amus~Davis, and M.~H. Hamidian,
  Science {\bf 364},  976  (2019).

\bibitem{cuprates_CDW_science}
E.~H. da~Silva~Neto, P. Aynajian, A. Frano, R. Comin, E. Schierle, E. Weschke,
  A. Gyenis, J. Wen, J. Schneeloch, Z. Xu, S. Ono, G. Gu, M. Le~Tacon, and A.
  Yazdani, Science {\bf 343},  393  (2014).

\bibitem{condmat2030026}
G. Campi, A. Ricci, N. Poccia, M. Fratini, and A. Bianconi, Condensed Matter
  {\bf 2},    (2017).

\bibitem{Bakr:2009p2641}
W.~S. Bakr, J.~I. Gillen, A. Peng, S. F{\"o}lling, and M. Greiner, Nature {\bf
  462},  74  (2009).

\bibitem{PhysRevLett.114.213002}
M.~F. Parsons, F. Huber, A. Mazurenko, C.~S. Chiu, W. Setiawan, K.
  Wooley-Brown, S. Blatt, and M. Greiner, Phys. Rev. Lett. {\bf 114},  213002
  (2015).

\bibitem{Cheuk2015}
L.~W. Cheuk, M.~A. Nichols, M. Okan, T. Gersdorf, V.~V. Ramasesh, W.~S. Bakr,
  T. Lompe, and M.~W. Zwierlein, Phys. Rev. Lett. {\bf 114},  193001  (2015).

\bibitem{quantumgasmicroscope_BlochPRL}
A. Omran, M. Boll, T.~A. Hilker, K. Kleinlein, G. Salomon, I. Bloch, and C.
  Gross, Phys. Rev. Lett. {\bf 115},  263001  (2015).

\bibitem{quantumgasmicroscope_review}
C. Gross and I. Bloch, Science {\bf 357},  995  (2017).

\bibitem{quantumgasmicroscope_spinresolved}
M. Boll, T.~A. Hilker, G. Salomon, A. Omran, J. Nespolo, L. Pollet, I. Bloch,
  and C. Gross, Science {\bf 353},  1257  (2016).

\bibitem{Gorshkov:2010p1052NO}
A.~V. Gorshkov, M. Hermele, V. Gurarie, C. Xu, P.~S. Julienne, J. Ye, P.
  Zoller, E. Demler, M.~D. Lukin, and A.~M. Rey, Nature Physics {\bf 6},  289
  (2010).

\bibitem{symmps}
S. Paeckel and T. Köhler, SymMPS, \url{https://www.symmps.eu}, accessed:
  2019-12-29.

\bibitem{PhysRevLett.92.256401}
H. Benthien, F. Gebhard, and E. Jeckelmann, Phys. Rev. Lett. {\bf 92},  256401
  (2004).

\end{thebibliography}

\end{document}